\documentclass[fleqn,usenatbib]{rasti}

\usepackage{newtxtext,newtxmath}
\usepackage{orcidlink}
\usepackage{bm}
\usepackage{float}
\usepackage{soul}

\usepackage[T1]{fontenc}
\usepackage{balance}
\usepackage{booktabs}

\usepackage{tikz}
\usetikzlibrary{shapes.geometric, arrows.meta, positioning, fit, backgrounds, calc}

\usepackage{comment}
\usepackage{scalerel}
\usepackage{algorithm}     
\usepackage{algpseudocode} 
\DeclareRobustCommand{\VAN}[3]{#2}
\let\VANthebibliography\thebibliography
\def\thebibliography{\DeclareRobustCommand{\VAN}[3]{##3}\VANthebibliography}

\usepackage{graphicx}	
\usepackage{float} 
\usepackage{subcaption} 

\usepackage{amsmath}	
\newcommand{\minus}{-}

\newcommand*\diff{\mathop{}\!\mathrm{d}}

\DeclareMathOperator{\Tr}{Tr}

\newcommand{\Tsys}{T_{\rm sys}}
\newcommand{\wn}{\hat{\mathit{w}} }
\newcommand{\LscaleG}{g}
\newcommand{\deltaG}{\hat{\epsilon}}

\newcommand*\NCov[1]{\mathop{}\!\mathbf{N}_#1}
\newcommand{\Ncov}{\mathbf{N}}

\newcommand{\logposterior}{\mathcal{P}_{\rm post}}
\newcommand{\logprior}{\mathcal{P}_{\rm prior}}
\newcommand{\loglikeli}{\mathcal{L}}

\newcommand{\be}{\begin{equation}}
\newcommand{\ee}{\end{equation}}
\newcommand{\bea}{\begin{align}}
\newcommand{\eea}{\end{align}}

\title[Joint Calibration and Map-Making]
{Joint Bayesian calibration and map-making for intensity mapping experiments}

\author[Z. Zhang et al.]{
Zheng Zhang\orcidlink{0000-0002-9154-2803}$^{1}$
\thanks{E-mail: zheng.zhang@manchester.ac.uk (ZZ)},
Philip Bull\orcidlink{0000-0001-5668-3101}$^{1,2}$,
Mario G. Santos$^{2}$,
and 
Ainulnabilah Nasirudin$^{1}$ \orcidlink{0000-0003-2213-4547}
\\
$^{1}$Jodrell Bank Centre for Astrophysics, University of Manchester, Manchester, M13 9PL, United Kingdom\\
$^{2}$Department of Physics and Astronomy, University of Western Cape, Cape Town 7535, South Africa\\
}

\date{Accepted XXX. Received YYY; in original form ZZZ}

\pubyear{\the\year{}}

\begin{document}
\label{firstpage}
\pagerange{\pageref{firstpage}--\pageref{lastpage}}
\maketitle

\begin{abstract}
    Line-intensity mapping (LIM) is an emerging cosmological technique that traces large-scale structure through the integrated spectral-line emission of unresolved sources. Reconstructing unbiased sky maps requires careful joint treatment of instrumental calibration and map-making, a task made challenging by time-varying receiver gains, thermal drifts, and correlated $1/f$ noise intrinsic to single-dish radio telescopes. We present a Bayesian framework for joint calibration and map-making using Gibbs sampling, giving access to the full joint posterior of calibration and sky map parameters. Our data model is grounded in the radiometer equation, capturing the coupling between noise level and system temperature without assuming a fixed noise amplitude. Gain and system temperature are estimated via an iterative generalised least squares (GLS) scheme, while absolute flux calibration is achieved either with external calibrators or via known signal injections such as noise diodes. We further introduce a $1/f$ noise model that avoids spurious periodic correlations arising from the common assumption of a diagonally structured noise covariance in the frequency domain. The workflow is implemented in an efficient software package using the Levinson algorithm and a polynomial emulator to reduce computational cost. Demonstrated on simulations representative of MeerKLASS single-dish observations, the framework generalises to other single-dish surveys and to cross-correlation and interferometric data.
\end{abstract}




\section{Introduction}

Line-intensity mapping (LIM) is a powerful technique for probing the large-scale structure of the Universe. It involves measuring the integrated emission from unresolved sources across cosmic volumes. 
By targeting specific spectral lines, such as the 21\,cm hyperfine transition of neutral hydrogen \citep{2010Natur.466..463C,pritchard201221, bull2015late} or molecular lines like CO \citep{righi2008carbon}, LIM can efficiently map the three-dimensional distribution of matter over a wide range of redshifts, with access to larger volumes and unresolved faint galaxy populations that are missed by galaxy surveys.

Calibration and map-making are foundational steps of data processing in intensity mapping experiments, enabling the translation of raw detector signals into scientifically meaningful sky maps. 
Traditional approaches often decouple these processes, first calibrating the instrument's gain and noise properties before reconstructing the sky map using linear or maximum-likelihood solvers \citep[e.g.][]{tegmark1997make, sutton2010fastDescarte,dore2001mapcumba, keihanen2005madam}.
However, this sequential treatment can propagate calibration errors into the final map and neglect important interdependencies between gain variations, system temperature fluctuations, and noise statistics. 
These issues are especially pronounced in low-frequency radio experiments, where $1/f$ noise, thermal drifts, and other instrumental systematics can significantly impact data quality and bias sky reconstruction \citep{bigot2015simulations, harper2018impact, 2020MNRAS.491.4254C, hu20211, matshawule2021h, li2021h, irfan2024mitigating}.

Bayesian methods allow for the joint, global analysis of LIM data, rigorously propagating uncertainties and dependencies while incorporating prior knowledge. This approach yields exact posterior distributions and enables straightforward marginalisation over nuisance parameters. 
Recent advances in hierarchical Bayesian methods, particularly ultra-high-dimensional implementations of Gibbs sampling, have enabled joint estimation of calibration parameters and sky signals by iteratively sampling from their conditional posteriors \citep[e.g.][]{wandelt2004global}. This approach has been successfully applied to Cosmic Microwave Background (CMB) datasets through pipelines such as 
{\tt BeyondPlanck} and {\tt Cosmoglobe}, where calibration, noise modelling, and component separation are treated in a unified framework \citep{galloway2023beyondplanck, eskilt2023cosmoglobe}. These developments offer a promising path for extending similar techniques to 21\,cm cosmology and other intensity mapping experiments.

However, replicating the success of joint Bayesian analysis in intensity mapping experiments is far from straightforward. There are two main reasons for this. First, compared with the CMB, the foreground components in intensity mapping are significantly brighter than the target signal \citep[see e.g.][]{spinelli2022skao}. Within the required dynamic range, these foregrounds cannot be reliably captured by simple empirical models, making standard component separation techniques from the CMB context less directly applicable. Second, intensity mapping experiments often contend with poorly understood or experiment-specific systematics, such as instrument-induced spectral structures \citep{2015ApJ...804...14T, 2019ApJ...884..105K, cunnington202121, 2021MNRAS.508.5556M, 2024MNRAS.534.2653M, 2024MNRAS.534.3349C}, calibration drifts \citep{harper2018impact, 2020MNRAS.491.4254C, li2021h, irfan2024mitigating}, and environmental effects like groundspill \citep{2021MNRAS.505.3698W} and radio-frequency interference \citep{2015PASA...32....8O, 2018MNRAS.479.2024H, 2022MNRAS.510.5023W, 2025MNRAS.536.1035E}.

Given the complexity of these challenges, a fully end-to-end Bayesian framework would be rather complex and unwieldy, with many model components. Instead, a modular Bayesian pipeline -- dividing the problem into stages -- may offer a more tractable solution for the time being. In what follows, we specialise to autocorrelation or `single-dish' intensity mapping experiments, which measure the total power from their receivers and use the resulting time-ordered data to make three-dimensional maps of line brightness temperature. Recent examples of such experiments include Green Bank Telescope (GBT; \cite{2010Natur.466..463C, 2013ApJ...763L..20M}), Parkes Radio Telescope \citep{2018MNRAS.476.3382A}, and Five-hundred-meter Aperture Spherical Telescope (FAST; \cite{2023ApJ...954..139L}) single-dish observations with the 21cm line, the MeerKAT Large Area Synoptic Survey (MeerKLASS; \cite{2016mks..confE..32S, 2021MNRAS.505.3698W}), also with the 21\,cm line, and the CO Mapping Array Project (COMAP) with CO lines \citep{2022ApJ...933..184F}. The data reduction and analysis steps for these types of experiments can be split into two major stages:
(A) joint calibration and map-making from time-ordered data to a three-dimensional sky map, and
(B) component separation and scientific inference from the resulting 3D map.
The present work focuses on the implementation and demonstration of the feasibility of a fully Bayesian approach to stage (A).
A complementary Bayesian approach to (B) is presented in G.~Murphy et al. ({\it in prep.})

A concrete and well-characterised example of stage~(A) is
the calibration and map-making pipeline developed for
MeerKLASS \citep{santos2017meerklass, wang2021h}.
MeerKLASS is the MeerKAT Large Area Synoptic Survey, which
uses the 64-dish MeerKAT array in single-dish mode to map the redshifted 21\,cm signal across the southern sky.
The pipeline of \citet{wang2021h}, described in detail in
Section~\ref{sec:conventional}, operates in two sequential
steps.
A brief tracking observation of a bright, compact point
source (e.g.\ 3C\,273 or Pictor~A) before and after each
field scan establishes the absolute flux scale and bandpass
shape, and calibrates the power of a periodic noise diode
injection.
During the $\sim\!90$-min field scan itself, the noise diode
fires every $\sim\!20$\,s into each receiver, providing a
regular internal gain reference that stabilises the
calibration against slow receiver drifts.
The time-varying gain is modelled as a fourth-order Legendre
polynomial in time, and the sky and instrument temperature
components are estimated jointly via a Bayesian
maximum-\textit{a-posteriori} (MAP) estimator.
This pipeline has demonstrated the feasibility of single-dish HI intensity mapping with MeerKAT, and enabling
the first cross-correlation detections of the HI~intensity
signal \citep{cunnington2023h}.
More recently, the MeerKLASS deep-field campaign
\citep{Barberi2025} has increased the observation time per dish to 62~hours, reaching a thermal noise floor of approximately $1.21$~mK.  
This brings the subtle calibration systematics, which are related to the thermal noise level, to the forefront.
The present work is motivated precisely by these challenges. 
Based on the \cite{wang2021h} framework, we embed the generalised time-ordered data model within a fully Bayesian Gibbs sampling scheme. This allows us to address the limitations of the conventional pipeline in this higher sensitivity regime.
Although the implementation and demonstration are tailored to MeerKLASS-type observations, the methodology can be readily generalised to other single-dish and interferometric experiments.

In designing our model and workflow, we identified two common pain points in standard calibration practices.
First, noise modelling is often oversimplified: many existing frameworks either assume a fixed noise level during calibration or neglect the coupling between noise and system parameters. This contradicts the radiometer equation, which predicts that the noise variance scales multiplicatively with both the temperature of the system and the gain \citep{burke2019introduction}. Such simplifications can lead to biased parameter estimates \citep{winkel2012unbiased}, especially in high-dynamic-range observations.
Second, a major challenge in auto-correlation time-ordered data (TOD) is disentangling stochastic gain fluctuations, commonly referred to as $1/f$ noise, from the underlying sky signal. Unlike cross-correlation systems (interferometers), where correlated noise between detectors can be mitigated, auto-correlation measurements are inherently more sensitive to gain instabilities, as they depend on the squared response of a single detector. These stochastic gain variations couple multiplicatively with both system temperature and sky signal, introducing correlated noise and spectral leakage that contaminate both calibration and map reconstruction.
Conventional methods either ignore these effects or approximate $1/f$ noise in the Fourier domain with diagonal covariance matrices \citep{bigot2015simulations, harper2018impact, li2021h, ihle2023beyondplanck}, a simplification that introduces unphysical periodic correlations in the time domain. 
Such approaches lead to a barrier to understanding the bias in system temperature estimates and the propagation of uncertainties.

In this paper, we present a unified framework for joint calibration and map-making that rigorously addresses these limitations. 
At its core is a data model incorporating multiplicative noise derived directly from the radiometer equation, which self-consistently captures the interplay between time-varying gain, system temperature, and noise amplitude. 
Crucially, our formalism explicitly models stochastic gain variations with the time-domain parameters to be sampled alongside the smooth gain variation and system temperature. 
We avoid the common but flawed assumption of diagonal noise covariance in the Fourier domain, which artificially imposes periodic correlations [See \cite{tegmark1997cmb} for a detailed discussion]. Instead, we derive a time-domain correlation model for $1/f$ noise that preserves realistic long-range temporal correlations without spectral leakage or edge artifacts.

The global joint Bayesian workflow involves exploring a high-dimensional posterior distribution arising from the simultaneous inference of sky signal parameters and a large set of instrumental parameters, such as per-antenna and per-time-chunk gain values, as well as system temperature values. This high dimensionality poses significant computational and algorithmic challenges. To enable efficient sampling of the high-dimensional posterior distributions, we develop an iterative generalised least-squares (GLS) sampling method. 
This approach accelerates convergence by iteratively refining estimates of the effective additive noise covariance, which allows a fast linear sampling of smooth gain and system temperature parameters even for the multiplicative noise model. 
By treating stochastic gain as a correlated noise process, our framework not only mitigates $1/f$ contamination but also quantifies its uncertainty, ensuring robust error propagation into the final map. Computational scalability is achieved through the Levinson algorithm \citep{levinson1949wiener}, which reduces the complexity of sampling noise parameters from $\mathcal{O}(N^3)$ to $\mathcal{O}(N^2)$, ensuring scalability for large auto-correlation datasets.

Absolute flux scale calibration can be implemented as strong priors on certain sky pixels in the workflow.
While demonstrated here for auto-correlation systems, the framework is readily adaptable to cross-correlation or interferometric measurements by neglecting noise correlations and incorporating additional multiplicative gain terms.

This work bridges a critical gap in auto-correlation data processing, where the combined effects of multiplicative noise and stochastic gain variations have historically limited calibration accuracy and map fidelity. By unifying the treatment of these effects within a statistically rigorous Gibbs sampling framework, we enable high-precision recovery of both instrument parameters and sky signals.
To demonstrate this generic Bayesian workflow, we apply the pipeline to a simulated MeerKLASS-type survey \citep{santos2017meerklass}.

The remainder of this paper is structured as follows: Section~\ref{sec: model} presents the models of the data and each of its components and the intrinsic degeneracy of the model; Section~\ref{sec: framework} details the Gibbs sampling framework, including the noise parameter sampling and the iterative GLS method for sampling the system temperature and smooth gain parameters; Section~\ref{sec: example} validates the framework through simulations, presents and discusses the map results; and Section~\ref{sec: conclusion} concludes with broader implications for radio astronomy and future extensions.

\section{Statistical model}
\label{sec: model}

\subsection{Basic Model: The Radiometer Equation and Gain Variation}
Unlike conventional methods that employ either a given thermal noise level or treat it as an independent parameter, our data model directly incorporates the radiometer equation, representing thermal noise as a multiplicative term rather than as an additive form.

\paragraph*{The Radiometer Equation.} We adopt a per-frequency data model for time-ordered auto-correlation data (TOD), in order to avoid imposing an explicit model for frequency correlations.
Although frequency correlations (of $1/f$ noise) are known to exist, they are often not accurately characterised by simple analytical models \citep{keshner19821}. 
On the other hand, including frequency correlations can greatly improve constraints on the noise model parameters as data from many frequency channels can be combined to constrain a small number of noise model parameters. We do not consider frequency correlations in what follows.
The basic model is
\be
    \label{eq: basic model}
    d(t_a) = 
    G(t_a) \Tsys(t_a)
    \left(1 + \wn \right),
\ee
where $a=0, \dots, n-1$ is the time index of the TOD with  $n$ the number of data points, $G$ is the time-dependent gain, $\Tsys$ is the system temperature,
and $\wn$ is the stochastic thermal noise fluctuation, which is typically Gaussian white noise with distribution
\begin{align}
    \label{eq: white noise}
    \wn &\sim \mathcal{N}\left(0, \NCov{w}\right),
    &
    \NCov{w}&=
    \sigma_w^2 \mathbf{I},
\end{align}
where $\mathbf{I}$ is an identity matrix of size $N$ and
\be
\sigma_w=\frac{1}{\sqrt{\mathcal{T} \Delta\nu} }
\label{eq: radiometer}
\ee
with $\Delta\nu$ the frequency channel width and $\mathcal{T}$ the integration time.
This is known as the radiometer equation \citep[see e.g.][]{burke2019introduction}.

\paragraph*{Gain fluctuations.}
The time-dependent gain,  $G(t_a)$, accounts for all gain contributions, including stochastic variations, and is modeled as follows:
\be
\label{eq: full gain model}
    G(t_a) = \LscaleG(t_a)(1 + \deltaG),
\ee
where $\LscaleG(t_a)$ represents the large timescale evolution of the gain, and
$\deltaG$ denotes the stochastic zero-mean variations. 
In most cases, the stochastic gain variations $\deltaG$ are modelled as $1/f$ noise (sometimes called flicker noise), while $\LscaleG$ is treated as a fairly smooth function to account for time-dependent gain variations that cannot be captured by the $1/f$ model. However, it is important to note that these noise models are usually only an empirical description of the noise power spectral density (PSD), far from being a complete stochastic model. 
To obtain complete noise statistics for data analysis, it is common practice to assume both Gaussianity and stationarity of the $1/f$ noise \citep{bigot2015simulations, harper2018impact, li2021h, ihle2023beyondplanck}. 
(Caveat: see \cite{ninness2002estimation} for a theoretical review, which shows that no general conclusion can be drawn regarding the Gaussianity of electronic $1/f$ noise.)
We also adopt these assumptions, although we stress that they must be implemented carefully to avoid introducing spurious noise structures. In Section~\ref{sec: flicker model} we elaborate on this point and present a proper $1/f$ model.

\paragraph*{On the form of gain variation.}
The classical $1/f$ noise model is a phenomenological (though physically motivated) law that is widely observed in electronic systems \citep{keshner19821}: in its conventional form, $1/f$ noise is understood to describe gain fluctuations across all timescales, and the full time-dependent gain is expressed as a constant reference gain multiplied by a fractional gain variation. In this formulation, modelling the effect as an additive $1/f$ noise process is formally equivalent.

However, in practical analyses [e.g. \cite{wang2021h}], this model is often generalised to the form given in Eq.~(\ref{eq: full gain model}), in which the reference gain itself is allowed to vary slowly in time. This generalisation relaxes the assumption that the true gain evolution must strictly follow a stationary empirical $1/f$ process, and instead permits more realistic departures from the idealised model.

Conceptually, the necessity of this generalisation can be understood from the following two perspectives:
\begin{enumerate}
    \item The power-law structure may be a good approximation at certain scales, but may not be consistent across all scales. For example, suppose a gain system consists of three-stage amplifiers in tandem, each with perfect $1/f$ noise. Then the total gain as the product of the three could appear as a deviated power-law PSD. See Appendix~\ref{appendix: a gain model} for a detailed discussion. 

    \item As the time scale increases, the corresponding number of samples decreases, leading to greater sampling variance and greater uncertainty in estimating the statistical structure of the PSD at large scales. As a result, even for perfect $1/f$ noise, the estimated power spectrum may deviate from power-law behaviour at lower temporal frequencies. 
    Conversely, the sparse sampling of the PSD at low Fourier frequencies makes reliably inferring the true spectral behaviour (if it is scale-dependent) difficult.
\end{enumerate}

\subsection{Conventional Calibration and Map-making}
\label{sec:conventional}

The data model of Equations~(\ref{eq: basic model})--(\ref{eq: full gain model}) 
forms the foundation of
the calibration and map-making pipeline used by MeerKLASS
\citep{wang2021h}, which serves as our primary reference
throughout this work.
In this section we briefly describe the key model choices and
the calibration and map-making workflow of \citet{wang2021h}.

\paragraph*{Two-step calibration strategy.}
Each MeerKLASS observation consists of a $\sim\!90$-min scan
at fixed elevation, during which the dishes sweep the survey
field back and forth in azimuth at
$5\,\mathrm{arcmin\,s}^{-1}$ \citep{wang2021h}.
Before and after the scan, a $\sim\!15$-min tracking
observation of a bright, compact calibrator source is
performed.
These tracking data serve a dual purpose: they constrain the
absolute flux scale and the frequency-dependent bandpass
shape of each dish, and they calibrate the noise diode
injection power.
The pointing locations used during the track are chosen to
sample the beam at the source position and at several offset
positions, providing sufficient information to separate the
point-source signal, the diffuse sky model, and the
instrumental terms.
During the scan itself, no bright calibrator source is in the
field of view, so the absolute calibration obtained from the
tracking step is propagated into the scan using the noise
diode as an internal transfer standard.

\paragraph*{Noise diode as an internal gain reference.}
Each MeerKAT receiver is equipped with a noise diode that
periodically injects a broadband signal of known temperature
$T_\mathrm{nd}(\nu)$ into the signal path
\citep{wang2021h, Barberi2025}.
In the MeerKLASS pilot survey \citep{wang2021h}, the diode
fires for 1.8\,s once every 20\,s.
The calibrated injected temperature, from the tracking step, enters the system temperature as an additive component
that switches on and off according to the known injection
pattern.
Within each injection cycle, comparing the diode-on and
diode-off time samples constrains the instantaneous gain to
better than $\sim\!1\%$, anchoring the gain solution at
regular $\sim\!20$\,s intervals and suppressing slow
receiver gain drifts over the full scan duration.
Because the noise diode amplitude is known from tracking
and the injection timing is recorded, its contribution can
be modelled deterministically, with only the fraction of the reference temperature treated as a free parameter
\citep{wang2021h}.

\paragraph*{Smooth gain model.}
Between noise diode injections, the smooth gain component
$g(t_a)$ in Eq.~(\ref{eq: full gain model}) is assumed to evolve slowly.
\citet{wang2021h} parameterises it as a
low-order Legendre polynomial in time,
\begin{equation}
    g(t_a) = \sum_{n=0}^{N_{\rm order}} a_n\, P_n(x_a),
    \qquad
    x_a = \frac{2t_a - t_\mathrm{min} - t_\mathrm{max}}
               {t_\mathrm{max} - t_\mathrm{min}}
    \in [-1,\,1],
    \label{eq:gain_legendre}
\end{equation}
where $P_n$ is the Legendre polynomial of degree $n$,
$t_\mathrm{min}$ and $t_\mathrm{max}$ are the start and end
times of the scan, and $\{a_n\}$ are the free coefficients.
The Legendre basis is preferred over an ordinary monomial
basis because the orthogonality of $\{P_n\}$ on $[-1,1]$
ensures that the coefficients $\{a_n\}$ remain approximately
uncorrelated for uniformly sampled data. This makes the
constrained linear fit better conditioned numerically
\citep{wang2021h}.
The polynomial order is chosen to be the minimum
sufficient to represent the observed gain variations without
overfitting.
In the MeerKLASS context, \citet{wang2021h} verified
empirically that $N = 4$ is adequate for a single scan:
higher orders absorb noise fluctuations rather than signal,
yielding unstable coefficients.
This is physically consistent with the MeerKAT receiver
stability: the dominant gain drifts occur on timescales
comparable to or longer than the $\sim\!200$\,s azimuth
stripe duration.
This means that four polynomial modes are sufficient to track the variation across the full 90~min scan, without introducing unnecessary model complexity.

\paragraph*{System temperature components.}
The system temperature, $T_\mathrm{sys},$ comprises several components, which are also treated independently.
The receiver temperature, $T_\mathrm{rec}$, which encompasses the thermal noise of the receiver electronics, is assumed to be smooth over time and is modelled using a third-order Legendre polynomial \citep{wang2021h}.
A Gaussian prior centred on a laboratory measurement of $T_\mathrm{rec}(\nu)$ with a standard deviation of 
$0.5\,T^\mathrm{ref}_\mathrm{rec}(\nu)$ is used to regularise the fit \citep{wang2021h}.
The elevation-dependent terrestrial emission
$T_\mathrm{el}$, including atmospheric emission and
ground spill, is approximately constant during a
fixed-elevation MeerKLASS scan, and can be represented as a
constant or a very low-order polynomial in time.
The sky contribution is either
fixed to a known template (e.g. a global sky model for the
diffuse component) during calibration, or constrained by the
absolute flux calibration against the tracking source.
All these components are fitted jointly with the gain in a
single Bayesian MAP optimisation
\citep{wang2021h}.\footnote{During the tracking step, the calibrator flux model and beam model provide additional constraints, whereas the noise diode injections are the primary source of relative calibration during the scan.}

\paragraph*{Conventional treatment of $1/f$ noise and
map-making.}
In the conventional approach, and in the MeerKLASS pipeline
specifically, the $1/f$ noise is not explicitly
modelled during calibration; instead, the noise level during
the MAP fit is set to the white-noise floor
$\sigma_\mathrm{w} = (\mathcal{T}\,\Delta\nu)^{-1/2}$, assuming that the $1/f$ knee lies below the noise
diode cadence frequency \citep{wang2021h}.
After gain calibration, the sky temperature at each time sample is estimated by subtracting the fitted instrument components from the calibrated data and allocating the result to a sky pixel,
\begin{equation}
    T_\mathrm{sky}(t_a, \nu)
    =
    \frac{d(t_a,\nu)}{g(t_a,\nu)}
    - T_\mathrm{el}(t_a,\nu)
    - T_\mathrm{nd}(t_a,\nu)
    - T_\mathrm{rec}(t_a,\nu).
    \label{eq:tsky_conventional}
\end{equation}
The sky map $\hat{T}_\mathrm{sky}(\hat{n},\nu)$ is then
formed by a weighted average of all time samples falling
within each pixel, using a Zenith Equal Area (ZEA)
projection at $0.3^\circ$ pixel scale \citep{wang2021h}.
We note that, at this map-making step, the calibrated parameters $\{g, T_\mathrm{el},
T_\mathrm{nd}, T_\mathrm{rec}\}$ are treated as exact when
forming the map, and the $1/f$ noise structure is not
accounted for in the averaging weights.

\paragraph*{Limitations motivating this work.}
Simplifications in the conventional pipeline become
increasingly consequential as observations push to higher
sensitivity.
First, treating calibration and map-making sequentially
means that the posterior uncertainty in the calibration
parameters is never propagated into the sky map, and any
bias in the MAP gain estimate propagates unchecked into the
final product.
Second, fixing the noise amplitude to the white-noise level
during the MAP estimation decouples it from the system
temperature being simultaneously fitted, violating the
self-consistency of the radiometer equation [Eq.~(\ref{eq: radiometer})], which
predicts that $\sigma_\mathrm{w} \propto T_\mathrm{sys}$.
This coupling is especially significant when the noise diode
signals and sky components span a wide dynamic range
\citep{burke2019introduction}.

\subsection{Statistical Models for This Work}
\label{sec: model of this  work}

In this section, we describe the statistical model adopted in this work. This model addresses both limitations by jointly sampling all signal and noise parameters within a Gibbs sampling framework. It builds on the fundamental components of the model described in Section~\ref{sec:conventional}.

\subsubsection{Smooth gain variation}

To improve the accuracy of the gain model, $\LscaleG(t_a)$ is typically fitted with a smooth time-dependent function.
This function can be parameterised using the first few Legendre polynomials \citep{wang2021h}, which in linear algebra terms can be written as
\begin{equation}
    \label{eq: parameterisation TildeG}
    \Vec{\LscaleG} = \mathbf{U}_{\rm g} \, \boldsymbol{p}_{\rm g},
\end{equation}
where $\boldsymbol{p}_{\rm g}$ is the vector of polynomial coefficients, and $\mathbf{U}_{\rm g}$ is the design matrix formed from evaluations of the Legendre polynomials. 
The order of the fiducial polynomial is set to $N_{\rm order} = 4$ in this work, in accordance with the MeerKLASS standard (see Section~\ref{sec:conventional}). We note that the polynomial order can be adjusted as required. Adopting a linear model is instructive, as it enables the use of a linear solver for Gibbs sampling, significantly enhancing computational tractability \citep{wandelt2004global, kennedy2023statistical}.

\subsubsection{Flicker noise model}
\label{sec: flicker model}

\begin{figure*}
    \centering
    \includegraphics[width=\linewidth]{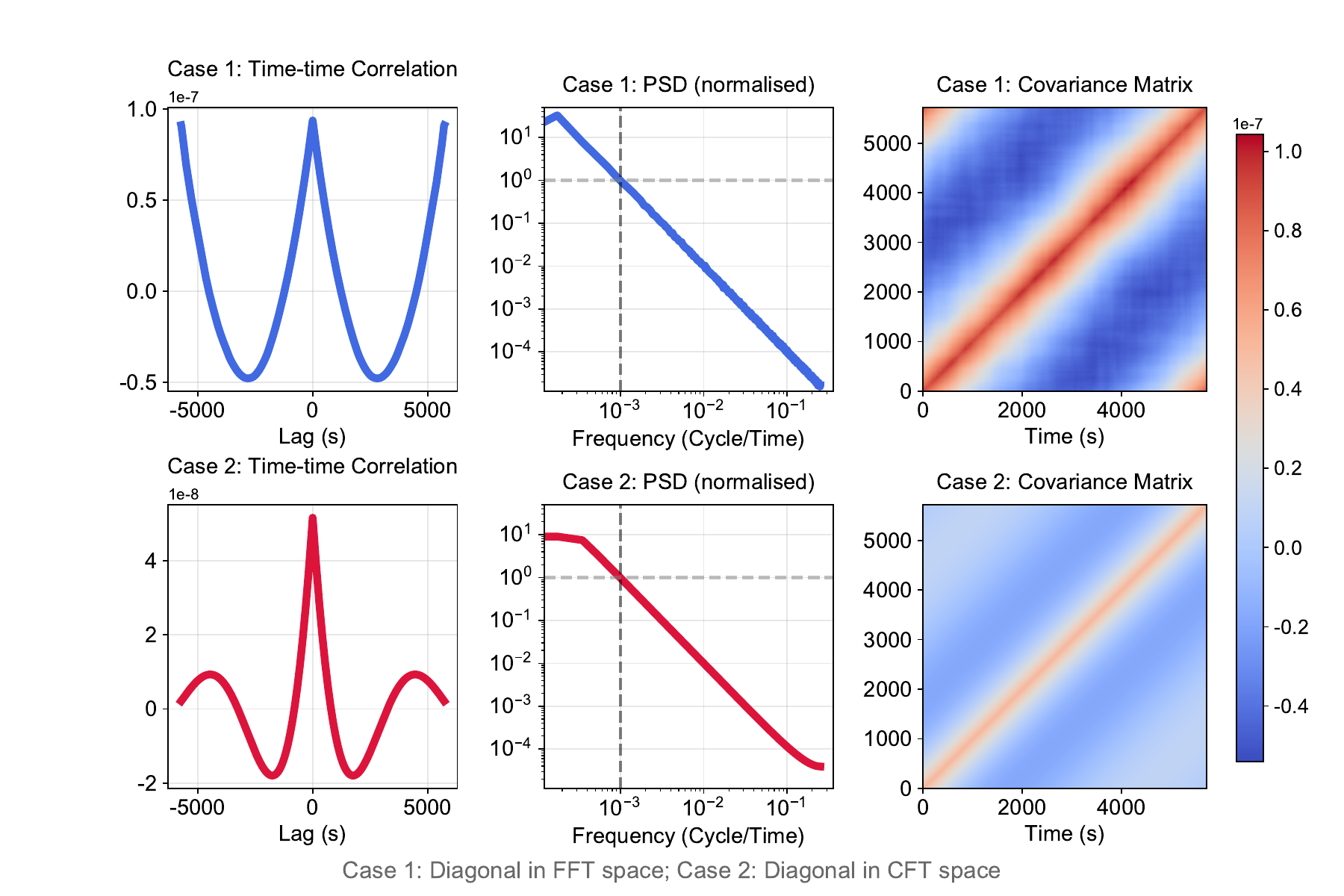}
    \caption{
    Comparison between the conventional $1/f$ noise model [Uncorrelated DFT modes (Case 1); see Appendix~\ref{Appendix: PSD}] and our analytical approach (Case 2; see Section~\ref{sec: flicker model}).
    \textbf{Left}: Time-domain correlation function.
    \textbf{Middle}: Power Spectral Density (PSD; defined with DFT), normalised by that of white noise. The vertical line marks the knee frequency ($1$\,mHz), where the flicker noise power matches that of white noise (shown by the horizontal line).
    \textbf{Right}: Time-time covariance matrix.
    This figure highlights the consequences of periodicity assumptions in DFT-based $1/f$ noise models. In Case 1, the diagonal covariance in DFT space leads to artificial periodic and symmetric correlations in the time domain. In contrast, our model (Case 2) yields a covariance consistent with the underlying Gaussian statistics and statistical homogeneity of the noise.
    Both models share the same PSD power law and knee frequency. For Case 2, the parameters are $f_c = 1.099 \times 10^{-3}$ (rad/s) and $f_0 = 1.335 \times 10^{-5}$ (rad/s).
    }
    \label{fig: flicker model comparisons}
\end{figure*}
On shorter timescales than the smooth gain model covers, we introduce a stochastic flicker noise model. The stochastic fractional gain variation, $\deltaG$, is typically characterised by a $1/f$ noise spectrum with a power-law PSD 
\footnote{See Appendix~\ref{Appendix: PSD} for a proper definition and normalisation in the finite discrete frequency domain.}
\footnote{Temporal frequencies in this work are expressed as angular frequencies, with units of radians per second. (We adopt the convention to keep the brevity of the analytic correlation function.) The only exception is the knee frequency, traditionally given in cycles per second (Hz), a convention we maintain.}
\be
    \label{eq: PSD deltaG}
    P_{\deltaG}(f)
    =
    \begin{cases}
        0, &  |f| < f_c,\\
        \left({f_0}/{|f|}\right)^\alpha, &  |f| \geq f_c,
    \end{cases}
\ee
where $f$ represents the temporal frequency, $f_0$ is the reference frequency, and $\alpha$ denotes the power index.
To prevent the ``infrared catastrophe,'' we introduce $f_c$ as the lower bound of the power-law PSD.

In this model, the power spectrum below the cutoff frequency $f_c$ vanishes. In other words, the model neglects contributions to the temporal gain variance from time scales longer than $1/f_c$. This does not conflict with a true flicker noise random process, given the finite TOD length. On the other hand, the smooth gain model [Eq.~(\ref{eq: parameterisation TildeG})] can capture the variance contribution from the largest scales. In fact, $f_c$ can either be treated as a parameter or set as a constant, but we choose $f_c$ to be approximately the inverse of the time scale of the TOD period, or in other words,
the lowest frequency mode resolvable within a single TOD of $N$ samples at cadence $\Delta t$. Fluctuations on timescales longer than the total scan duration cannot be constrained by the data, so this choice provides a natural lower bound without extrapolating the $1/f$ power law beyond the reach of the measurement.

In measurement problems, it is common to assume that the $1/f$ noise is Gaussian and stationary. 
The Gaussianity means that the stochastic gain variation is completely characterised by the two-point statistics, i.e., correlation function or noise covariance:
\be
    \deltaG \sim \mathcal{N}\left(0, \NCov{\text{corr}} \right),
\ee
where $\NCov{\text{corr}}$ is the time-time covariance matrix of the correlated noise.
The further assumption on ``stationarity'' implies that the noise in the Fourier domain can be described as an uncorrelated multivariate Gaussian. Equivalently, in real space, the covariance matrix does not change with time.
The assumptions together with the PSD model for $1/f$ noise provide a complete statistical description necessary for solving the measurement equations.

In many implementations, this model is represented in a diagonal form within the Discrete Fourier transform (DFT) space [see e.g. \cite{li2021h, harper2018impact, ihle2023beyondplanck}], which can introduce artificial periodic and symmetric correlations. 
The issue arises not from the flicker noise model itself, but from the periodic flicker noise assumption inherent in the DFT. In our paper, we deviate from this conventional model and instead use a model that aligns with the assumed noise statistics. 

Less abstractly, we propose a flicker noise model with the covariance diagonal in continuous Fourier transform (CFT) space. 
Using the Wiener-Khinchin theorem, we define the noise correlation function directly from the analytic PSD:
\be
\label{eq: 1/f noise correlation}
(\NCov{\text{corr}})_{aa'}  = \xi(t_{a'}-t_a) \equiv  \int \frac{\diff{f}}{2\pi} P_{\deltaG}(f) e^{ i f(t_{a'}\minus t_a)} ,
\ee
where we have defined the two-point autocorrelation function of gain variation, $\xi$. By substituting Eq.~(\ref{eq: PSD deltaG}) we get the following analytical form of $\xi$
\footnote{This correlation function can be implemented in Python with {\tt mpmath}, but we use \textsc{MomentEmu} \citep{zhang2025general} to create an emulator, which speeds up the evaluation by approximately 1,720 times compared to the original {\tt mpmath} code.}
\begin{align}
&\xi(\tau)
=\frac{1}{\pi \tau}  \Theta_0^\alpha \, \mathrm{Re}\left[ \Gamma(\mu, i \Theta_c) \,e^{\minus i\frac{\pi}{2}\mu } \right] ,  & \text{for } \tau > 0, 
\label{eq: flicker correlation function}\\
&\mu = 1 - \alpha , \quad \Theta_0= \tau f_0,  \quad
\Theta_c =  \tau f_c, & \nonumber
\end{align} 
where $\tau = t_{a'}-t_a$ represents the time separation between data points; $\Gamma$ is the incomplete gamma function and ``$\mathrm{Re}[A]$'' is the real part of $A$. 
For $\tau < 0$ we have $\xi(\tau)=\xi(-\tau)$;
for $\tau=0$ the integral can be evaluated directly, which gives 
$\xi(0)= \frac{1}{\alpha-1} \frac{f_c}{\pi}\left(\frac{f_0}{f_c}\right)^\alpha$.
Using this correlation function, we can compute the proper time domain covariance matrix, which is then used to define the likelihood or simulate $1/f$ noise.
Figure~\ref{fig: flicker model comparisons} compares our analytical model with the conventional approach, using parameters representative of MeerKLASS: the cutoff frequency is set to $f_c = 1/T_{\rm obs}$, where $T_{\rm obs}$ is the duration of a single scan, and the knee frequency is $f_0 = 0.001$,Hz, a typical value identified for MeerKAT receivers in \citet{li2021h}.
An additional discussion on full gain modelling can be found in Appendix~\ref{appendix: a gain model}.

\subsubsection{System temperature model}
\label{sect: system temperature}
The system temperature can be modelled as the sum of several components: sky emission, receiver temperature, and contributions from ground and atmospheric effects \citep{wang2021h}. For measurements using internal calibration sources, we also need to account for the intentional signal injection, but with known specific characteristics.

Without loss of generality, the complete model can be expressed as:
\begin{multline}
    T^{\rm sys}_{j,x}(t_a, \nu_b) 
    =
    \sum_{X} B^X_{x}(t_a, \nu_b)\ast T^{\text{sky},X}(\nu_b)  +  \mathit{T}^{\rm el}_{x}(t_a, \nu_b) \\
        + T_{j,x}^{\rm nd}(t_a, \nu_b) 
        + T_{j,x}^{\text{rec}}(t_a, \nu_b)  ,
\end{multline}
where $j$ indexes the antenna and $x$ is the feed; together they specify the receiver. 
$T^{\text{sky},X}$ represents the polarised sky emission, with $X$ denoting $I, Q, U, V$, the Stokes parameters. 
$\mathit{T}^{\rm el}_{x}$ is the elevation-dependent terrestrial emission, including the atmospheric
emission and the ground pickup.
$T_{j,x}^{\rm nd}$ represents an artificially injected signal that is typically used as an internal calibration source. It is specifically introduced to represent the temperature contribution from a noise diode. 
A real-world example of this implementation can be found in MeerKAT's\footnote{MeerKAT is a 64-dish radio telescope array, one of whose science goals is 21\,cm intensity mapping using its single-dish mode \citep{wang2021h}.} periodic noise diode injection system \citep{wang2021h}.
The remaining component is the receiver temperature, $T_{j,x}^{\text{rec}}$, which is simply assumed to be a smooth function of time. It thus functionally absorbs uncaptured effects or offsets in models.

It is important to note that while the Stokes $I$ and $V$ beams remain constant in celestial coordinates, the $Q$ and $U$ beams, which are not scalar fields, change as the antenna pointing changes. Therefore, the `$\ast$' in the above equation should not be understood as convolutions with constant beams, and careful definitions of the $Q$ and $U$ beams should be taken into account (see \cite{zhang:tel-04750721} for a detailed discussion).
In this paper, we consider only the Stokes-$I$ sky for convenience, as this is sufficient to demonstrate the analysis framework we propose. 
Based on this, the polarisation cases can be extended directly, by adding three more components linearly to the model.

For nuisance components such as ground spillover, it is not necessary to explicitly disentangle contributions from different polarisations. 
We simply assume that these components vary smoothly over time and model them independently for each receiver and TOD, using separate parameters.  
Therefore, although the spillover is intrinsically polarised and different polarimeters probe distinct temporal variations, explicit modelling of its intrinsic polarisation structure is not required since the contamination is fitted independently for each TOD (and thus for each receiver).

Finally, we adopt a system temperature model that represents all components linearly:
\be
\label{eq: system temp model}
\begin{split}
    \boldsymbol{T}_{\rm sys}
    &=
    \boldsymbol{T}_{\rm cel} 
    +
    \boldsymbol{T}_{\rm nd} 
    +
    \boldsymbol{T}_{\rm rec} 
    +
    \boldsymbol{T}_{\rm el} \\
    &=
    \mathbf{U}_{\rm cel} \, \boldsymbol{p}_{\rm cel} 
    +
    \mathbf{U}_{\rm nd} \, \boldsymbol{p}_{\rm nd} 
    +
    \mathbf{U}_{\rm rec} \, \boldsymbol{p}_{\rm rec} 
    +
    \mathbf{U}_{\rm el} \, \boldsymbol{p}_{\rm el} \\
    &\equiv
    \mathbf{U}_{\rm cel} \, \boldsymbol{p}_{\rm cel} 
    +
    \mathbf{U}_{\rm loc} \, \boldsymbol{p}_{\rm loc} 
\end{split}
\ee
where the celestial sky, $\boldsymbol{T}_{\rm cel}$, (consisting of all components stationary on the celestial sphere) is parameterised by its individual pixels. $\boldsymbol{T}_\text{el}$ can be modelled as a smooth function of elevation using e.g., a Legendre polynomial, while $\boldsymbol{T}_\text{rec}$ can be modelled with a smooth function of time. 
If, depending on the scanning strategy, the elevation of the pointing centre is itself a smooth function of time, then the $\boldsymbol{T}_\text{el}$ sequence is also smooth with respect to time.
Given these considerations, we can use a {single combined term -- modelled, for example, with a Legendre polynomial -- to represent the smooth time-dependent component, effectively capturing the combined contribution of $\boldsymbol{T}_\text{rec}$ and $\boldsymbol{T}_\text{el}$.
In the final line of Eq.~(\ref{eq: system temp model}), we group the system temperature parameters into two components: $\boldsymbol{p}_{\rm cel}$, which contains the parameters common to all TOD sets, and $\boldsymbol{p}_{\rm loc}$, which captures the residual, independent parameters specific (or `local') to each TOD set.
Less abstractly, celestial parameters describe the desired sky map and are therefore shared across all TOD sets, while local parameters (such as receiver temperature, and noise diode amplitude) are specific to each individual scan.

\subsubsection{Model degeneracies and absolute calibration}
In previous sections, we have described our general data model, which makes optimal use of linear modelling. 
We now inspect possible model parameter degeneracies:
\begin{enumerate}
    \item Large-scale gain and flicker noise parameters

    \item Flux scale degeneracy between gain and system temperature.
    
    \item Degeneracy between system temperature components. For example absolute offsets (baseline) of both the measured sky and the receiver temperature.
\end{enumerate}
To mitigate these degeneracies, known as absolute calibration, we typically require precise knowledge of specific aspects of $\Tsys$. This knowledge can be acquired in at least two ways:
    \begin{itemize}
        \item \textit{External calibration}: Known $T_{\rm sky}$ parameters (pixels) for specific calibration sources;
        \item \textit{Internal calibration}: Knowledge of the absolute level of other $\Tsys$ components. 
    \end{itemize}
    These constraints can be incorporated by placing priors on the corresponding components.
    As an example, in our subsequent demonstrations, we assume prior knowledge of specific sky pixel(s) used as external calibration source(s).

\section{Joint analysis framework}
\label{sec: framework}

\subsection{Probability functions and Gibbs sampling}
It is usually the case that both the stochastic gain ($\hat{\epsilon}$) and the thermal fluctuation ($\hat{w}$) are small quantities so that their product term is trivial and can be omitted. 
The data model is then rewritten as
\be
    d(t_a) \simeq 
    \LscaleG(t_a)
    \Tsys(t_a) 
    \left[ 1  +  \wn  + \deltaG \right].
\ee

For ease of formalising the probability function, we define the scaled, centred data as
\be
    \label{eq: residual noise definition}
    n(t_a) \equiv \frac{d(t_a)}{\LscaleG(t_a)
    \Tsys(t_a) } - 1
    \simeq 
    \wn  + \deltaG  .
\ee
We also introduce the column vector notation $\boldsymbol{n}$ for the time sequence of all scaled data points.  
Assuming independent Gaussian statistics for both $\hat{\epsilon}$ and $\hat{w}$,
the likelihood function of the data model is given by
\begin{equation}
        L(\boldsymbol{n} | \LscaleG, \Tsys, \Ncov) = (2\pi)^{-\frac{N}{2}}
        | \Ncov |^{-\frac{1}{2}}
        \exp{\left[-\frac{1}{2}\boldsymbol{n}^{T}\mathbf{N}^{-1}\boldsymbol{n}\right]},
\end{equation}
where $\mathbf{N}$ is the total noise covariance matrix, given as the sum of the two components
\be
\mathbf{N} = \NCov{w} + \NCov{\text{corr}},
\label{eq: total noise covariance}
\ee
and $|\mathbf{N}|$ represents the determinant of $\mathbf{N}$.
Note that the normalisation term, $|\mathbf{N}|^{-1/2}$, should be retained to fit or sample the noise parameters. 

Maximizing the likelihood is effectively minimizing the negative logarithmic likelihood function defined as
\be
\label{eq: likelihood function}
\mathcal{L} = -2\ln{L}   = \ln{|\mathbf{N}|} + \Tr(\mathbf{N}^{\minus \! 1}\mathbf{D}),
\ee
where $\mathbf{D} = \boldsymbol{n} \, \boldsymbol{n}^\top$, and we have omitted the constant term in the second equation.
Similar to $\loglikeli$, we also define the negative logarithmic prior ($\logprior$) and posterior ($\logposterior$) so that Bayes' Theorem can be written as 
\be
\label{eq: logpost}
    \logposterior 
    =
    \logprior + \loglikeli 
\ee
where we have dropped the logarithmic evidence, as we have specified the model to be used.

For complex probability functions such as Eqs.~(\ref{eq: likelihood function}, \ref{eq: logpost}) with numerous parameters direct estimation becomes impractical. Therefore, we use sampling-based approaches, specifically the Gibbs sampling method \citep{GibbsSampling}. This method iteratively samples from conditional distributions (denoted as $p(x|y)$) to approximate the joint distribution. Specifically, we follow these Gibbs sampling steps: 
\begin{subequations}
\begin{align}
    \boldsymbol{p}_{\rm g}^{(i+1)} \leftarrow& \, p (\boldsymbol{p}_{\rm g} |\boldsymbol{d}, \boldsymbol{p}_{\rm cel}^{(i)}, \boldsymbol{p}_{\rm loc}^{(i)}, \boldsymbol{p}_{\rm n}^{(i)})  
    \label{eq: sampling gain}\\
    \boldsymbol{p}_{\rm loc}^{(i+1)} \leftarrow& \, p (\boldsymbol{p}_{\rm loc} |\boldsymbol{d}, \boldsymbol{p}_{\rm cel}^{(i)}, \boldsymbol{p}_{\rm g}^{(i+1)}, \boldsymbol{p}_{\rm n}^{(i)})  
    \label{eq: sampling p loc}\\
    \boldsymbol{p}_{\rm n}^{(i+1)}\leftarrow& \, p (\boldsymbol{p}_{\rm n} |\boldsymbol{d}, \boldsymbol{p}_{\rm cel}^{(i)}, \boldsymbol{p}_{\rm loc}^{(i+1)}, \boldsymbol{p}_{\rm g}^{(i+1)})  
    \label{eq: sampling noise}\\
    \boldsymbol{p}_{\rm cel}^{(i+1)} \leftarrow& \, p (\boldsymbol{p}_{\rm cel} |\boldsymbol{d}, \boldsymbol{p}_{\rm loc}^{(i+1)}, \boldsymbol{p}_{\rm g}^{(i+1)},  \boldsymbol{p}_{\rm n}^{(i+1)})  
    \label{eq: sampling p sky}
\end{align} 
\end{subequations}
where $\boldsymbol{p}_{\rm cel}$, $\boldsymbol{p}_{\rm g}$, and $\boldsymbol{p}_{\rm n}$ denote the celestial parameters, other $T_{\mathrm{sys}}$ parameters associated with each TOD, and the noise parameters ($f_0$ and $\alpha$) for each TOD, respectively. 
For clarity, Figure~\ref{fig:gibbs_flowchart} presents a flowchart that provides more details of the full Gibbs sampler.
The first three samplers can be run independently for each TOD set, whereas the final sampler (for the sky maps) must be applied jointly across all TOD sets.

The essential idea of Gibbs sampling is that we expect the conditional probability functions to have a simpler form, making them numerically tractable. Indeed, as we will see in sections \ref{sec: ite GLS sampler}, both the gain and system temperature samplers can be implemented as multivariate Gaussian distributions in the coefficients of linear basis functions. 
However, sampling the noise parameters (section~\ref{sec: noise sampler}) does not result in a similarly linear problem. As an alternative, we employ an MCMC sampler. Nonetheless, we can leverage the structure of the flicker noise covariance (real symmetric Toeplitz\footnote{But not cyclic.}) to significantly reduce computational complexity.

\subsection{Iterative GLS sampler for the smooth gain and system temperature parameters}
\label{sec: ite GLS sampler}

  \begin{figure}                                                                                   
      \centering                                                                                   
      
\begin{tikzpicture}[
    node distance=0.7cm,
    >={Stealth[length=2.5mm]},
    init/.style={rectangle, rounded corners=3pt, draw=black, fill=gray!15,
                 text width=6cm, minimum height=0.8cm, align=center, font=\small},
    proc/.style={rectangle, rounded corners=3pt, draw=black, fill=blue!10,
                 text width=6cm, minimum height=0.8cm, align=center, font=\small},
    sync/.style={rectangle, rounded corners=3pt, draw=black, fill=orange!15,
                 text width=6cm, minimum height=0.8cm, align=center, font=\small},
    decision/.style={diamond, draw=black, fill=yellow!12,
                     aspect=3.5, align=center, font=\small, inner sep=0pt},
    arrow/.style={->, thick},
    groupbox/.style={draw=black!50, dashed, rounded corners=5pt, inner sep=15pt},
]

\node[init] (init) {%
    \textbf{Initialise} parameters:\\[2pt]
    $\boldsymbol{p}_\mathrm{sky}^{(0)}$, \;
    $\{\boldsymbol{p}_{\mathrm{loc},j}^{(0)}\}$, \;
    $\{\boldsymbol{p}_{g,j}^{(0)}\}$, \;
    $\{\boldsymbol{p}_{n,j}^{(0)}\}$
};

\node[proc, below=1.2cm of init] (gain) {%
    \textbf{(a)} Sample gains $\boldsymbol{p}_{g,j}^{(i)}$\\[2pt]
    {\footnotesize $\boldsymbol{p}_{g,j}^{(i)} \sim P\!\left(\boldsymbol{p}_{g,j} \;\middle|\; \boldsymbol{d}_j,\, T_{\mathrm{sys},j}^{(i-1)},\, \boldsymbol{p}_{n,j}^{(i-1)}\right)$}
};

\node[proc, below=0.5cm of gain] (tloc) {%
    \textbf{(b)} Sample local temperature $\boldsymbol{p}_{\mathrm{loc},j}^{(i)}$\\[2pt]
    {\footnotesize $\boldsymbol{p}_{\mathrm{loc},j}^{(i)} \sim P\!\left(\boldsymbol{p}_{\mathrm{loc},j} \;\middle|\; \boldsymbol{d}_j,\, \boldsymbol{p}_{\mathrm{sky}}^{(i-1)},\, \boldsymbol{p}_{g,j}^{(i)},\, \boldsymbol{p}_{n,j}^{(i-1)}\right)$}
};

\node[proc, below=0.5cm of tloc] (noise) {%
    \textbf{(c)} Sample noise $\boldsymbol{p}_{n,j}^{(i)}$\\[2pt]
    {\footnotesize $\boldsymbol{p}_{n,j}^{(i)} \sim P\!\left(\boldsymbol{p}_{n,j} \;\middle|\; \boldsymbol{d}_j,\, T_{\mathrm{sys},j}^{(i)},\, \boldsymbol{p}_{g,j}^{(i)}\right)$}
};

\node[sync, below=1.6cm of noise] (tsky) {%
    \textbf{(d)} Sample sky temperature $\boldsymbol{p}_{\mathrm{sky}}^{(i)}$\\[2pt]
    {\footnotesize $\boldsymbol{p}_{\mathrm{sky}}^{(i)} \sim P\!\left(\boldsymbol{p}_{\mathrm{sky}} \;\middle|\; \{\boldsymbol{d}_j\},\, \{\boldsymbol{p}_{\mathrm{loc},j}^{(i)}\},\, \{\boldsymbol{p}_{g,j}^{(i)}\},\, \{\boldsymbol{p}_{n,j}^{(i)}\}\right)$}
};

\node[decision, below=0.8cm of tsky] (converge) {$i < N_\mathrm{samp}$};

\node[init, below=0.8cm of converge] (output) {%
    \textbf{Output}: posterior samples\\[2pt]
    $\boldsymbol{p}_\mathrm{sky}^{(i)}$, \;
    $\{\boldsymbol{p}_{\mathrm{loc},j}^{(i)}\}$, \;
    $\{\boldsymbol{p}_{g,j}^{(i)}\}$, \;
    $\{\boldsymbol{p}_{n,j}^{(i)}\}$
};

\draw[arrow] (init) -- (gain);
\draw[arrow] (gain) -- (tloc);
\draw[arrow] (tloc) -- (noise);
\draw[arrow] (noise) -- node[right, font=\footnotesize, text=black!70]{sync across ranks} (tsky);
\draw[arrow] (tsky) -- (converge);
\draw[arrow] (converge) -- node[right, font=\footnotesize]{no} (output);

\coordinate (topright)    at ($(gain.east) + (1.5, 0)$);
\coordinate (bottomright) at (topright |- converge.east);
\draw[arrow] (converge.east) -- node[above, font=\footnotesize]{yes} (bottomright);
\draw[thick] (bottomright) -- (topright)
    node[midway, fill=white, font=\footnotesize,
         draw=black, rounded corners=2pt, inner sep=3pt]{$i = i+1$};
\draw[arrow] (topright) -- (gain.east);

\begin{scope}[on background layer]
    \node[groupbox, fill=blue!3, fit=(gain)(tloc)(noise),
          label={
          [font=\footnotesize, anchor=north east, text=black!70]%
          north east: 
          \textit{for each TOD\, $k$ (parallel over MPI ranks)}}] {};
\end{scope}

\begin{scope}[on background layer]
    \node[groupbox, fill=orange!5, fit=(tsky),
          label={[font=\footnotesize, anchor=north east, text=black!70]%
          north east:\textit{global (MPI allreduce)}}] {};
\end{scope}

\end{tikzpicture}

      \caption{Flowchart of the Gibbs sampling procedure. 
      Note that there is no strict preference for the order of the different Gibbs sampling steps; in practice, however, the first step is determined by the setup of initial conditions: Parameters without specified initial values should be sampled first.
      Also, step (b) can be incorporated into step (d) as part of a large joint linear sampling task. In this task, the local system temperature parameters from all TOD sets are sampled alongside the sky parameters.
      }                
      \label{fig:gibbs_flowchart}                                                                  
  \end{figure} 

In this section, we present a generic approach that can be applied to both the gain and system temperature parameters, i.e., the linear model coefficients in Eqs.~(\ref{eq: sampling gain}), (\ref{eq: sampling p loc}) and (\ref{eq: sampling p sky}).
For a general multivariate Gaussian, sampling the linear model parameters of its mean can be transformed into a linear system problem. However, unlike such problems, our data model employs a multiplicative noise term rather than an additive one. The advantage of this approach is that it accurately captures the coupling between the thermal noise level and the system temperature. However, the inherent noise structure with parameter-dependent noise variance violates the exogeneity assumption \citep[see e.g. Chapter~4 of ][]{greene2000econometric} of Maximum Likelihood methods, also known as \textit{Generalized Least Squares} (GLS), leading to significant biases in estimation. To address this, we use an iterative GLS method, which essentially transforms the multiplicative noise into a consistent additive noise.

\subsubsection{Formulation}
Specifically, whether sampling $\LscaleG$ or $\Tsys$, the data model can be expressed as 
\begin{equation}
    \boldsymbol{d}' = ( \mathbf{U} \boldsymbol{p} + \boldsymbol{\mu}) \circ (1 + \boldsymbol{n}),
\end{equation}
with $\boldsymbol{d}'$ the data vector, $\boldsymbol{p}$ the parameter vector, $\mathbf{U}$ the projection matrix,  and $\boldsymbol{p}$-independent vector $\boldsymbol{\mu}$. The ``$\circ$'' means elementwise multiplication.
The noise vector $\boldsymbol{n}$ is assumed zero-mean Gaussian with known covariance $\mathbf{N}$, as defined in Eq.~(\ref{eq: total noise covariance}).

We rewrite the original multiplicative noise model into an additive noise model:
\begin{equation}
    \boldsymbol{d}' - \boldsymbol{\mu} = \mathbf{U} \boldsymbol{p} + (\mathbf{U} \boldsymbol{p} + \boldsymbol{\mu}) \circ \boldsymbol{n}.
\end{equation}
The rearrangement results in a form that resembles the commonly used data model:
\begin{equation}
    \boldsymbol{d}'' = \mathbf{U} \boldsymbol{p} + \boldsymbol{\delta}, \quad \boldsymbol{\delta} \sim \mathcal{N}(0, \mathbf{\Sigma}),
\end{equation}
where $\boldsymbol{d}''=\boldsymbol{d}'-\boldsymbol{\mu}$ and the noise covariance structure is given by:
\begin{equation}
    \mathbf{\Sigma} = \text{diag}(\mathbf{U} \boldsymbol{p} + \boldsymbol{\mu}) \, \mathbf{N} \, \text{diag}(\mathbf{U} \boldsymbol{p}+ \boldsymbol{\mu}).
\end{equation}
Note that the noise covariance $\mathbf{\Sigma}$ is correlated with the parameters $\boldsymbol{p}$, rendering standard GLS biased. Instead, we first estimate $\mathbf{\Sigma}$ using an iterative GLS approach. Then we sample $\boldsymbol{p}$ (see section~\ref{sec: sampling gain and system temperature params} and \ref{sec: sampling sky params}).


To estimate the covariance of the effective additive noise term, we solve for the GLS estimate of $\boldsymbol{p}$ iteratively:
\begin{enumerate}
    \item Initialize $ \boldsymbol{p}^{(0)} $ using Ordinary Least Squares (OLS):
    \begin{equation}
        \boldsymbol{p}^{(0)} = (\mathbf{U}^\top \mathbf{U})^{-1} \mathbf{U}^\top \boldsymbol{d}''.
    \end{equation}
    \item At iteration $k$, compute 
    \begin{equation}
    \label{eq: ite GLS noise cov}
        \mathbf{\Sigma}^{(k)} = \text{diag}(\mathbf{U} \boldsymbol{p}^{(k)} + \boldsymbol{\mu}) \, \mathbf{N} \, \text{diag}(\mathbf{U} \boldsymbol{p}^{(k)} + \boldsymbol{\mu}).
    \end{equation}
    \item Solve the GLS system:
    \begin{equation}
    \label{eq: ite GLS linear solve}
        \left[\mathbf{U}^\top \left(\mathbf{\Sigma}^{(k)}\right)^{-1} \mathbf{U}\right] \boldsymbol{p}^{(k+1)} = \mathbf{U}^\top \left(\mathbf{\Sigma}^{(k)}\right)^{-1} \boldsymbol{d}''.
    \end{equation}
    \item Repeat until convergence: \( \| \boldsymbol{p}^{(k+1)} - \boldsymbol{p}^{(k)} \| < \text{tol} \).\footnote{The default tolerance adopted in \texttt{hydra-tod} is $10^{-10}$, with a maximum of 100 iterations. In this work, we have verified that further tightening the tolerance or increasing the maximum number of iterations does not produce any statistically significant change in the map-making results, at least for the purpose of demonstrating the methodology. For completeness, we note that importance weighting can be applied in post-processing to correct for any residual approximation, although we find this to be unnecessary at the accuracy level considered here.}
\end{enumerate}
The final converged solution gives the estimated parameter $\boldsymbol{p}_{\text{GLS}}$, and thus the estimated noise covariance, $\mathbf{\Sigma}$:
\begin{equation}
\label{eq: final Sigma}
    \mathbf{\Sigma} = \text{diag}(\mathbf{U} \boldsymbol{p}_{\rm GLS} + \boldsymbol{\mu}) \, \mathbf{N} \, \text{diag}(\mathbf{U} \boldsymbol{p}_{\rm GLS} + \boldsymbol{\mu}).
\end{equation}
Note that solving Eq.~(\ref{eq: ite GLS linear solve}) requires the inverse of $\mathbf{\Sigma}^{(k)}$. This involves inverting the diagonal matrices in Eq.~(\ref{eq: ite GLS noise cov}) and multiplying by the precomputed inverse of $\mathbf{N}$, which remains constant when the flicker noise parameters are fixed.

\subsubsection{Sampling gain and local system temperature parameters}
\label{sec: sampling gain and system temperature params}
The idea behind sampling in probabilistic analysis is to draw random model parameters such that the sample distribution reflects the underlying probability density function.
A special case arises when the conditional probability distribution is multivariate Gaussian; in this case, solving the system of measurement equations becomes a GLS problem. By adding a random noise realisation to the linear system, the noise covariance is propagated into the parameter space, effectively yielding samples that follow the desired distribution.
To support this linear sampling method, priors are typically modelled as a multivariate Gaussian (specified by the mean $\Bar{\boldsymbol{p}}$ and the covariance matrix $\mathbf{C}$), which can be interpreted as an additional independent noise contribution to the effective linear system.
We refer to the steps in the Gibbs sampler that operate on such multivariate Gaussian distributions as \textit{Gaussian Constrained Realisation} (GCR) steps \citep{wandelt2004global, eriksen2008joint, kennedy2023statistical}.

In the framework of the iterative GLS sampler, once the effective additive noise covariance $\mathbf{\Sigma}$ is estimated, we sample $\boldsymbol{p}$ from its conditional posterior distribution by solving the following GCR equation
\begin{multline}
\label{eq: linear sampler}
\left(
\mathbf{C}^{-1}
+
\mathbf{U}^\top \mathbf{\Sigma}^{-1} \mathbf{U}
\right) \boldsymbol{p}_{\text{sample}}
= \\
\mathbf{U}^\top 
\left(
\mathbf{\Sigma}^{-1} \boldsymbol{d}''
+
\mathbf{\Sigma}^{-\frac{1}{2}} \boldsymbol{\omega}
\right)
+
\mathbf{C}^{-1}\Bar{\boldsymbol{p}}
+
\mathbf{C}^{-\frac{1}{2}} \boldsymbol{\eta},
\end{multline}
where $\boldsymbol{\omega}$ is an uncorrelated unit Gaussian random draw with the same dimension as the data vector, while $\boldsymbol{\eta}$ is another independent Gaussian random draw but with the dimension of the parameter vector.

More specifically, the $(i+1)$-th sample of the gain parameters $\boldsymbol{p}^{(i+1)}_{\rm g}$ is obtained from the conditional posterior~(\ref{eq: sampling gain}) by solving Eq.~(\ref{eq: linear sampler}) 
for $\boldsymbol{p}_{\text{sample}}$ with
\begin{align*}
    &\boldsymbol{d}'' = \boldsymbol{d} / \boldsymbol{T}_{\rm sys}^{(i)}, \quad\quad
    \mathbf{U}  = \mathbf{U}_{\rm g}, \quad
    \mathbf{C}  = \mathbf{C}_{\rm g}, \quad
    \Bar{\boldsymbol{p}}  = \Bar{\boldsymbol{p}}_{\rm g}, \\
    &\mathbf{\Sigma} \xleftarrow{\text{Eq.~(\ref{eq: final Sigma})}} \boldsymbol{p}^{(i+1)}_{\rm g, GLS} \xleftarrow{\text{iterative GLS}} \boldsymbol{T}_{\rm sys}^{(i)}, f_0^{(i)}, \alpha^{(i)} ,
\end{align*}
where $\boldsymbol{T}_{\rm sys}^{(i)} = \mathbf{U}_{\rm cel} \, \boldsymbol{p}_{\rm cel}^{(i)} 
+
\mathbf{U}_{\rm loc} \, \boldsymbol{p}_{\rm loc}^{(i)} $ and $\mathbf{N}^{(i)}$ represents the sample of system temperature after the $i$-th Gibbs sampling step.

The $(i+1)$-th sample of the local system temperature parameters $\boldsymbol{p}_{\rm loc}$ is obtained from the conditional posterior~(\ref{eq: sampling p loc}) by solving Eq.~(\ref{eq: linear sampler}) 
for $\boldsymbol{p}_{\text{sample}}$ with
    \begin{align*}
        &\boldsymbol{d}'' = \boldsymbol{d} / \boldsymbol{g}^{(i+1)} - \mathbf{U}_{\rm cel} \boldsymbol{p}^{(i)}_{\rm cel}, \quad\quad
        \mathbf{U}  = \mathbf{U}_{\rm loc}, \quad
        \mathbf{C}  = \mathbf{C}_{\rm loc}, \quad
        \Bar{\boldsymbol{p}}  = \Bar{\boldsymbol{p}}_{\rm loc}, \\
        &\mathbf{\Sigma} \xleftarrow{\text{Eq.~(\ref{eq: final Sigma})}} \boldsymbol{p}^{(i+1)}_{\rm loc, GLS} \xleftarrow{\text{iterative GLS}} \boldsymbol{g}^{(i+1)}, \boldsymbol{p}^{(i)}_{\rm cel}, f_0^{(i)}, \alpha^{(i)} ,
    \end{align*}
where $\boldsymbol{g}^{(i+1)} = \mathbf{U}_{\rm g} \, \boldsymbol{p}_{\rm g}^{(i+1)} $ represents the smooth gain sample after the Gibbs sampling step of Eq.~(\ref{eq: sampling gain}).

As these sampling steps for different TOD sets are independent, they can be carried out in parallel.

\subsubsection{Sampling celestial sky parameters}
\label{sec: sampling sky params}

Since all TOD sets share the sky parameters $\boldsymbol{p}_{\rm cel}$, the sampler must combine the groups under the assumption that the sky temperature is consistent across these observations.
The noise in different TOD sets is modelled independently. 
This is because different scans are taken at different times and under different receiver conditions. Treating them independently is therefore both physically motivated and numerically natural: in particular, each noise covariance matrix has size $N{\rm time} \times N_{\rm time}$, and its manipulation dominates the cost of noise parameter sampling regardless of how many TOD sets are present.

As a result, Eq.~(\ref{eq: linear sampler}) can naturally be extended to accommodate multiple TODs:
\begin{multline}
\label{eq: GCR sampler joint solver}
\left(
\mathbf{C}^{-1}
+
\sum_{j}
{\mathbf{U}_{j}}^\top \mathbf{\Sigma}_{j}^{-1} \mathbf{U}_{j}
\right) \boldsymbol{p}_{\text{sample}}
= \\
\sum_{j}
{\mathbf{U}_{j}}^\top 
\left[
\mathbf{\Sigma}_{j}^{-1} \boldsymbol{d}''_{j}
+
\mathbf{\Sigma}_{j}^{-\frac{1}{2}} \boldsymbol{\omega}_{j}
\right]
+
\mathbf{C}^{-1}\Bar{\boldsymbol{p}}
+
\mathbf{C}^{-\frac{1}{2}} \boldsymbol{\eta},
\end{multline}
where $j$ is the index of the TOD set within the group. 

The specific implementation of the Gibbs sampling step [Eq.~(\ref{eq: sampling p sky})] is given by
\begin{align*}
    &\boldsymbol{d}_j'' = \boldsymbol{d}_j / \boldsymbol{g}_j^{(i+1)} - \mathbf{U}_{\rm loc,j} \boldsymbol{p}^{(i+1)}_{\rm loc, j}, \quad
    \mathbf{U}  = \mathbf{U}_{\rm cel}, \quad
    \mathbf{C}  = \mathbf{C}_{\rm cel}, \quad
    \Bar{\boldsymbol{p}}  = \Bar{\boldsymbol{p}}_{\rm cel}, \\
    &\mathbf{\Sigma} \xleftarrow{\text{Eq.~(\ref{eq: final Sigma})}} \boldsymbol{p}^{(i+1)}_{\rm cel, GLS} \xleftarrow{\text{iterative GLS}} \boldsymbol{g}_j^{(i+1)}, \boldsymbol{p}^{(i+1)}_{\rm loc, j}, f_{0,j}^{(i+1)}, \alpha^{(i+1)}_j .
\end{align*}
We note that the dimension of this linear system matches that of $\boldsymbol{p}_{\rm cel}$ and is independent of the number of TOD sets. 
While the dimension of the linear system in Eq.~(\ref{eq: GCR sampler joint solver}) is independent of the number of TOD sets, the total computational cost of the full Gibbs iteration scales linearly with the number of TOD sets, as each requires independent sampling of its gain, local $T_{\rm sys}$, and noise parameters. Since these per-TOD steps are conditionally independent given the celestial parameters, they can be executed in parallel across MPI ranks, so the wall-clock time per iteration need not increase with the number of TOD sets.

\subsection{Flicker noise parameter sampler}
\label{sec: noise sampler}
\begin{figure}
    \centering
    \includegraphics[width=\linewidth]{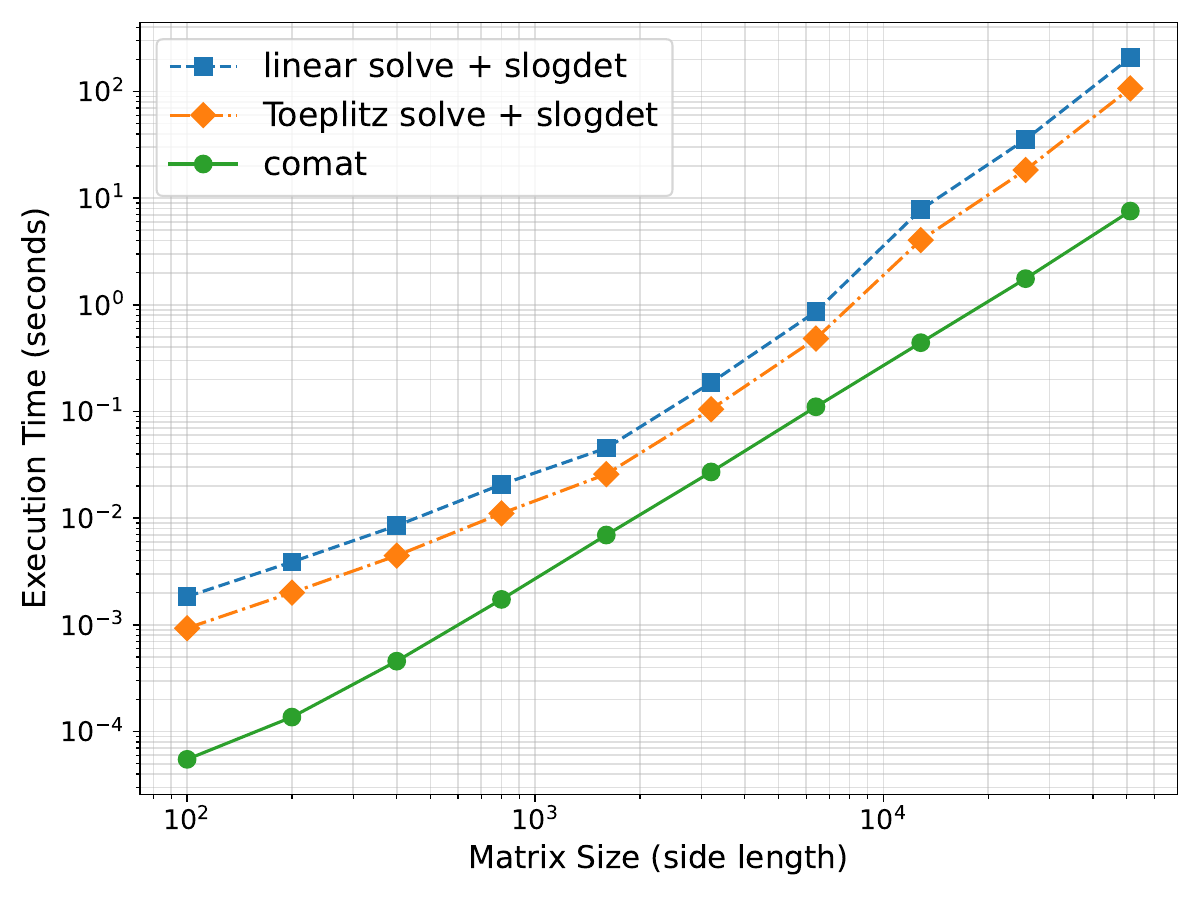}
    \caption{Execution time comparison of three computational methods for evaluating the log-likelihood with real symmetric Toeplitz noise covariance matrices.
    We compare three implementations, listed in order of decreasing computational complexity:
    (1) using {\tt NumPy}'s {\tt slogdet} and {\tt solve};
    (2) using {\tt NumPy}'s {\tt slogdet} combined with {\tt SciPy}'s {\tt solve\_toeplitz};
    (3) using our own {\tt comat} implementation.
    As shown, {\tt comat} outperforms the other methods in terms of execution time. This benchmark analysis was performed on a Mac Studio equipped with an Apple M3 Ultra chip.
    }
    \label{fig: comat efficiency}
\end{figure}

In this section, we discuss how to sample the two noise parameters $\{f_0, \alpha\}$ for each TOD set independently [see Eq.~(\ref{eq: sampling noise})], when the sample of smooth gain and system temperature are provided. 
The main effort to sample the noise parameters is to calculate the logarithmic determinant and the quadratic form involving $\Ncov^{\minus 1}$ in Eq.~(\ref{eq: likelihood function}). 

At first glance, the computation appears intractable for large data sizes, since it involves frequent computation of the inverse and determinant of a non-diagonal matrix, which would seem to imply a computational complexity of $\mathcal{O}(N^3)$. However, by recognising the unique structure of $\Ncov$ [see Eq.~(\ref{eq: explicit form of Ncorr}) for an explicit expression], 
which has a single repeated element along each of its diagonals,
we can see that it can be transformed into a $\mathcal{O}(N^2)$-flops computation.
The quadratic form can be reformulated as the inner product of two vectors: $\Tr(\mathbf{N}^{\minus \! 1}\mathbf{D}) = \boldsymbol{n}^\top \boldsymbol{x}$, where $\boldsymbol{x}$ is obtained by solving
$
\Ncov \boldsymbol{x} = \boldsymbol{n}.
$
\footnote{Here $\boldsymbol{n}$, as defined in Eq.~(\ref{eq: residual noise definition}), is the residual of the data with the gain and system temperature sample.}
This linear equation can be solved quickly with $\mathcal{O}(N^2)$ time complexity using the Levinson-Durbin recursion algorithm. The same recursion algorithm can be used to compute the logarithmic determinant term. 
We produce a fast dedicated code, {\tt comat}\footnote{\href{https://github.com/zzhang0123/comat}{https://github.com/zzhang0123/comat}}, for the joint computation of the log-determinant, $\ln{|\mathbf{N}|}$, and the quadratic form, $\Tr(\mathbf{N}^{\minus 1}\mathbf{D})$.
The algorithm is presented in Appendix~\ref{Appendix: numerical treatment Ncov}.
To benchmark computational efficiency, Figure~\ref{fig: comat efficiency} compares the execution time of {\tt comat} with two alternative implementations.
For a typical TOD length of a few thousand time samples, {\tt comat} enables the log-likelihood to be calculated in tens of milliseconds or less on an M3 Mac -- roughly ten times faster than other methods.

In addition, we present a perturbative method in Appendix~\ref{Appendix: perturbed matrix inverse and det} for computing the inverse and determinant of the noise covariance matrix. While this approach exhibits superior numerical performance, it is valid only under the assumption of very weak $1/f$ noise. Nevertheless, due to its computational advantages, we include it for reference. To accommodate a broader range of flicker noise scenarios, we adopt the Levinson algorithm as the primary method in this work.

\section{Demonstration}
\label{sec: example}

This section presents a series of analyses using simulated time-ordered data (TOD) to validate the proposed method. These analyses mimic MeerKLASS single-dish observing mode \citep{wang2021h}.
The performance of our joint calibration and map-making approach is evaluated under various experimental configurations (summarised in Section~\ref{sec: experimental setup}) and prior setups (summarised in Section~\ref{sec: prior setup}), with a detailed analysis presented in Section~\ref{sec: sim results}.

\begin{table}
  \centering
  \caption{Specifications of the simulation configuration}
  \label{tab:sim_params}
  \begin{tabular}{l c}
    \toprule
    Telescope location (lat, lon, height) & ($-30.713^\circ$, $21.443^\circ$, $1054$~m)\\
    Beam & Gaussian (FWHM$=1.1^\circ$) \\
    Sky/Beam pixelation & $N_{\rm side} = 64$ \\
    Scan speed (along azimuth) & $5$ arcmin/s \\
    Frequency & $750$~MHz \\
    Frequency resolution & 0.2~MHz \\
    Exposure time (per scan) & 1.5~hr \\
    Time resolution & 2~s \\
    \midrule
    Elevation (``setting'') & $41.5^\circ$\\
    Elevation (``rising'') & $40.5^\circ$ \\
    Az range (``setting'') & $(-60.3^\circ, -42.3^\circ)$\\
    Az range (``rising'') & $(43.7^\circ, 61.7^\circ$)\\
    Start time (``setting'') & ``2019-04-23 20:41:56''\\
    Start time (``rising'')  & ``2019-03-30 17:19:02''\\
    \midrule
    Gain coefficients (``setting'') & \{6.312, 0.420, 0.264, 0.056\}\\
    Gain coefficients (``rising'') & \{6.845, 0.142, 0.744, 0.779\}\\
    Diode injection & $15$~K (every $10$ data points) \\
    Residual $T_{\rm sys}$ coefficients & \{12.6, 0.5, 0.5, 0.5\} \\
    $1/f$ noise cutoff frequency $f_c$  & $1.099 \times 10^{-3}$ rad/s\\
    $1/f$ noise reference frequency $f_0$  & $1.335 \times 10^{-5}$ rad/s\\
    Effective knee frequency  & $1$~mHz \\
    $1/f$ noise PSD power index & $-2$ \\
    \bottomrule
  \end{tabular}
\end{table}

\begin{figure}
  \centering
  

  \begin{subfigure}[b]{\linewidth}
    \includegraphics[width=\linewidth]{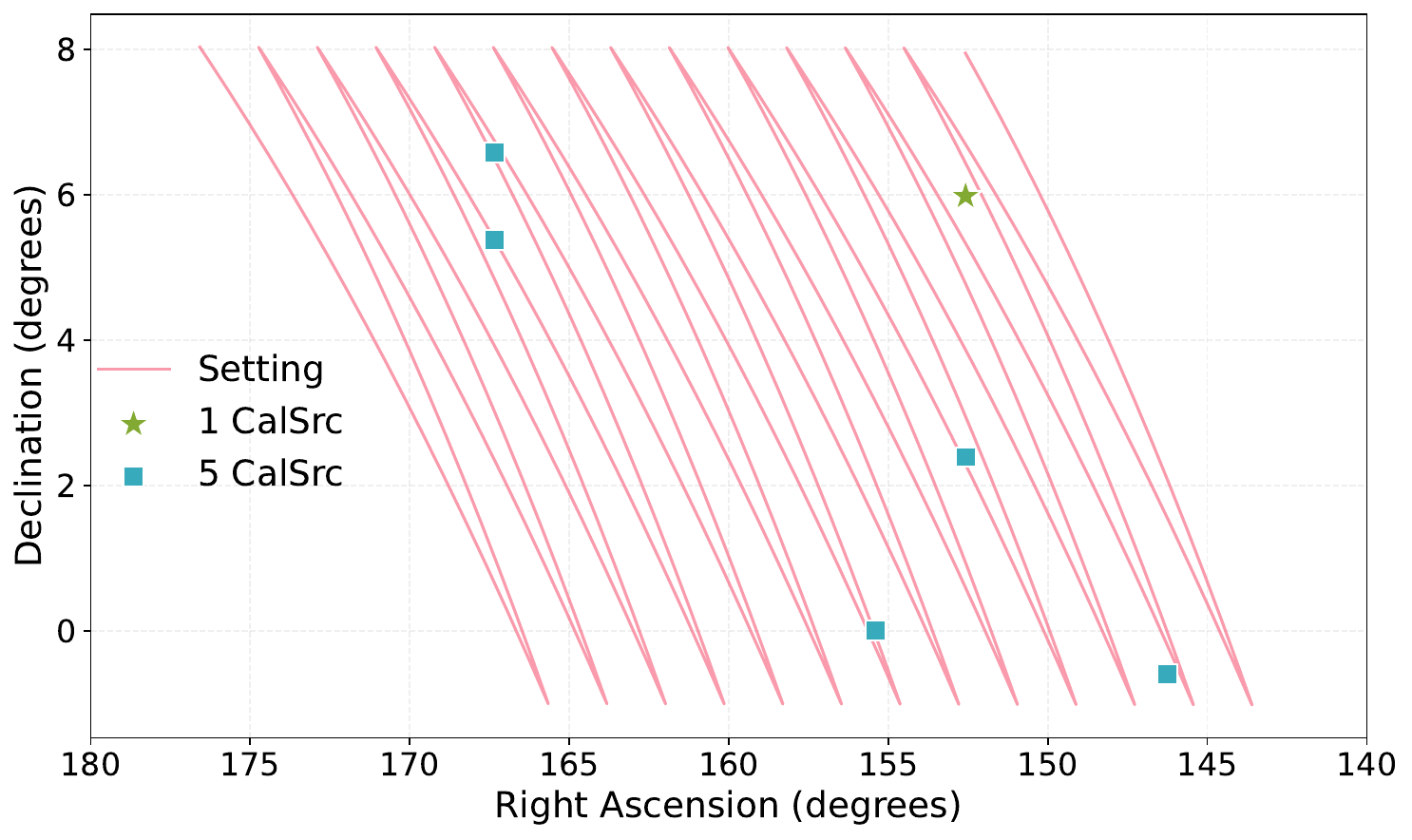}
    \caption{\textbf{$1\times$TOD}: Scan pattern and calibration sources.}
    \label{fig:sub_scan1}
  \end{subfigure}
  \vspace{1em}

  \begin{subfigure}[b]{\linewidth}
    \includegraphics[width=\linewidth]{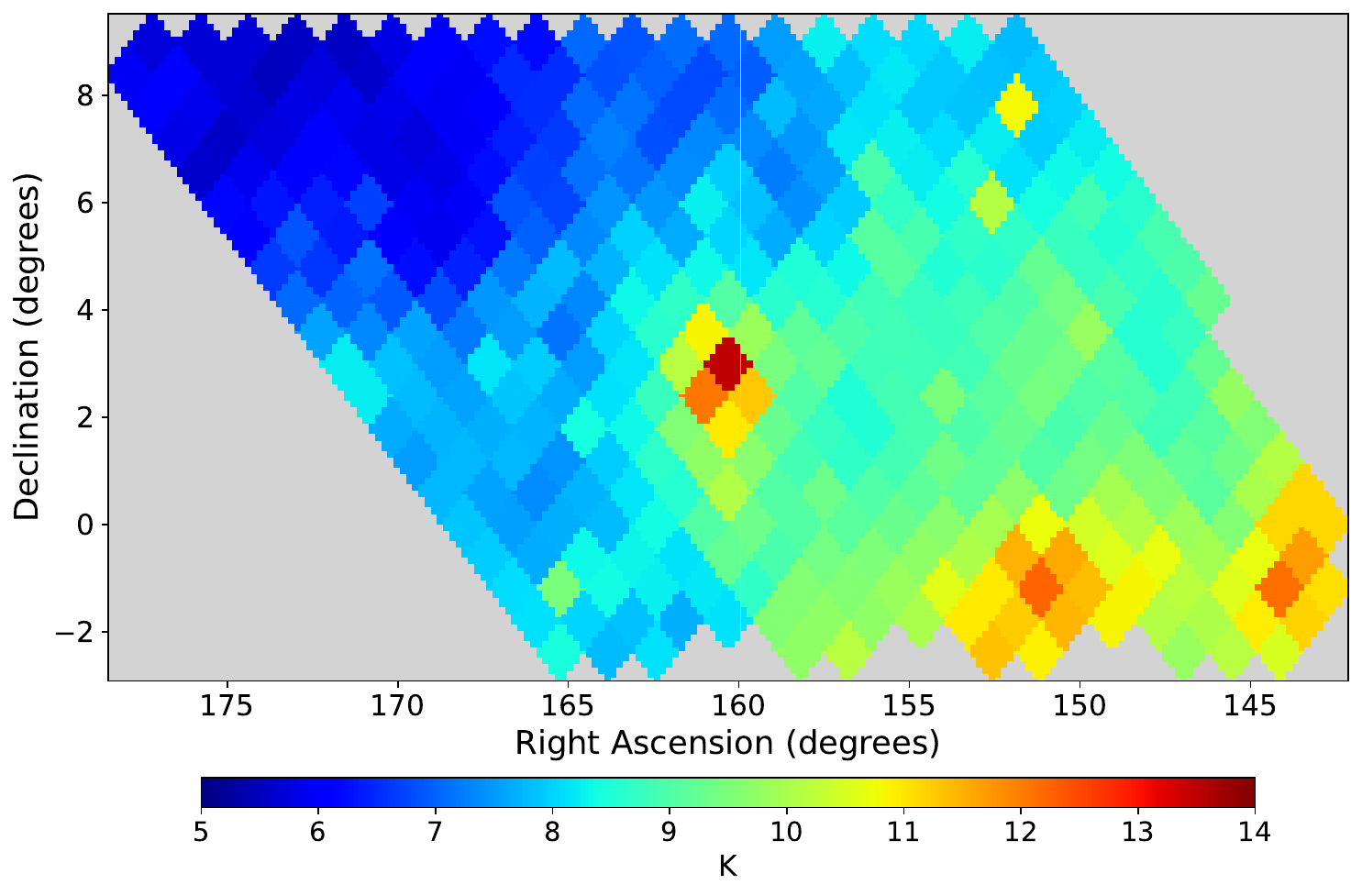}
    \caption{\textbf{$1\times$TOD}: Covered sky map.}
    \label{fig:sub_sky1}
  \end{subfigure}
  \hfill
  \begin{subfigure}[b]{\linewidth}
    \includegraphics[width=\linewidth]{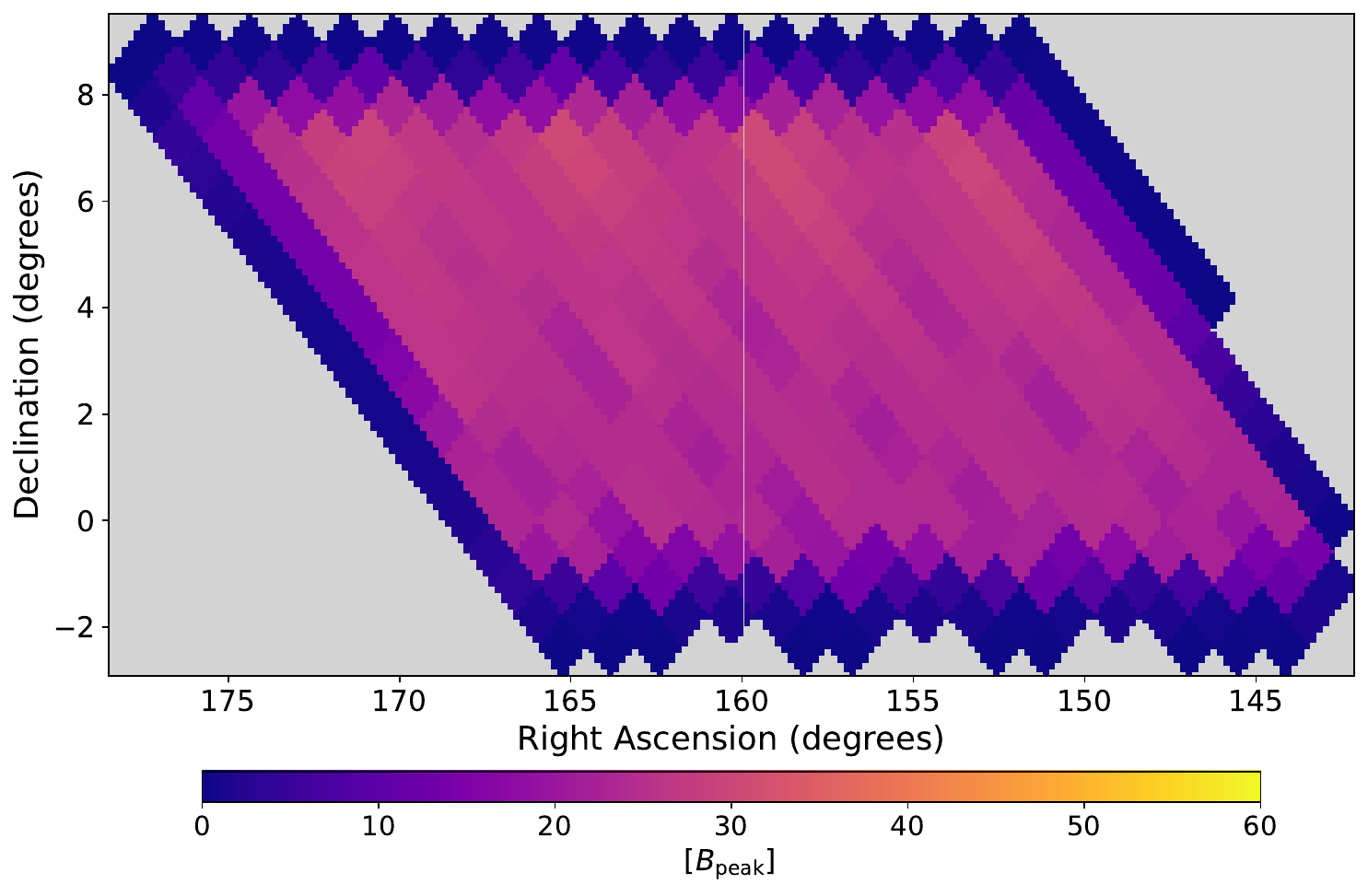}
    \caption{\textbf{$1\times$TOD}: 
    Integrated beam intensity $\sum_t B(p,t)$, where $B(p,t)$ is the beam response at pixel $p$ and time $t$, normalised so that the peak response $B_{\rm peak}=1$. 
    The value can be interpreted as the effective number of data points centred at that pixel in units of $B_{\rm peak}$. }
    \label{fig:sub_beam1}
  \end{subfigure}
  
  \caption{Scan pattern, covered sky and integrated beam intensity for the ``$1\times$TOD'' configuration.}
  \label{fig: 1TOD setup}
\end{figure}

\begin{figure}
  \centering
  

  \begin{subfigure}[b]{\linewidth}
    \centering
    \includegraphics[width=\linewidth]{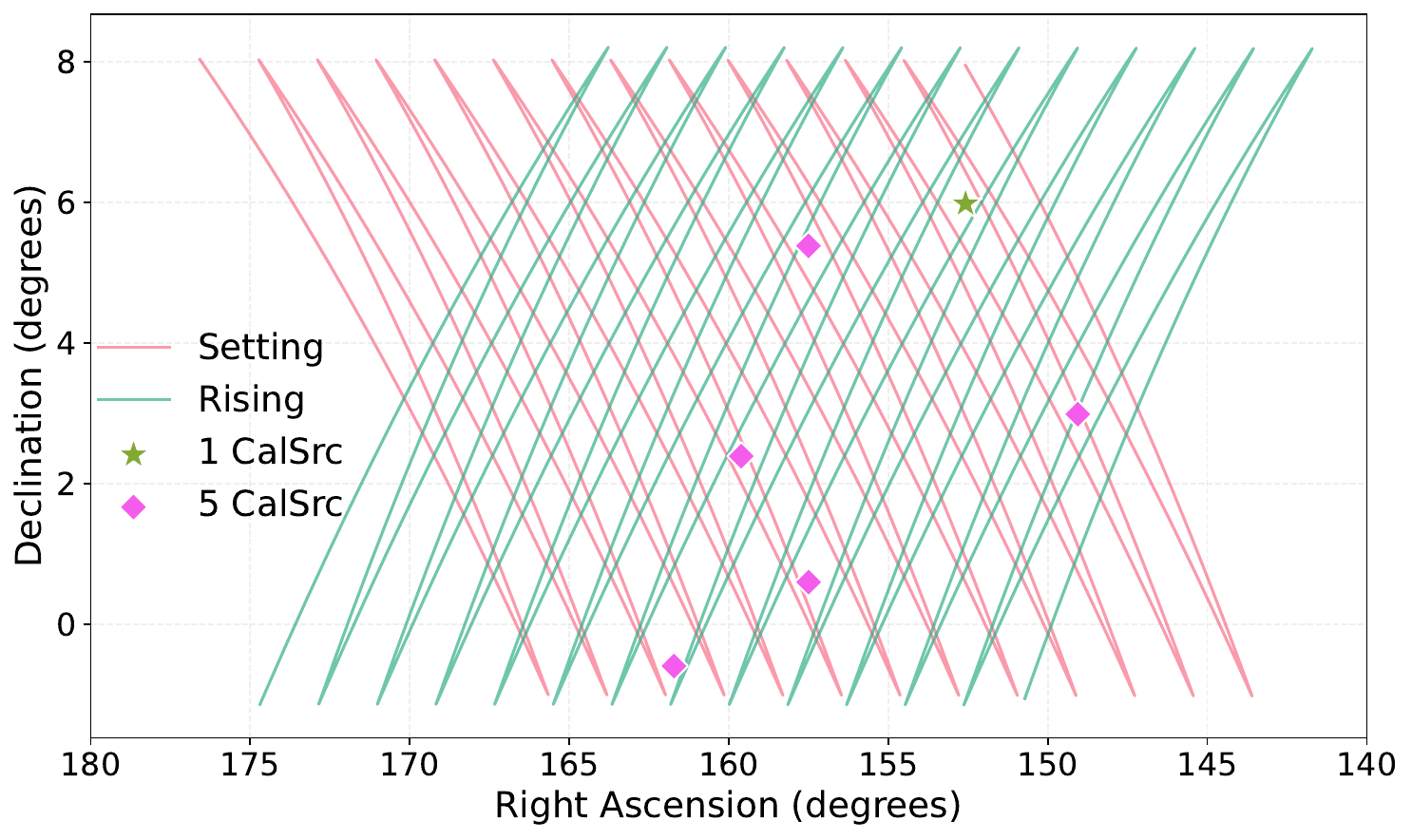}
    \caption{\textbf{$2\times$TOD}: Scan pattern and calibration sources.}
    \label{fig:sub_scan2}
  \end{subfigure}
  \vspace{1em}

  \begin{subfigure}[b]{\linewidth}
    \centering
    \includegraphics[width=\linewidth]{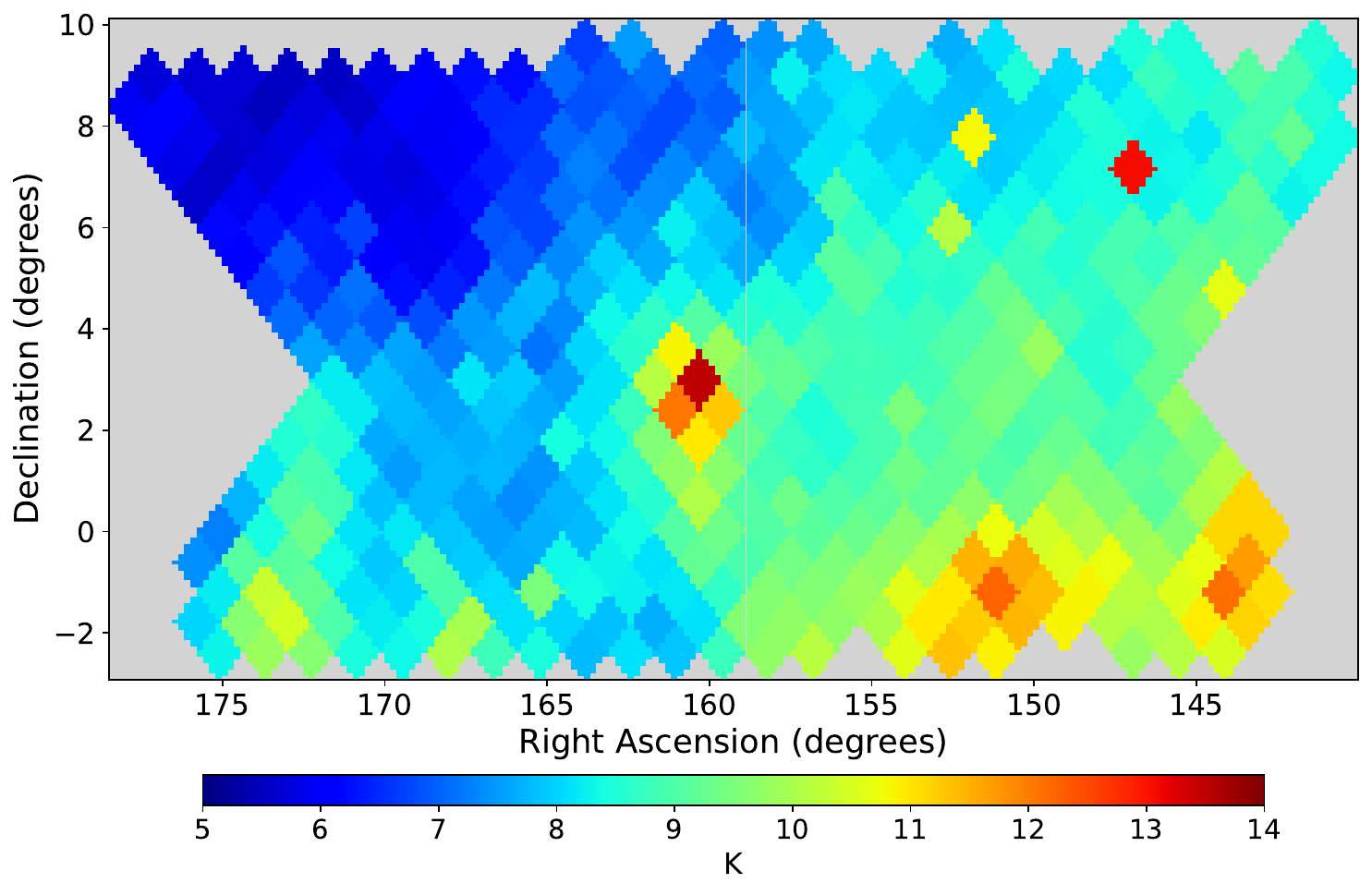}
    \caption{\textbf{$2\times$TOD}: Covered sky map.}
    \label{fig:sub_sky2}
  \end{subfigure}
  \hfill
  \begin{subfigure}[b]{\linewidth}
    \centering
    \includegraphics[width=\linewidth]{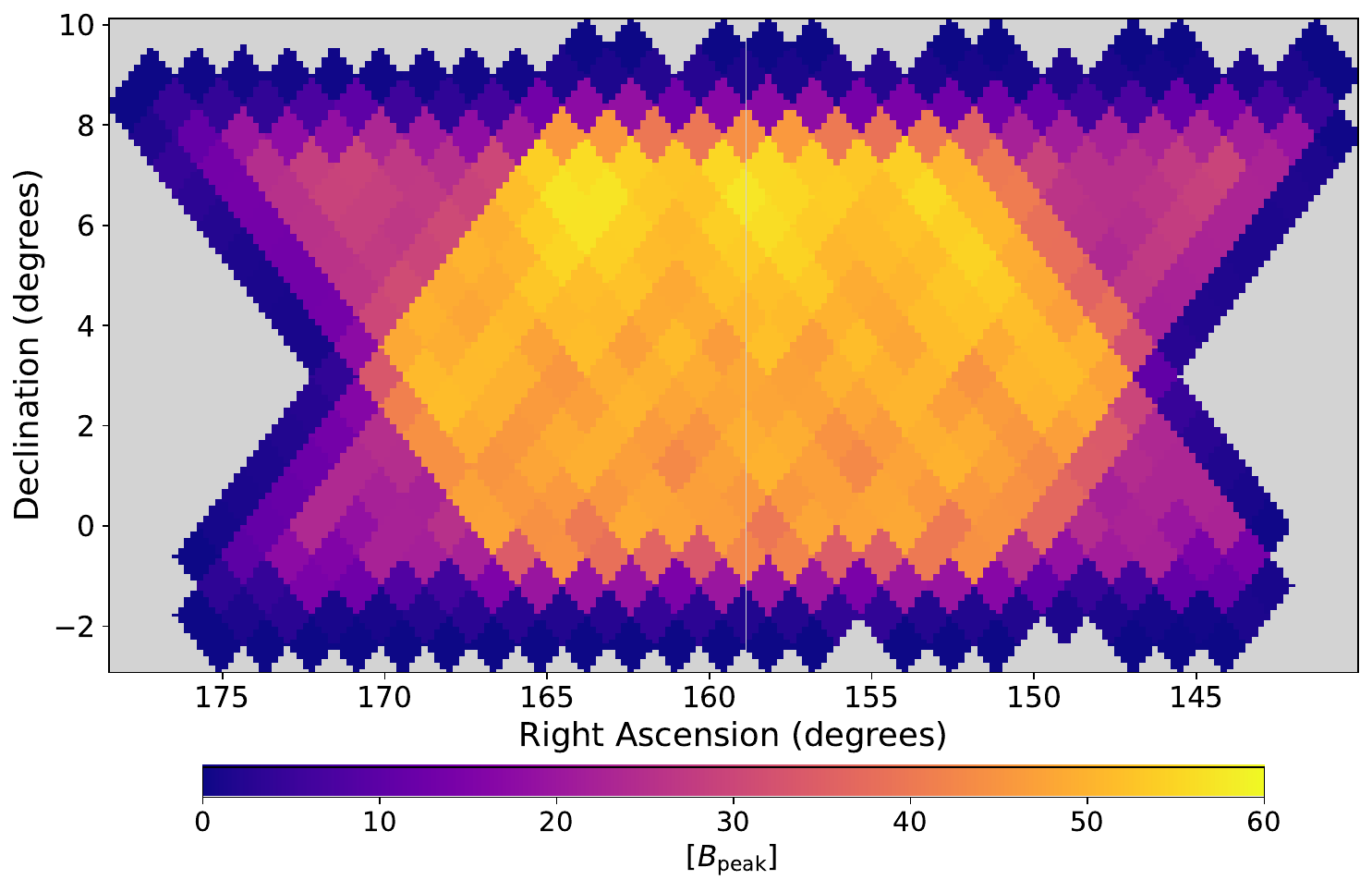}
    \caption{\textbf{$2\times$TOD}: 
    Integrated beam intensity $\sum_t B(p,t)$, where $B(p,t)$ is the beam response at pixel $p$ and time $t$, normalised so that the peak response $B_{\rm peak}=1$. The value at each pixel can therefore be interpreted as the effective number of data points centred at that pixel in units of $B_{\rm peak}$. Pixels in scan-overlap regions and near azimuth turnaround points accumulate contributions from multiple passes and therefore reach high values.
    }
    \label{fig:sub_beam2}
  \end{subfigure}
  
  \caption{Scan pattern, covered sky and integrated beam intensity for the ``$2\times$TOD'' configuration.}
  \label{fig: 2TOD setup}
\end{figure}

\subsection{Experimental setup}
\label{sec: experimental setup}
To emulate realistic observational conditions, we simulated two distinct time-ordered data (TOD) sets, each of which fully replicates MeerKLASS single-dish observing patterns. These are denoted as the `setting' and `rising' modes based on differences in scan strategy (as detailed in subsequent Section~\ref{sec: scan}).

Using these simulations, we designed two Bayesian analysis workflow scenarios:
\begin{itemize}
    \item \textbf{``$1\times$TOD'':} Only the ``setting''-mode TOD set is used; one TOD is used.
    \item \textbf{``$2\times$TOD'':} Both the ``setting'' and ``rising'' TOD sets are used; one TOD is used for each scan.
\end{itemize}
This design serves two purposes:
(1) Validating the workflow's capability to process both single and multiple TOD sets;
(2) Quantifying map-making accuracy by comparing results between the ``$1\times$TOD'' and ``$2\times$TOD'' configurations.
Further experimental details are provided in subsequent sections. 
Figures~\ref{fig: 1TOD setup} and \ref{fig: 2TOD setup} illustrate the two scenarios. The key parameters are summarised in Table~\ref{tab:sim_params}.

\subsubsection{Scanning strategy}
\label{sec: scan}

The MeerKLASS scanning strategy involves rapidly moving dishes in azimuth at a constant elevation to minimize variations from ground spill and airmass. Scanning is done at $5$~arcmin/s, ensuring the pointing remains within the primary beam during each $2$-second time dump. 
This allows coverage of ~$10^\circ$ in $100$ seconds, matching the timescale over which receiver gain remains relatively stable. 
Each scan stripe is $18^\circ$ wide, with back-and-forth slews taking about $200$ seconds. 

At a fixed elevation, each sky strip can be scanned using two methods: one when the field is rising and one when the field is setting.
These two scans overlap, as shown in Figure~\ref{fig: 2TOD setup}, providing improved sky coverage in the overlap region.
In this work, we simulate two TOD sets corresponding to the ``rising'' mode and the ``setting'' mode, respectively.

\subsubsection{Power beam and sky model}
We adopt a Gaussian beam with a full width at half maximum (FWHM) of $1.1^\circ$ to approximate the MeerKLASS single-dish beam profile. This beam model is applied consistently in both the TOD simulation and the construction of the linear operator that maps sky parameters (pixels) to the signal component in the data.

If the beam model is accurate, the resulting sky map will be fully deconvolved. Conversely, if the beam model is inaccurate\footnote{In the extreme case of assuming an idealised pencil beam, no deconvolution is performed at all.} -- which, without loss of generality, can be understood as the sky field absorbing a multiplicative distortion in Fourier space -- the recovered sky map will remain partially convolved (see, for example, the Appendix of \cite{matshawule2021h} for a discussion). This can be abstractly expressed as:\footnote{
We note that the equations are presented solely to give an intuitive picture of how beam modelling errors propagate into the recovered sky map; no explicit numerical deconvolution is performed in this work. In the Gibbs sampler, the beam enters as a linear forward operator mapping sky pixels to TOD samples, and sampling the sky conditioned on this operator implicitly accounts for the beam.
}
\be
\Tilde{B}(\ell) \Tilde{T}(\ell)
= \Tilde{B}_{\rm est}(\ell)\, \Tilde{E}(\ell)\, \Tilde{T}(\ell) \,
= \Tilde{B}_{\rm est}(\ell)\, \Tilde{T}_{\rm est}(\ell),
\ee
where $B$ and $T$ denote the true beam and sky temperature fields, respectively; $B_\mathrm{est}$ and $T_\mathrm{est}$ are their estimates; $\tilde{\cdot}$ denotes the Fourier transform; $\ell$ is the wavenumber; and $E$ is the effective error field defined by the above relation. The estimated sky $T_\mathrm{est}$ can be written as
\be
T_{\rm est}(\hat{n}) \equiv (T \ast E)(\hat{n})
\ee
is effectively the true sky field convolved with a kernel $E$. For intensity mapping applications, we assume that the beam model is sufficiently accurate for the purpose of science extraction.

The sky model consists of two components: (1) the diffuse emission is taken from the Global Sky Model \citep[][GSM]{de2008model} full-sky map at $750\,\mathrm{MHz}$; (2) the point source component is constructed by extending the GLEAM catalog \citep{hurley2017galactic}, originally covering the southern sky, to full-sky coverage via a simple mirroring approach, and then scale source fluxes to $750\,\mathrm{MHz}$ assuming a uniform spectral index of $-2.3$.
We represent the sky using the HEALPix pixelation scheme with $\texttt{Nside}=64$, which is sufficient for validating the analysis workflow. 
At each time sample, the beam is truncated at $3\sigma$ solely to identify the set of pixels receiving non-negligible signal; the full sky pixel set is the union of all per-sample covered pixels. For each data point, the beam intensity is then evaluated at every pixel in this set without further approximation. For the $2\times\mathrm{TOD}$ configuration, the telescope scanning at $5\,\mathrm{arcmin\,s}^{-1}$ yields approximately 5,700 time samples distributed across 473 sky pixels, ensuring the sky parameter estimation is well conditioned.

Figures~\ref{fig: 1TOD setup} and \ref{fig: 2TOD setup} show the sky coverage and integrated beam intensity for the $1\times$TOD and $2\times$TOD configurations, respectively.
Pixels in scan-overlap regions and near azimuth turnaround points accumulate beam contributions from multiple passes and therefore reach the highest integrated beam intensity, as shown in panel~(c) of Figures~\ref{fig: 1TOD setup} and \ref{fig: 2TOD setup}.
In practical applications, higher resolutions or lower truncation thresholds may be used when more independent measurements are available -- for instance, more overlapping TODs or data from multiple telescopes.

\subsubsection{Gain, noise diode and receiver temperature}

For some observations in the MeerKLASS survey, 
MeerKAT periodically fires noise diodes for $1.8$ seconds every $20$ seconds to provide a relative calibration reference \citep{wang2021h}.
Although our workflow does not rely on this feature, we incorporate the noise diode signal in the simulation to better reflect realistic observing conditions. For simplicity, we inject a $15$~K noise diode signal every ten data points, corresponding to a $20$-second cycle.

In the analysis, we assume that the timing of the noise diode injections is known and that the diode's temperature remains stable over time. However, the exact amplitude of the noise diode signal is treated as an unknown and is included as a free parameter to be fitted.

All residual contributions to the system temperature -- such as receiver noise, ground pickup, and atmospheric contamination -- are collectively referred to as the residual temperature (denoted by $T_{\rm res}$) and are modelled using four Legendre polynomial modes:
\begin{align}
    T_{\rm res} &= \sum_{n=0}^{3} c_n \, P_n(x),
    &
    x&= \frac{2(t_a - t_{\rm min})}{t_{\rm max}-t_{\rm min}} - 1,
\end{align}
with the coefficient values $c_n$ used in the simulation listed in Table~\ref{tab:sim_params}.
Similarly, for the smooth temporal gain variations, we do not impose a specific functional form beyond the assumption of temporal smoothness. These gain fluctuations are also modelled using Legendre polynomials [see Eq.~(\ref{eq: parameterisation TildeG})], with randomly assigned coefficients for simulation.

The flicker noise component is modelled in accordance with the Case~2 prescription shown in Figure~\ref{fig: flicker model comparisons}, with $\alpha=2$ and $\log_{10} f_0 \simeq -4.8746$, corresponding to a knee frequency of approximately 0.001 cycles per second.
This is comparable to the $1/f$ noise parameters measured in \cite{li2021h} and \cite{irfan2024mitigating}.

\begin{figure*}
    \centering
    \includegraphics[width=\linewidth]{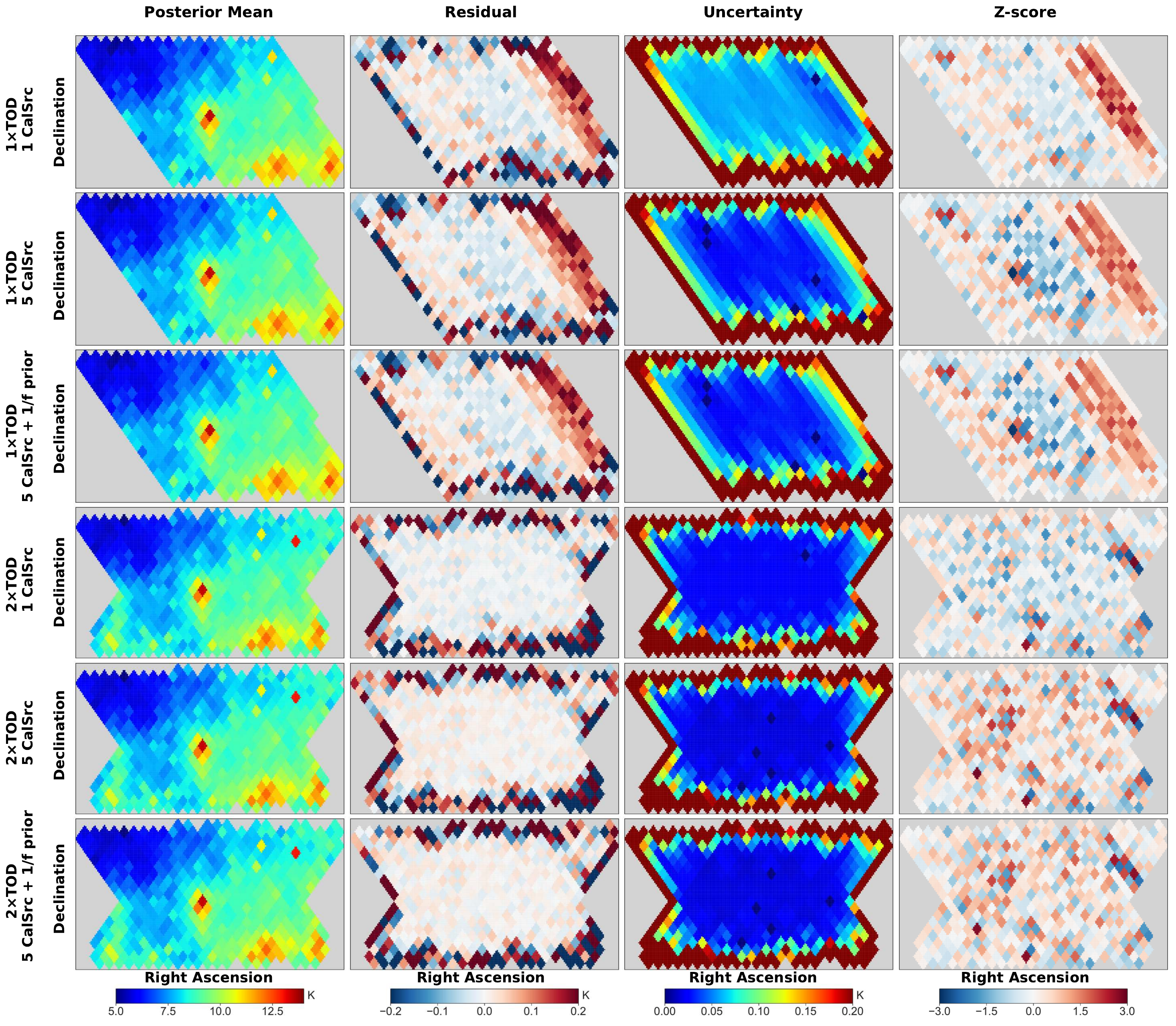}
    \caption{
      Estimated, residual, uncertainty, and Z-score maps in different experimental and prior setups.
      We compare results obtained using a single TOD set (``$1\times$TOD'') versus jointly analyzing two TOD sets (``$2\times$TOD'').
      We compare three different priors for each scenario: (1) 1 CalSrc, (2) 5 CalSrc, and (3) 5 CalSrc + $1/f$ prior  (see Section~\ref{sec: prior setup}).
      The different rows correspond to different setups.      
      \textbf{Posterior mean map} (left column): Posterior mean of sky parameters marginalized over other parameters, calculated as the sample average. 
      \textbf{Residual map} (middle-left column):
      Map showing the residual difference between the estimated sky signal and the true sky input used in simulations. 
      \textbf{Uncertainty map} (middle-right column): 
      Posterior standard deviation of sky parameters marginalized over other parameters, calculated as the sample standard deviation of the draws.
      The deep-blue pixels in the uncertainty map indicate the positions of the \textbf{calibration sources}, whose sky temperatures are tightly constrained by the strong prior imposed on external calibration references.
      \textbf{Z-score map} (right column):  defined as the residual map divided by the posterior standard deviation.
      }
    \label{fig: combined maps}
\end{figure*}

\begin{figure*}
    \centering
    \includegraphics[width=\linewidth]{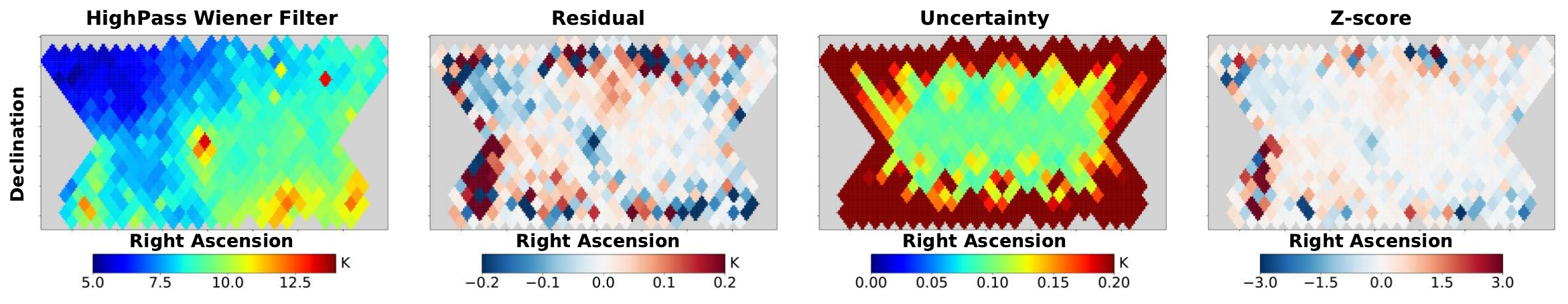}
    \caption{
    Map reconstructed using the high-pass + Wiener filter method. This method shows larger residuals (in the interior of the survey area) and higher uncertainty compared to the Bayesian approach.}
    \label{fig: wiener-filer maps}
\end{figure*}

\begin{figure*}
    \centering
    \includegraphics[width=\linewidth]{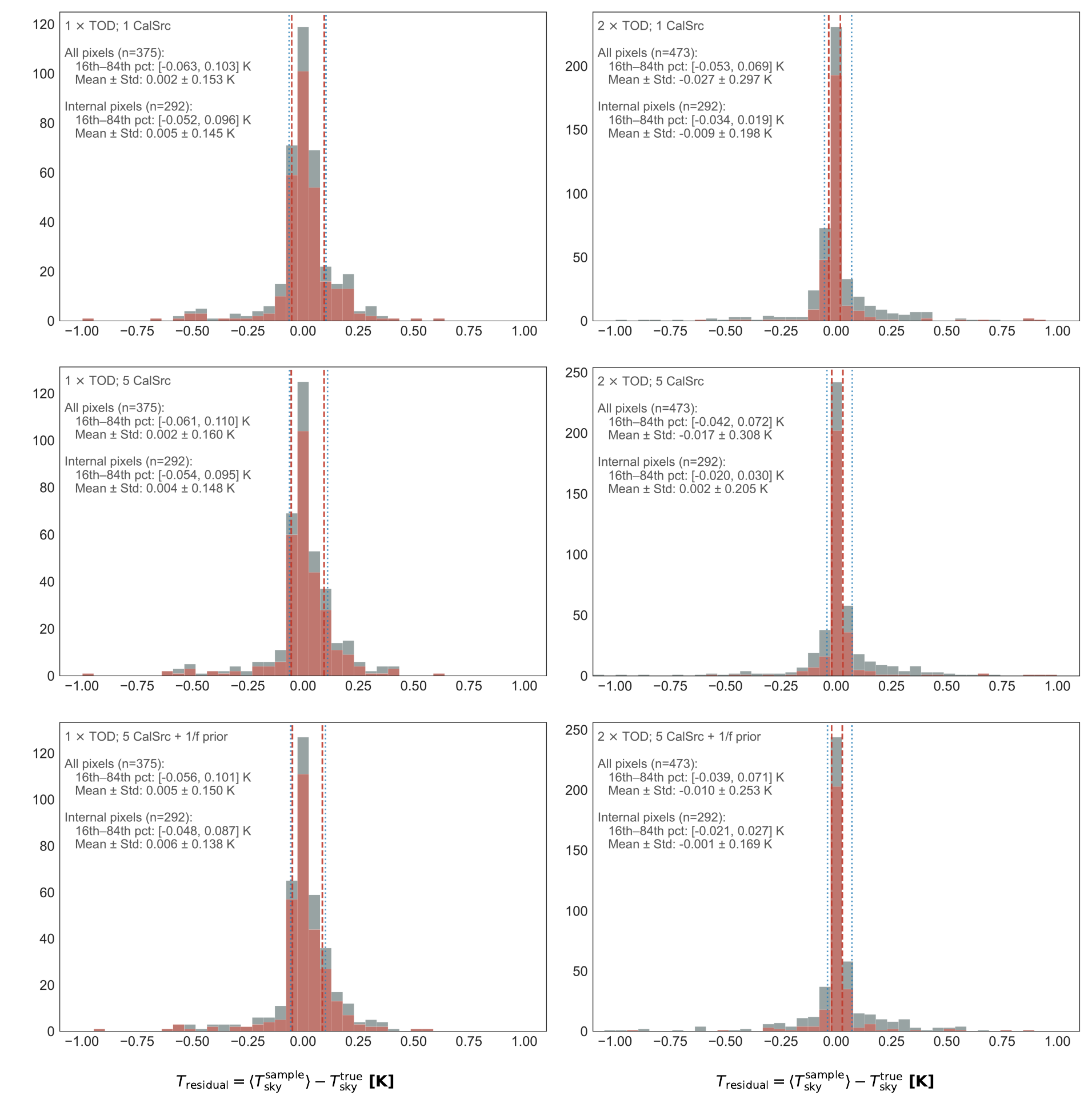}
    \caption{Distribution of residuals in the sky map estimate, defined as the difference between the posterior mean (the mean of the sampled sky pixels) and the true sky map. 
    The grey histogram (background) shows the residual distribution across all pixels in the experimental setup. The red histogram highlights the distribution of internal pixels, which are defined as pixels that appear in both the 1×TOD and 2×TOD cases. 
    Both histograms are overlaid with kernel density estimates (KDEs) for visual clarity.
    The vertical dotted lines mark the $16$th and $84$th percentiles, capturing the central $68$\% interval of the distribution.}
    \label{fig: combined hist}
\end{figure*}

\begin{figure*}
    \centering
    \includegraphics[width=0.9\linewidth]{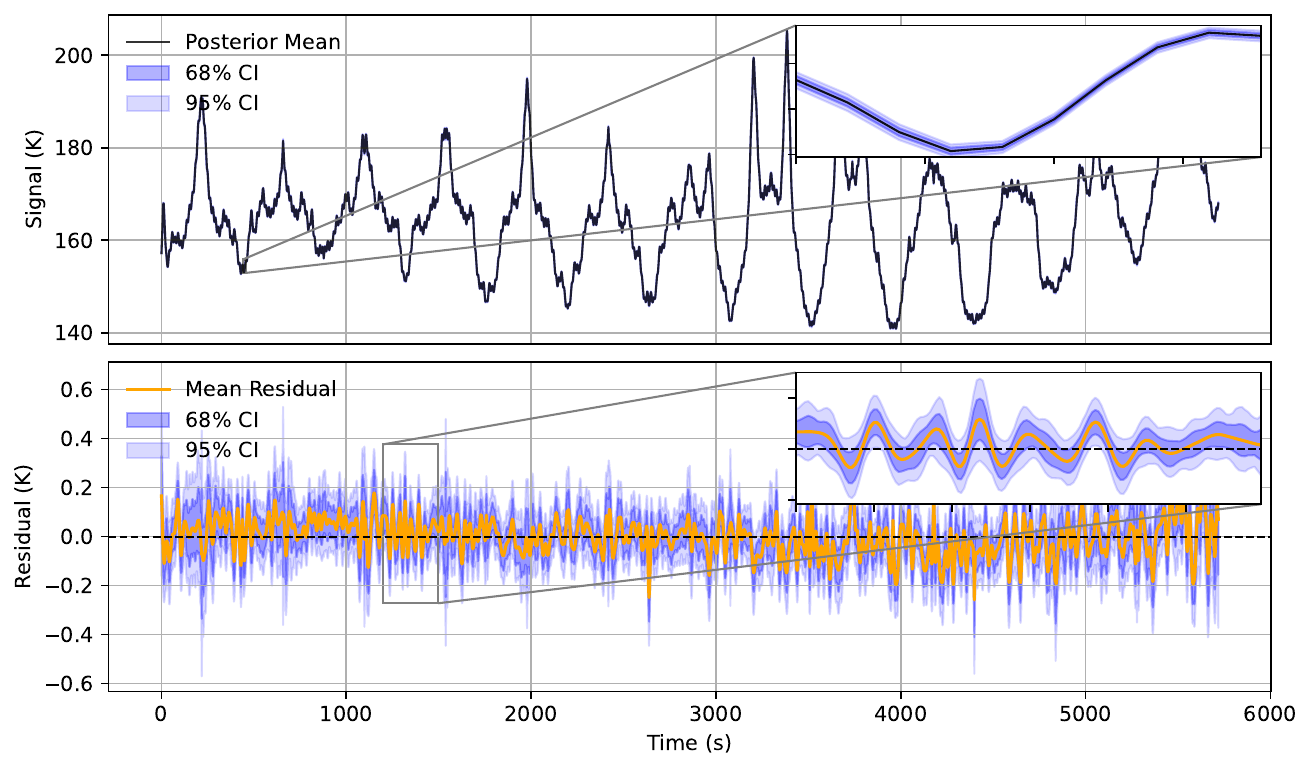}
    \caption{
    Reconstruction of the noise-free time-ordered data ($\mu=\LscaleG \Tsys$, i.e.,  the smooth TOD as the smooth gain multiplied by the system temperature) using posterior samples. The \textbf{top} panel shows the mean reconstructed model ($ \mu_{\rm est} \equiv \langle\mu_{\rm samples}\rangle$; black solid), and shaded regions indicating the 68\% and 95\% credible intervals. 
    The \textbf{bottom} panel shows the mean residual TOD (orange solid), given by $ \langle \Delta \mu_{\rm samples} \rangle \equiv \mu_{\rm est} - \mu_{\rm true}$. 
    The shaded regions indicate the 68\% and 95\% credible intervals of the residuals.
    The true TOD is taken from the ``setting'' mode simulation, with both noise and noise diode injection removed. The reconstructed TOD is generated using posterior samples from the ``$2\times$TOD; 1 CalSrc'' analysis.
    The inset window provides a zoomed-in view of a small segment, highlighting subtle variations in the reconstructed signal and uncertainties.
    This figure demonstrates the advantage of the Bayesian workflow, where ensemble sampling enables the construction of credible intervals for any observable.
    }
    \label{fig: posterior predicted TOD}
\end{figure*}

\begin{figure*}
    \centering
    \includegraphics[width=0.9\linewidth]{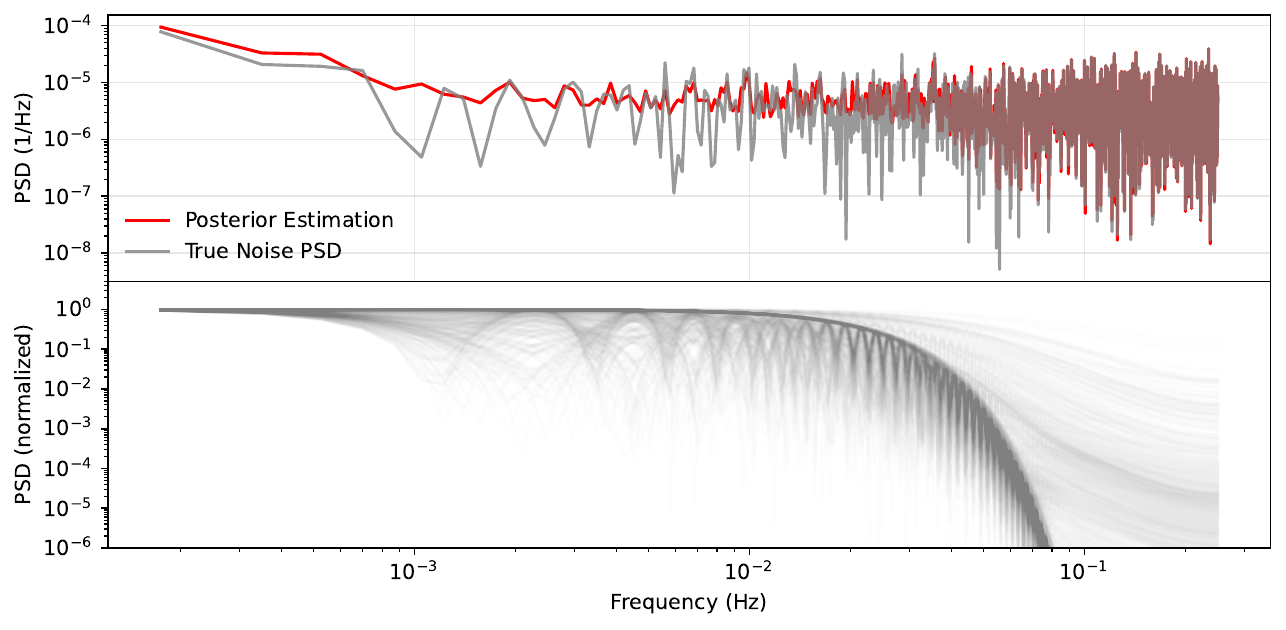}
    \caption{
    Comparison between the posterior mean noise PSD ($\langle \mathrm{PSD}(n_{\rm sample})\rangle$) and the true noise PSD ($\mathrm{PSD}(n_{\rm true})$), illustrating the consistency predicted by Eq.(\ref{eq: posterior noise}).
    \textbf{Top}: The posterior mean noise PSD (red), constructed from $n_{\rm sample}$ as defined in Eq.(\ref{eq: posterior noise}), compared against the true noise PSD (grey) used in the simulation [Eq.~(\ref{eq: residual noise definition})]. 
    \textbf{Bottom}: PSD of the individual $T_{\rm sys}$ basis functions evaluated per sky pixel, each curve corresponding to the contribution of a single pixel. The collective suppression of power at high frequencies reflects the intrinsic smoothness of the basis, and directly accounts for the high-frequency identity of the two curves seen in the top panel (see the discussion in Section~\ref{sec: sim results}). 
    }
    \label{fig: posterior predicted TOD PSD}
\end{figure*}

\subsection{Prior setup}
\label{sec: prior setup}
Having specified the data model, statistical framework, and observational configurations, we now introduce our assumptions about the measurement system necessary to complete the Bayesian analysis of the datasets.

We begin by outlining the fiducial prior setup, which serves as the foundation for all subsequent comparisons. Additional prior configurations are introduced as variations on this baseline to explore their impact on inference performance.

Given the structure of our modelling approach, certain priors are necessary for breaking parameter degeneracies, while others are intentionally avoided to maintain model generality. A key consideration is the degeneracy between the smooth gain fluctuations and the $1/f$ noise component, as the $1/f$ model can in principle absorb large-scale temporal structures. To break this degeneracy, we introduce minimal prior information guided by practical considerations.

In the fiducial scenario, we assume approximate prior knowledge of the smooth component of the gain, motivated by the fact that gain amplitudes can often be constrained using noise diode signals or external calibration procedures. Specifically, we impose a Gaussian prior with a standard deviation of $10\,\%$ on the gain coefficients.
The prior standard deviation is thresholded at $0.1$ for coefficients that are smaller than $1$. The same prior scheme is applied to the local system temperature parameters.
For the sky parameters, we assume rough prior knowledge of the sky signal, assigning each pixel a Gaussian prior with a standard deviation equal to 20\,\% of its sky temperature. 
However, we caution that the $20\%$ sky prior adopted here may not be sufficiently conservative for real observations, given the level of disagreement between commonly used sky models (see \cite{wilensky2025bayesian} for a discussion of model errors in popular sky models that could be greater than 20\,\%). In practice, we recommend deriving the prior mean and width directly from the data -- for instance, by using the output of a linear map-making pipeline as the prior mean or by running the workflow iteratively, with the posterior sky map from an initial broad-prior pass informing a tighter prior in subsequent iterations.

Building upon the fiducial scenario, we consider the following three enhanced prior setups:
\begin{itemize}
    \item \textbf{1 CalSrc: }The fiducial setup plus a strong prior on the temperature of a single pixel, used to establish the absolute flux scale.
    \item \textbf{5 CalSrc: }The fiducial setup plus strong priors on five widely separated pixels, providing more temporal functional information for the sky component.
    \item \textbf{5 CalSrc + $1/f$ Prior: }The above configuration further augmented with a strong prior on the power-law index of the $1/f$ noise component.
\end{itemize}
A strong prior on a single pixel is equivalent to selecting a bright point source as a flux scale calibrator while also assuming an accurate beam model.
Each of these prior setups is evaluated under both the $1\times$TOD and $2\times$TOD observational configurations. The resulting analyses and comparative performance assessments are presented in Section~\ref{sec: sim results}.

\subsection{Results}
\label{sec: sim results}

In this section, we present the map results of joint Bayesian analyses for six toy model scenarios, defined by two experimental configurations and three prior setups. Each scenario is labeled using a two-part notation -- for example, ``(1×TOD; 1 CalSrc)'' -- where the first part denotes the experimental configuration (using a single TOD set), and the second indicates the prior setup (one calibration source prior).
Each experimental scenario is analysed using $2,000$ iterations of the Gibbs sampler. Each iteration takes approximately $12$ seconds on a Mac with an M3 Ultra chip. Full details of the numerical setup are given in Appendix~\ref{sec: sampler numerical spec}.

Our analysis focuses on two key aspects. First, we examine the impact of different experimental setups on map-making performance, specifically comparing the use of a single TOD set versus two overlapping TOD sets. Second, we investigate how varying prior assumptions influence the outcome of the Bayesian inference.

\begin{figure*}
    \centering
    \includegraphics[width=0.83\linewidth]{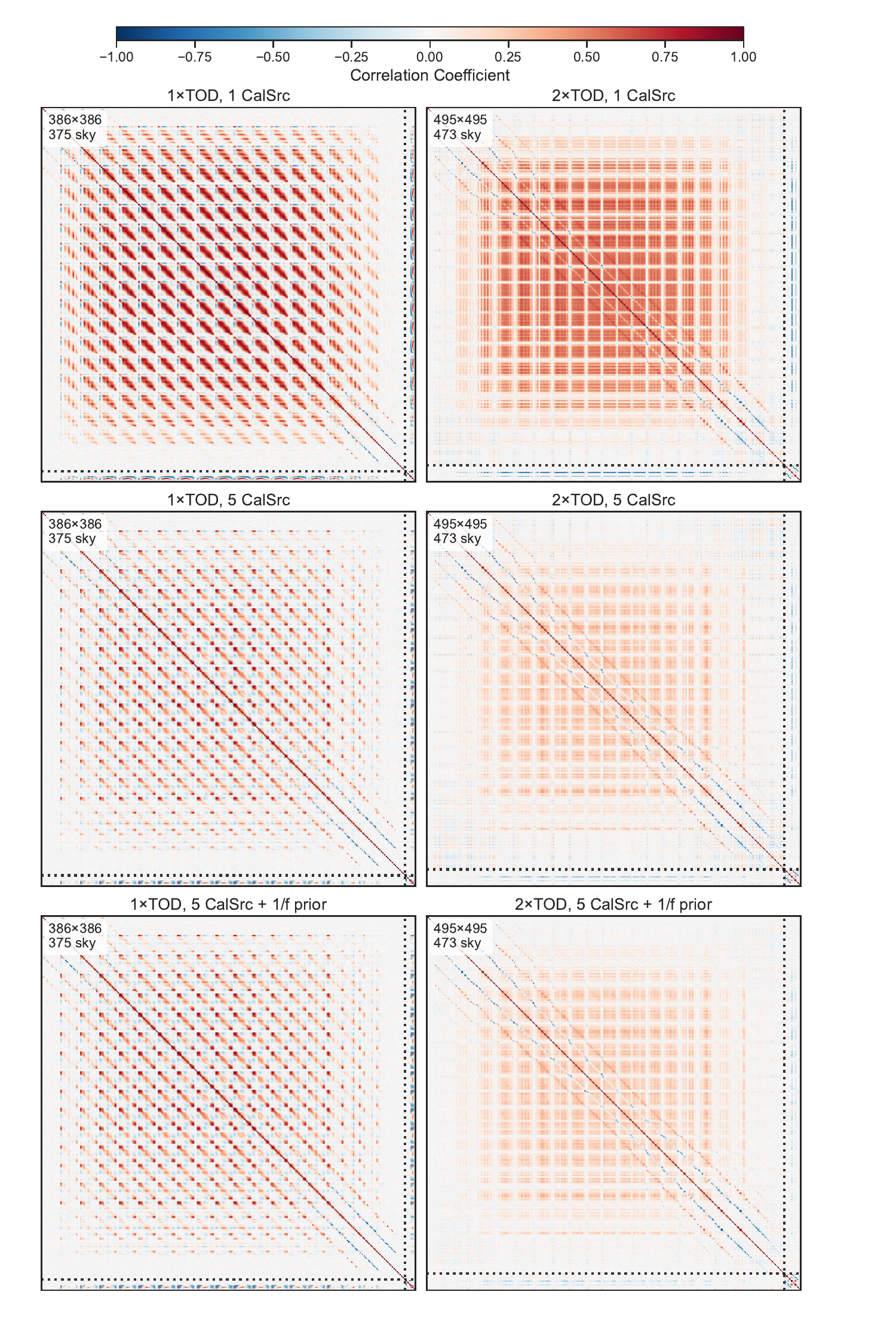}
    \vspace{-3em}
    \caption{
    Correlation matrices of sampled parameters for different experimental setups. Each matrix is organised such that the origin is at the top-left corner, and the parameters are concatenated in the following order: celestial sky map pixels ($p_{\mathrm{cel}}$), gain coefficients ($p_g$), local $\Tsys$ parameters ($p_{\mathrm{loc}}$), and $1/f$ noise parameters. A dotted line separates the sky parameters from the instrumental and noise parameters, partitioning each matrix into four blocks. The top-left block shows the sky–sky correlations, the bottom-right block shows correlations between the instrumental and noise components, and the off-diagonal rectangular blocks show the cross-correlations between the sky and instrumental parameters. 
    The total number of TOD-specific parameters -- comprising the gain coefficients, local system temperature components, and noise parameters -- scales linearly with the number of TOD observations. Consequently, the 1$\times$TOD configuration has 11 such parameters, while the 2$\times$TOD configuration has 22, since each additional TOD contributes an independent set of per-observation instrument parameters.
    }
    \label{fig: correlation matrix}
\end{figure*}

\begin{figure*}
    \centering
    \includegraphics[width=0.85\linewidth]{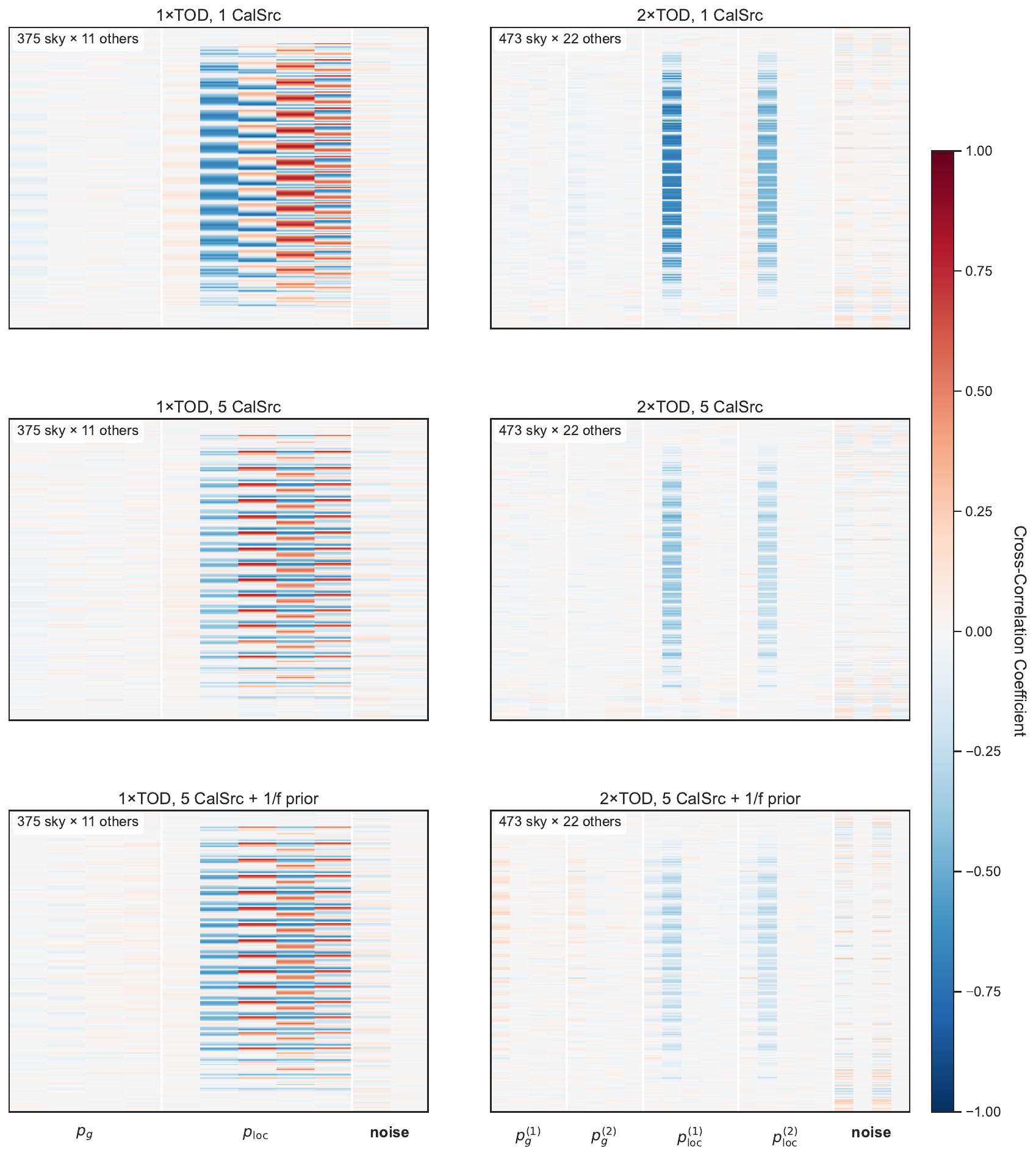}
    \caption{Cross-correlation coefficients between sky parameters and other model parameters across different analysis configurations. Each panel shows the correlation structure between sky parameters (vertical axis; each row is for a single pixel) and instrumental/noise parameters (horizontal axis), with white vertical lines delineating parameter groups: gain parameters ($p_g$), local system temperature parameters ($p_{\rm loc}$), and noise parameters ($\log_{10} f_0$, $\alpha$). The six panels compare single vs. double TOD analysis with varying numbers of calibration sources and prior constraints.
    }
    \label{fig: sky-others cross correlation}
\end{figure*}

Figures~\ref{fig: combined maps} show, for each scenario, the estimated sky map (computed as the sample average, representing the posterior mean of the sky parameters marginalised over all other parameters), the residual map (defined as the difference between the estimated and true sky maps), the uncertainty map (given by the sample variance), and
the Z-score map (defined as the residual normalised by the posterior standard deviation). For a well-calibrated posterior, $Z$ should follow a standard normal distribution $\mathcal{N}(0,1)$, with values significantly exceeding unity indicating either a biased estimate or underestimated uncertainty.
The stripy structure visible in the residual and Z-score maps (but absent from the uncertainty map) for the $1\times\mathrm{TOD}$ case reflects a systematic bias in the sky estimate for those pixels, likely arising from degeneracies between the sky signal and the gain or nuisance $T_{\rm sys}$ components; this is discussed further in Section~\ref{sec: 1TOD vs 2TOD}.

For comparison, Figure~\ref{fig: wiener-filer maps} shows the traditional Wiener and high-pass filter maps. We will discuss the comparison in detail in Section~\ref{sec: method comparison}.
We note that edge pixels are less constrained in both Bayesian and Wiener-filter map-making. While this can be formalised within the Fisher information framework, we offer an intuitive explanation here. The Fisher information coupling two sky pixels $p_1$ and $p_2$ is
\begin{equation}
\begin{split}
    \mathcal{I}(p_1, p_2)
    &= \mathbb{E}\!\left[\frac{\partial \ln \mathcal{L}}{\partial p_1}\frac{\partial \ln \mathcal{L}}{\partial p_2}\right] \\
    &= \sum_{i,j}\mathbb{E}\!\left[\left(\frac{\partial \ln \mathcal{L}}{\partial T_i^{\rm sys}}\right)
    \left(\frac{\partial \ln \mathcal{L}}{\partial T_j^{\rm sys}}\right)\right]
    \frac{\partial T^{\rm sys}_i}{\partial p_1}\frac{\partial T^{\rm sys}_j}{\partial p_2},
\end{split}
\end{equation}
where $\partial T_i^{\rm sys}/\partial p_a$ denotes the beam response of data point $i$ at pixel $p_a$. 
Edge pixels receive lower beam weight throughout the observation, so the corresponding diagonal elements of the Fisher information matrix are small, leading to larger posterior uncertainties. A rigorous treatment would require inverting the full Fisher information matrix, which is omitted in this work.

We further examine the bias in the sky map estimates.
The residual, defined as the posterior mean minus the true sky temperature, directly measures the pixel-level bias, and its distribution across the map is shown in Figure~\ref{fig: combined hist}. \footnote{In this paper, we present only the plots related to the reconstructed maps. Histogram plots of all nuisance parameters, along with the full posterior samples for all parameters across the six analyses, are available at: \href{https://github.com/zzhang0123/hydra-tod}{https://github.com/zzhang0123/hydra-tod}} 
To isolate the bias from the additional scatter introduced by edge-pixel uncertainty, the red histograms show only pixels common to both the 1$\times$TOD and 2$\times$TOD scan footprints; we refer to these as \textit{internal pixels}, which have comparatively small posterior uncertainties.
The figure reveals both the typical bias magnitude and any systematic offset across the six configurations.
A clear improvement is visible when comparing 2$\times$TOD to 1$\times$TOD results, which is discussed in detail in Section~\ref{sec: 1TOD vs 2TOD}; the effect of different prior choices is analysed in Section~\ref{sec: prior comparison}.

A key advantage of the Bayesian workflow is that ensemble sampling allows us to construct credible intervals for any observable. To illustrate this, we reconstruct the noise-free TOD $\mu_{\rm sample}$, defined as the product of the smoothed gain and system temperature, and examine its residuals with respect to the true noise-free TOD $\mu_{\rm true}$. 
Figure~\ref{fig: posterior predicted TOD} presents this comparison, with the shaded band indicating the posterior credible interval on $\mu_{\rm sample}$.
For clarity, we have removed the contribution from the noise diode\footnote{Note that the noise diode injection does not play any special role in the sampling mechanism itself, although in practice it may serve as a primary source of prior information about the gain.}, as its frequent injection every $20$ seconds introduces strong striping that obscures the visual structure in the plots. 
In this figure, the residual panel shows the difference between the true TOD and the reconstructed TOD. 
For the corresponding sky residuals at this level of TOD residuals, see row 4 (2×TOD; 1 CalSrc) in Figure~\ref{fig: combined maps}.

As an illustrative diagnostic of the posterior sampling, we interpret the residual between the posterior reconstruction and the truth shown in Figure~\ref{fig: posterior predicted TOD}. Specifically, we address the origin of the correlated noise feature visible in the residual, and ask what noise behaviour should be expected in a posterior reconstruction of the TOD.
We offer a concise intuitive explanation as follows. From our data model [see Eq.(\ref{eq: residual noise definition})], the quantity $n_{\rm sample} \equiv d/\mu_{\rm sample} - 1$ is expected to behave as Gaussian noise composed of white noise and $1/f$ noise. This expression can be rewritten as
\begin{equation}
n_{\rm sample} \simeq \frac{\mu_{\rm true} (1+n_{\rm true} ) }{\mu_{\rm sample} }- 1,
\end{equation}
from which we obtain
\begin{equation}
n_{\rm sample} \simeq
n_{\rm true}
-
\frac{\Delta\mu_{\rm sample}}{\mu_{\rm true}},
\label{eq: posterior noise}
\end{equation}
where $\Delta\mu_{\rm sample} \equiv \mu_{\rm sample} - \mu_{\rm true}$. Accordingly, the posterior noise PSD is expected to be statistically consistent with the true noise PSD. In Figure~\ref{fig: posterior predicted TOD PSD}, we compare $\langle\mathrm{PSD}(n_{\rm sample})\rangle$ with $\mathrm{PSD}(n_{\rm true})$ and confirm their consistency.
We observe that, at high wavenumbers, the posterior noise PSD and the true noise PSD appear to be identical. 
This behaviour follows directly from the structure of the TOD model. The linear basis for the system temperature comprises low-order Legendre polynomials for the nuisance component and temporal cross-beam functions evaluated per pixel. This basis set has an intrinsic smoothness lower bound, which manifests in Fourier space as collective suppression of power at high frequencies, as shown in the bottom panel of Figure~\ref{fig: posterior predicted TOD PSD}. Consequently, $\Delta\mu_{\rm sample}/\mu_{\rm true}$ does not contribute to the PSD at high frequencies, and $\mathrm{PSD}(n_{\rm sample})$ becomes dominated by $\mathrm{PSD}(n_{\rm true})$.

Another advantage of the Bayesian approach is its ability to reveal parameter correlations. In Figure~\ref{fig: correlation matrix}, we present and compare the correlation matrices across different experimental configurations. 
Note that, because we are indexing sky pixels embedded in the HEALPix ring scheme, pixels close in latitude are separated by approximately the length of the involved longitudes at that latitude. Consequently, we observe periodic off-diagonal strips indicating strong correlation.
For accurate map reconstruction, we focus particularly on the cross-correlations between sky parameters and all remaining (nuisance) parameters (See Figure~\ref{fig: sky-others cross correlation}, which is a zoomed-in version of the off-diagonal blocks in Figure~\ref{fig: correlation matrix}). Since the sky parameters are fundamentally independent of instrument or noise parameters, any observed correlations must arise from model degeneracies. We observe that by imposing informative priors and incorporating additional independent datasets, these degeneracies -- and hence the undesired correlations -- can be effectively mitigated. 

Figure~\ref{fig: sky-others cross correlation} shows some interesting collective structures that require interpretation.
The most interpretable feature is the first column of blue stripes, corresponding to the DC (zero frequency) mode of $T_\mathrm{rec}$: it reflects the anticorrelation between the receiver temperature offset and the sky signal, a direct manifestation of the $T_\mathrm{rec}$--$T_\mathrm{sky}$ degeneracy. The alternating correlation/anti-correlation (red/blue) pattern in the higher-order Legendre modes arises from the interplay between two orderings: the Legendre modes vary smoothly and change sign $n$ times along the scan time sequence, while the sky pixels are indexed in the HEALPix row-by-row scheme, which does not coincide with the scan order. As the scan sweeps back and forth across the field, the correlation between each Legendre mode and a given sky pixel oscillates in sign, and the mismatch between the temporal polynomial ordering and the spatial pixel ordering produces the fine periodic stripe pattern seen in the figure. 

Below, we present a comparative analysis of the map-making performance across different scenarios. The corresponding posterior statistics for all other nuisance parameters -- including the posterior means and 68\% credible intervals -- are provided in Appendix~\ref{append: nuisance results}. 
For the $1\times\mathrm{TOD}$ scenario, we present corner plots where space permits; for the $2\times\mathrm{TOD}$ scenario, we provide a summary table of posterior statistics.

\subsubsection{``$1\times\textrm{TOD}$'' vs ``$2\times\textrm{TOD}$''}
\label{sec: 1TOD vs 2TOD}
A clear observation is that the $2\times$TOD maps are significantly more accurate than those from the $1\times$TOD analysis. Both the error (or residual) and uncertainty maps exhibit a strong anti-correlation with the integrated beam intensity. This is expected: a stronger beam response to a given sky pixel implies that more information is collected from that direction, resulting in reduced uncertainty.

In both configurations, we observe that pixels near the edge of the covered sky patch consistently show higher uncertainty. This suggests that in practical scientific applications, it may be beneficial to exclude or down-weight these edge pixels from further analysis. 

Another notable point is the presence of large-scale structures in the error map of the $1\times$TOD case (e.g., extended blue or red regions in the first three residual maps in Figure~\ref{fig: combined maps}). These patterns indicate that the sky component in the TOD has inadvertently absorbed temporal functional behavior that does not belong to it -- possibly due to gain variations or other components of the system temperature.
The same conclusion can be drawn from a comparison of the two columns in Figure~\ref{fig: sky-others cross correlation} that the correlation between sky pixels and nuisance parameters is reduced in the $2\times$ TOD analysis.

This analysis also serves as a validation of our workflow's capability to perform joint inference over multiple TOD sets. In the $2\times$TOD case, the issue is mitigated, which will be discussed in more detail below.

\begin{figure*}
    \centering
    \includegraphics[width=\linewidth]{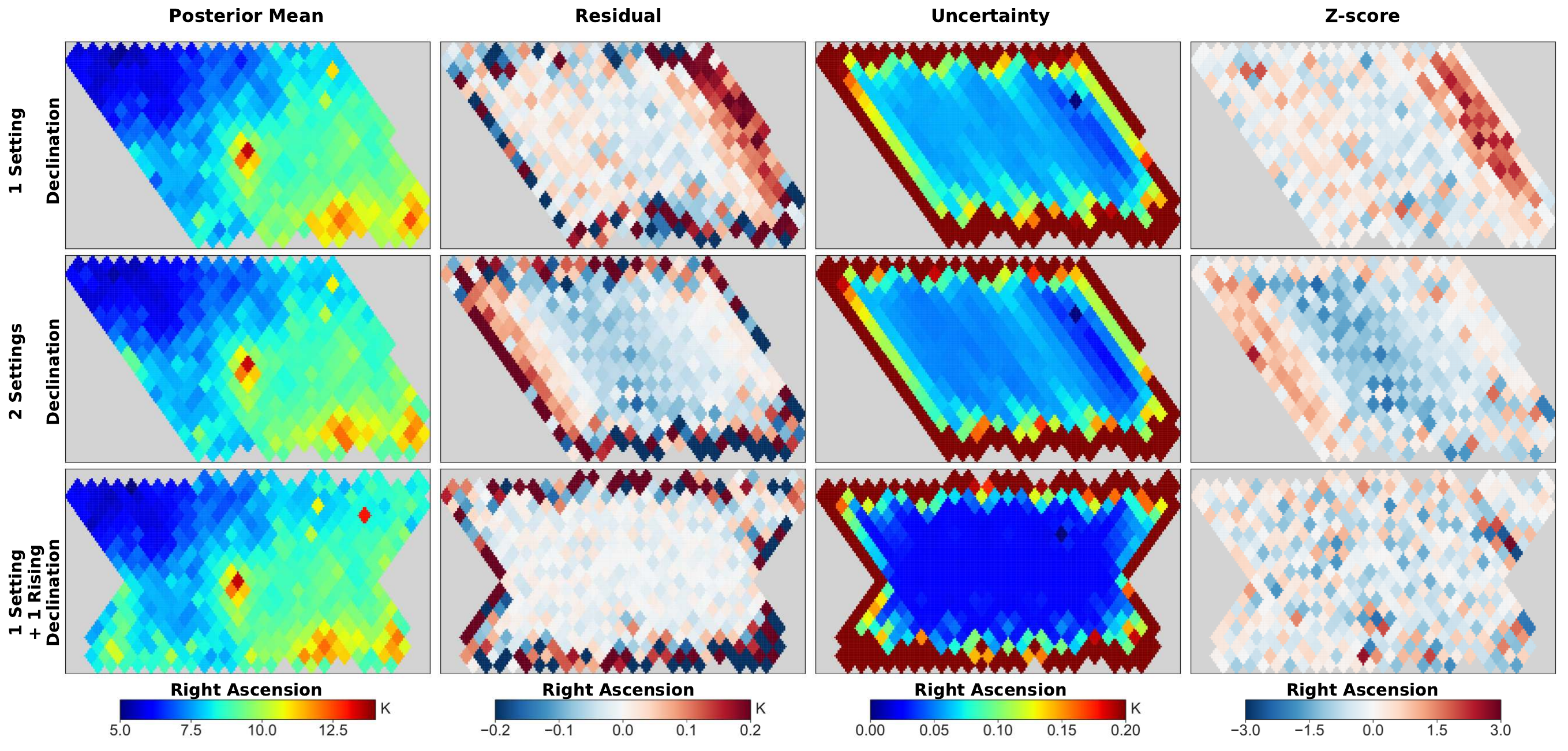}
    \caption{Sky map reconstructions for three scan configurations: \textit{1 setting} (single setting-mode TOD), \textit{2 setting} (two independent setting-mode TODs), and \textit{1 setting $+$ 1 rising} (one setting- and one rising-mode TOD).}
    \label{fig:map_comparison}
\end{figure*}

\paragraph*{What Caused the Improvement in the Map?}
The direct comparison between the $1\times\mathrm{TOD}$ and $2\times\mathrm{TOD}$ scenarios demonstrates that the latter yields a markedly improved sky map reconstruction. We now ask what factor underlies this improvement. In other words, does adding more data help simply by raising the signal-to-noise ratio, or does the new scan pattern, providing a genuinely different linear combination of sky degrees of freedom, play the decisive role?

To disentangle these two effects, we perform a controlled sequential comparison involving three mapping configurations. The first is the $1\times\mathrm{TOD}$ result already presented, hereafter referred to as \textit{1 setting}. The second uses two noise-independent realisations of the same scan in setting mode, denoted \textit{2 setting}; this configuration doubles the data volume while keeping the scan pattern fixed. The third is the $2\times\mathrm{TOD}$ configuration introduced earlier, combining one setting-mode and one rising-mode scan, denoted \textit{1 setting $+$ 1 rising}.

Figure~\ref{fig:map_comparison} compares the three configurations. The results show that it is the introduction of a new scan geometry, rather than a mere increase in signal-to-noise ratio, that is responsible for the map-making improvement. A systematic investigation of how scanning strategy shapes map-making performance and the underlying mechanisms is beyond the scope of this work. Here we offer a brief intuitive conjecture: observing the same patch of sky with different scan orientations likely induces different degeneracy structures between the nuisance parameters and the sky signal. Combining the two scans therefore suppresses certain systematic biases that neither scan alone can resolve.

\subsubsection{``1 CalSrc'' vs ``5 CalSrc'' vs ``5 CalSrc + $1/f$ prior''}
\label{sec: prior comparison}
We compare three progressively stronger priors: ``1 CalSrc'', ``5 CalSrc'', and ``5 CalSrc + $1/f$ prior''. The ``1 CalSrc'' case serves as the baseline, as at least one calibration source is always required to fix the flux scale of the map. The ``5 CalSrc'' scenario uses five spatially distributed calibration pixels. This setup is motivated by the idea that providing more large-scale information about the sky component in the TOD may help mitigate large-scale bias in the estimated map.
The ``5 CalSrc + $1/f$ prior'' case further incorporates a strong prior on the power-law index of the $1/f$ noise PSD. This reflects a realistic situation in which some knowledge about the power-law behavior of the $1/f$ component is often available. Here, we aim to test whether such information can be effectively leveraged to reduce map bias.

By comparing the residual maps in Figures~\ref{fig: combined maps}, we find that strengthening the priors does not lead to a significant improvement in the overall accuracy of the reconstructed sky map (also see different rows of Figure~\ref{fig: combined hist}).
In particular, although stronger priors, corresponding to a larger number of calibration sources (compare GS1, GS5, and GSF5), significantly reduce the uncertainty level in the 1$\times$TOD case, the residuals are largely insensitive to this change. This indicates that the reconstruction bias is not limited by the calibration constraining power we chose, but rather by the intrinsic gain--$T_\mathrm{sys}$ and $T_\mathrm{rec}$--$T_\mathrm{sky}$ degeneracies inherent to a fixed scanning direction, which the additional calibration sources we use cannot break. These degeneracies are instead mitigated in the 2$\times$TOD case, where the second independent scan direction provides different dependencies which help mitigate the bias.
However, Figure~\ref{fig: sky-others cross correlation} and \ref{fig: GS1 corner}-\ref{fig: GSF5 corner} show that stronger priors do help reduce the correlation between sky pixels and nuisance parameters. That said, the impact of prior strengthening is notably weaker than the effect of simply incorporating an additional TOD dataset.

\subsection{A Comparison with Linear Map-Making Using High-Pass and Wiener Filtering}
\label{sec: method comparison}

In this section, we present a simple, conventional map-making approach and compare its performance with our Bayesian method.

Using the same simulated dataset, we adopt a ``high-pass filter + Wiener filter'' scheme with the following assumptions:
\begin{enumerate}
    \item TOD is first calibrated using a perfectly known DC (direct-current, i.e. zero-frequency) gain mode.
    \item A high-pass filter is applied to remove large-scale fluctuations in the TOD. The cutoff frequency is 0.001 Hz, which is the knee frequency of the flicker noise in the simulated data. Therefore, we can ignore any residual correlated noise.
    \item The noise variance in the filtered data is estimated, and the sky map is reconstructed using a Wiener filter.
    \item The same $\Tsys$ priors are adopted for the Wiener filter, i.e., Gaussian priors with 10\% uncertainty.
\end{enumerate}	
The resulting map is shown in Figure~\ref{fig: wiener-filer maps}. Compared to the full Bayesian approach, this method exhibits larger residuals, particularly in the large-scale structures within the central survey region, as well as a noisier uncertainty map. Within the survey area, the Bayesian method uniformly shows low bias and reduced uncertainty.

In summary, although the Wiener filter uses exact knowledge of the DC gain and knee frequency, our Bayesian mapping method still outperforms it when the prior system temperature is the same.
A detailed comparison with this and other conventional map-making methods is beyond the scope of this work. This is an encouraging sign that our Bayesian approach is making good use of prior information and the model structure to enhance the recovery of the true temperature field from the same time-ordered data, however.

\section{Conclusions and Discussions}
\label{sec: conclusion}
In this paper, we present a workflow for joint Bayesian gain calibration, $1/f$ noise characterisation, and mapmaking.  
The primary goal of this workflow is to disentangle instrumental gain and noise contributions from the raw time-ordered data, thereby enabling the construction of a reliable intensity map that can serve as the basis for subsequent scientific analysis.

Designing a fully Bayesian workflow -- from model formulation to numerical implementation -- requires a series of principled decisions. To conclude, we highlight several such decisions and the rationale behind them.

A natural question is: \textbf{Why adopt a joint Bayesian approach in the first place?}
The motivation lies in the coupling nature of calibration and map-making. Treating these steps jointly allows us to propagate uncertainties coherently, account for degeneracies explicitly, and mitigate biases introduced by sequential processing (see Section~\ref{sec: method comparison} for a comparison with a traditional high-pass + Wiener filter approach). In particular, the Bayesian framework allows marginalisation over nuisance parameters (e.g., gain fluctuations), leading to more robust inferences about the sky signal, as well as detailed and explicit control over prior assumptions, such as the accuracy of point source calibration models, noise diode stability, etc. The priors also provide a natural way of incorporating external datasets, such as reference sky models.


One might also ask: \textbf{Why use a highly degenerate model with multiple nuisance parameters?}
The model’s degeneracy actually reflects genuine ambiguities in our knowledge of the system (typically, the artificial separation of large-scale gain variation and $1/f$ gain variation). We intentionally avoid ``reducing this degeneracy for the sake of numerical performance''. Instead, we rely on experimental constraints -- such as distinct temporal behaviors of different components -- to naturally break degeneracies. When this is insufficient, we explore the use of informative priors through numerical tests.
In the MeerKLASS-type ``rapid scan'' example presented in this paper, we demonstrate that moderate prior knowledge of the gain and sky structure is sufficient to break key degeneracies and recover accurate sky maps.

To make inference feasible in this high-dimensional setting, where there are 495 parameters in total in the $2\times$TOD setup, we introduced an \textbf{iterative Generalised Least Squares (GLS) sampler}, i.e., an iterative version of the Gaussian constrained realisations approach.
In the Gibbs sampling steps for the system temperature parameters and the smooth gain parameters, the iterative GLS sampler estimates the noise covariance matrix as the best-fit noise model to the data, conditioned on the current values of all other parameters. This estimated noise covariance is then used to sample the linear parameters in the model. In essence, the method provides a Gaussian approximation to the conditional posterior distribution, where noise and signal parameters are otherwise tightly coupled.


\vspace{0.5em}
In addition to the Bayesian considerations discussed above, we briefly outline several potential extensions to this work and highlight some of the caveats.

First, our current implementation focuses on a single-frequency workflow. 
It can be naturally extended to include multiple TOD chunks from multiple dishes at the same frequency. 
Such an approach would allow joint inference across dishes. 
The method could also be extended along the frequency axis without specifying the 1/f noise-noise covariance. 
For example, one could divide the full band into sub-bands, and assume that the 1/f noise parameters remain constant across those frequencies if the sub-bands are narrow enough.  
Two possible schemes arise in this context: (1) joint inference with shared 1/f noise parameters but otherwise independent parameter sets for each sub-band; or (2) one may perform averaging across frequency, thereby reducing the effective white noise level -- though this does not suppress the 1/f component itself -- before creating a map of the averaged sky.
In terms of numerical tractability, Scheme (2) is straightforward: treating each sub-band independently results in a parallel frequency extension with no additional computational cost per channel. Scheme (1) requires a joint sampling step for the shared $1/f$ noise parameters across all sub-bands. A general correlated treatment across the full band would increase the expense of inverting the full noise covariance. 
And, to render it tractable, one may have to use for example the cyclic approximation of the noise covariance to invert it effeiciently in Fourier space.

Another important consideration is polarisation leakage \citep[see e.g.][]{cunnington202121, nunhokee2017constraining}. While this work models only the total intensity sky, any practical intensity mapping experiment must contend with leakage from polarised sky signals. Accurately characterising the system temperature component from the full Stokes sky requires modelling the antenna $E$-field beam and synthesizing the polarised power beams onto the Stokes $Q/U/V$ basis (or equivalently, $E/B/V$) in the sky coordinate frame. Although circular polarisation $V$ is often negligible, the leakage from $Q/U$ can contaminate intensity measurements if unaccounted for. A full-Stokes formalism is therefore essential for precision intensity mapping. The inference framework developed in this work is fully compatible with full-Stokes map-making.
Extending to full Stokes is numerically tractable under the assumption that the beam patterns for all Stokes parameters are known accurately. The sky signal then enters the system temperature model as a set of independent linear coefficients (one for each pixel, frequency and Stokes parameter), leaving the structure of the Gibbs sampling steps unchanged. The computational cost increases moderately with the number of Stokes parameters included, scaling linearly for a sparse beam matrix in the conjugate gradient solver, while the noise and gain sampling steps remain unchanged.

Radio frequency interference (RFI) presents another caveat. We do not explicitly model RFI in this work. Unless RFI manifests in a form that can be absorbed into the smooth temporal basis used for $T_{\rm sys}$, additional modelling would be necessary, or alternatively, the contaminated TOD segments can be flagged. Flagging can be efficiently implemented via a diagonal selection matrix without increasing the computational cost of the quadratic term in the log-likelihood, meaning that the Gibbs steps for gain and $T_{\rm sys}$ remain unaffected. However, the modified noise covariance introduced by flagging can affect the computation of the log-determinant. 
For contiguous or near-contiguous flagged segments, Levinson–Durbin algorithms remain applicable. 
In more irregular or extensive flagging cases, one may resort to perturbative methods or brute-force matrix inversion. If the effective dimensionality of the TOD is significantly reduced post-flagging, direct inversion may still be computationally tractable.

A further conceptual dependency of our method lies in the scanning strategy. The MeerKLASS scanning strategy used in our tests induces a linear discrimination between TOD response to different parameters, for example, sky pixels versus elevation-dependent terms, allowing effective separation using simple linear bases. This structural property may not generalise to arbitrary scanning strategies. In more complex cases, where linear decompositions fail, we may need to adopt explicit, nonlinear models for the receiver temperature, ground spillover, etc.

Finally, we note that our gain model assumes linearity, whereas significant non-linearities in the gain are known to occur for MeerKAT. One potential extension would be to include a simple nonlinear correction term in the gain model. While this could preserve the sampling strategy for noise and linear gain parameters in principle, it would break the Gaussian structure required by the iterative GLS step used for system temperature sampling. In such cases, alternative high-dimensional samplers, such as Hamiltonian Monte Carlo, may be required.

Looking ahead, this approach holds promise for a wide range of applications. It may be particularly valuable for experiments such as C-BASS \citep{jones2018cbass}, L-BASS \citep{zerafa2025Lbass}, COMAP \citep{cleary2022comap}, and other single-dish intensity mapping or global 21cm surveys, where high-fidelity map-making and robust noise modelling are crucial. This is especially true in regimes dominated by temporally correlated systematics, or where explicit models for the systematics (like ground spillover) are unavailable or incomplete.

\section*{Acknowledgements}

We are grateful to Clive Dickinson, Keith Grainge, Stuart Harper, Marta Spinelli, and Jingying Wang for helpful discussions.

This result is part of a project that has received funding from the European Research Council (ERC) under the European Union's Horizon 2020 research and innovation programme (Grant agreement No. 948764; ZZ, PB). 
ZZ also acknowledges support from the RadioForegroundsPlus project HORIZON-CL4-2023-SPACE-01, GA 101135036.

We acknowledge use of the following software: 
{\tt matplotlib} \citep{matplotlib}, {\tt numpy} \citep{numpy}, {\tt mpmath} \citep{mpmath}, {\tt scipy} \citep{2020SciPy-NMeth}, and {\tt emcee} \citep{foreman2013emcee}. 

\section*{Data Availability}
All code, data, and Jupyter notebooks necessary to reproduce the results presented in this paper are available in the associated GitHub repository: \href{https://github.com/zzhang0123/hydra-tod}{https://github.com/zzhang0123/hydra-tod}.

\section*{Conflict of Interest}
The authors declare no conflict of interest.



\bibliographystyle{rasti}
\bibliography{main}




\appendix

\section{Normalised Power Spectral Density}
\label{Appendix: PSD}

The total energy of a discrete-time signal $x[n]$ with sampling frequency $f_s$ is given by:
\be
E = \sum_{n=1}^{N} |x[n]|^2 \Delta t
\ee
where $N$ is the total number of data points, and we have included $\Delta t \equiv 1/ f_s$ to match the units of energy, that is, power $\times$ time.
Parseval's theorem states that the total energy in the time domain equals the total energy in the frequency domain. For the Discrete Fourier Transform (DFT) of $x[n]$, denoted by $\Tilde{x}[k]$, we have
\be
E = \left(\frac{1}{N}\sum_{k=1}^{N} |\Tilde{x}[k]|^2 \right)\Delta t
\equiv
\sum_{k=1}^{N} P[k],
\ee
where we have defined the power of the $k$th DFT frequency bin
\be 
P[k]\equiv\frac{\Delta t}{N}|\Tilde{x}[k]|^2.
\ee 
Then the power spectral density (PSD), defined as the power per unit frequency, is given by
\be 
P(f_k)\equiv \frac{P[k]}{\Delta f}=\frac{\Delta t}{N \Delta f }|\Tilde{x}[k]|^2 = 
\frac{|\Tilde{x}[k]|^2 }{f_s^2 },
\ee 
where $f_s = 1/\Delta t$ is the sampling frequency and $\Delta f = f_s / N$ is the DFT frequency resolution. 

Here we reiterate that, for dimensional consistency, we refer to the mean-squared DFT coefficient as the bin power, while the quantity defined over the continuum is referred to as the power spectral density.

\subsection*{White noise}

Given a sequence of white noise
\be
    x[n]\sim \mathcal{N}(0, \sigma^2)
    \;
    \Rightarrow
    \;
    \Tilde{x}[k] \sim \mathcal{N}\left(0, \frac{N}{2}\sigma^2\right) + i \mathcal{N}\left(0, \frac{N}{2}\sigma^2\right),
\ee
the expectation value of the bin power is 
\be
\left\langle{P^{\rm (wn)}[k]}\right\rangle = \sigma^2 \Delta t .
\ee
Consequently, the mean PSD of white noise is
\be
\left\langle{P^{\rm (wn)}(f_k)}\right\rangle = N \sigma^2 (\Delta t)^2 .
\ee

\subsection*{Conventional flicker noise PSD model}
In the conventional DFT-diagonal $1/f$ noise model, assuming that the flicker noise has the same power density as the white noise at the knee frequency $f_{\rm knee}$,
the power of the $k$-th DFT frequency bin of the flicker noise is given by
\be
\left\langle{P^{(1/f)}[k]}\right\rangle = \sigma^2 \Delta t \left(
\frac{f_{\rm knee}}{f_k}\right)^\alpha,
\ee
and the mean PSD of the flicker noise is
\be
\left\langle{P^{(1/f)}(f_k)}\right\rangle = N \sigma^2 (\Delta t)^2 \left(
\frac{f_{\rm knee}}{f_k}\right)^\alpha.
\label{eq: traditional PSD}
\ee
\section{An example full gain model}
\label{appendix: a gain model}
In this work, we use Eqs.~(\ref{eq: parameterisation TildeG}-\ref{eq: flicker correlation function}) as the full gain variation model. 
Although this approach of a smooth $\LscaleG(t_a)$ plus a perfect $1/f$-type $\deltaG$ inherently involves some degree of modelling uncertainty, we expect that it does not preclude meaningful scientific interpretation. This is because radio experiments are typically designed to ensure that $\deltaG$ remains sufficiently small, so these fluctuations do not significantly impact the main experimental objectives over relevant timescales.
For example, the MeerKLASS scanning strategy is designed to cover the relevant angular scales within the stability time scale of the instrument when the gains are approximately constant. 
Even if the empirical model does not perfectly capture the gain variations, its ability to approximate the primary structure of the gain provides valuable utility in data calibration and analysis.

There are of course other models (and they might be even better); see Appendix~\ref{appendix: a non-pl gain model} for another possible full gain model.
These discussions show that the (smooth) gain and $1/f$ noise are ad-hoc definitions depending on the specific data model and full-gain model. Consequently, they are also implicit assumptions for likelihood analysis.
Note that if the stochastic gain fluctuation is an important systematic for scientific extraction, one might want to test different full gain models and perform model selection with Bayesian evidence.
\subsection{An example of deviating power-law gain variation}
\label{appendix: a non-pl gain model}
In this Appendix, we give an example where the $1/f$ model (with a single power law) may not be accurate for the full temporal frequency domain.

We consider a gain system of three-stage amplifiers, each given as a classical $1/f$ model:
\begin{subequations}
\begin{align}
G_1 &= \Bar{G}_1 (1 + \hat{\epsilon}_1) \\
G_2 &= \Bar{G}_2 (1 + \hat{\epsilon}_2) \\
G_3 &= \Bar{G}_3 (1 + \hat{\epsilon}_3) 
\end{align}    
\end{subequations}
where $\Bar{G}_i$ are constants and $\hat{\epsilon}_i$ are perfect, independent zero-mean $1/f$ noise.
$\Bar{G}_i$ thus also denote the mean gains.
The overall gain, to the first order of $\hat{\epsilon}_i$, is then given by
\be
G = G_1 G_2 G_3 \approx \Bar{G}_1 \Bar{G}_2 \Bar{G}_3 \left(1 + \hat{\epsilon}_1 + \hat{\epsilon}_2 + \hat{\epsilon}_3\right).
\ee
The overall stochastic gain fluctuation is given as the sum of three independent $1/f$ noise. The PSD of the total gain variation, as the sum of three different power laws, is not a perfect power law in the full temporal frequency domain.

\subsection{Gain models linearly distinguishing scales}

In the above case, a single power law is not sufficient to characterise the PSD behaviour. However, according to the ``selection rule'', steeper (negative) power laws dominate at lower frequencies, while flatter power laws dominate at higher frequencies.
Therefore, it would be possible to assume a threshold frequency $f_n$ beyond which the PSD is still well characterised by a single power law.
The scale distinction can be directly realized by linear separation of the Fourier modes of the gain variation according to the threshold frequency.
Then the large-scale modes are resolved while the small-scale modes are fitted to a power law statistical model.
Less abstractly, we rewrite the gain variation as
\begin{align}
    \hat{\epsilon}' &\equiv
    \hat{\epsilon}_1 + \hat{\epsilon}_2 + \hat{\epsilon}_3
    = \epsilon_L + \deltaG 
\end{align}
where $\Tilde{\epsilon}$ is the large-scale component composed of discrete Fourier modes with the lowest $n$ Fourier frequencies
\be
    \epsilon_L(t_a) =\frac{1}{N}\sum_{k=0}^{n-1} g_k \,e^{i 2\pi \frac{k}{N} a} 
    .
\ee
On the other hand, $\deltaG$ takes into account the residual gain variation, understood as the contribution of all the remaining Fourier modes, which is stochastic and characterised by the flicker noise with the PSD 
\be
    P_{\deltaG}(f)
    =
    \begin{cases}
        0, &  |f| < f_n\\
        \left(\frac{f_0}{|f|}\right)^\alpha, &   |f|\geq f_n 
    \end{cases}
\ee
where $f_0$ and $\alpha$ are the $1/f$ noise parameters and $f_n$ is the threshold DFT frequency at which we separate the large-scale and small-scale gain modes.

In summary, the full gain is represented by
\be 
G = \Bar{G}_1 \Bar{G}_2 \Bar{G}_3 \left(1 + \epsilon_L + \deltaG \right).
\ee
It is instructive to compare the above model with the model of Eq.~(\ref{eq: full gain model}): the former uses an additive term ($ \epsilon_L$) to account for the deviation from the perfect power law, while the latter uses a multiplicative term ($\Tilde{G}$) to account for the deviation from the perfect power law.
For an even more general gain model, one can assume a model that includes both an additive and a multiplicative term:
\be
G = \Tilde{G} \left(1 + \epsilon_L + \deltaG \right).
\ee

\section{Bias of Flicker Noise Parameter Estimation}
\label{sec: noise estimation bias}

\begin{figure}
    \centering
    \includegraphics[width=\linewidth]{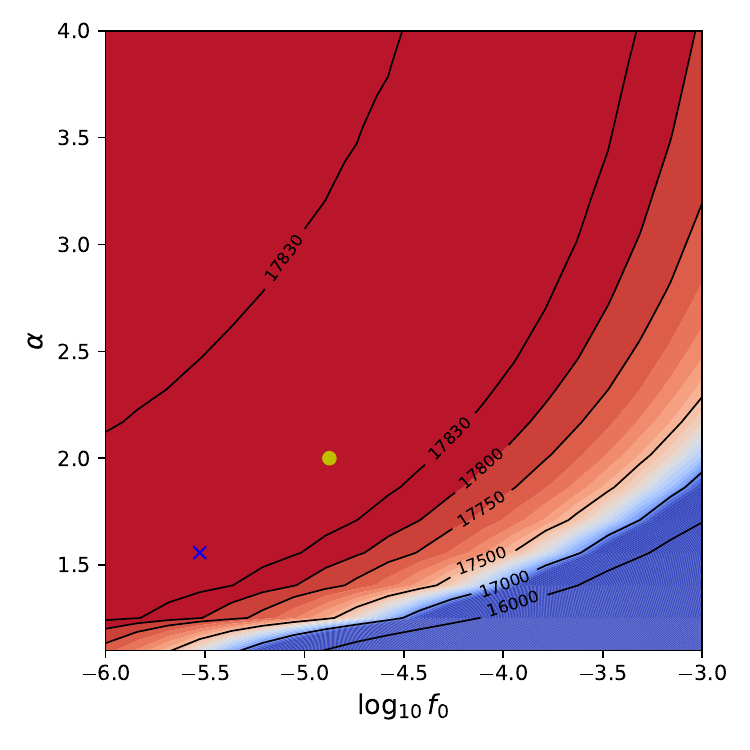}
    \caption{Heat map of the log-likelihood surface for the flicker noise parameters $\alpha$ and $\log_{10} f_0$, obtained from a simulated flicker noise sequence of length $3,000$ with true values $\alpha = 2$ and $\log_{10} f_0 = -4.875$. The marked points represent the true parameters (`o') and the maximum likelihood estimation (`x').}
    \label{fig: 2d flicker heatmap}
\end{figure}

\begin{figure}
    \centering
    \includegraphics[width=\linewidth]{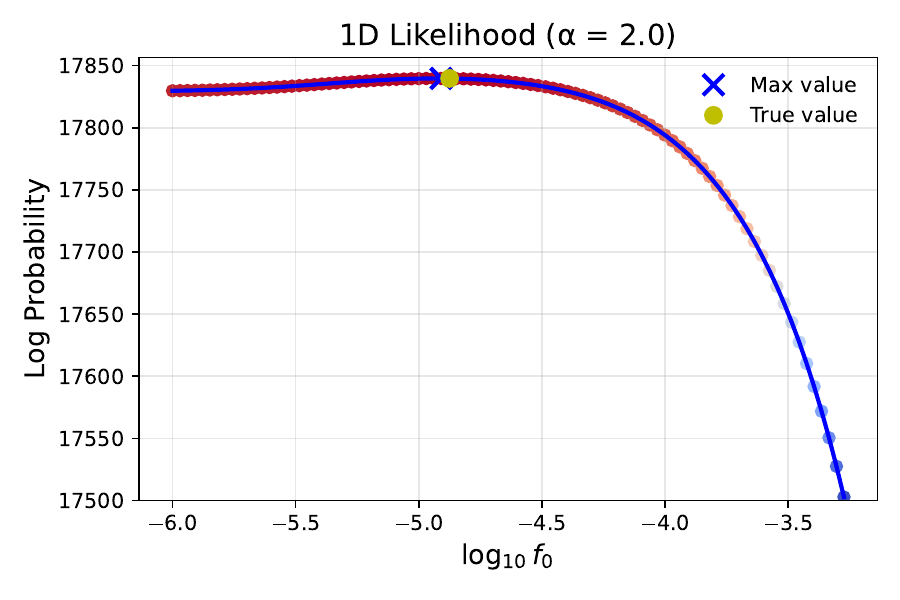}
    \caption{ Log-likelihood function of the flicker noise parameters $\log_{10} f_0$ for the fixed $\alpha=2$. It is obtained from a simulated flicker noise sequence of length $3,000$. }
    \label{fig: 1d flicker likelihood}
\end{figure}

Due to the limited length of the TOD samples and the strong nonlinearity of Eqs~(\ref{eq: 1/f noise correlation}) and (\ref{eq: likelihood function}), we expect the maximum likelihood estimation of the flicker noise parameters to be subject to finite-sample bias. 
To investigate this, we generated a flicker noise sequence of length $3,000$ and evaluated its log-likelihood as a function of $\log_{10}f_0$ and $\alpha$. The resulting two-dimensional heat map is shown in Figure~\ref{fig: 2d flicker heatmap}. There is a clear discrepancy between the true parameter values (marked ``o'') and the maximum-likelihood estimates (marked ``x''). However, if $\alpha$ is known, $\log_{10}f_0$ can be estimated with much smaller bias (see Figure~\ref{fig: 1d flicker likelihood}).

\section{Inverse and determinant of Noise Covariance}

In this section, we present two efficient methods for computing the inverse (or its associated quadratic form) and the determinant of the noise covariance matrix.
In Section~\ref{Appendix: numerical treatment Ncov}, we employ Levinson recursion to achieve an $\mathcal{O}(N^2)$ numerical implementation\footnote{\href{https://github.com/zzhang0123/comat}{https://github.com/zzhang0123/comat}}.
In Section~\ref{Appendix: perturbed matrix inverse and det}, we introduce an alternative approach that avoids explicit computation of the inverse or determinant. This method decomposes the noise covariance matrix into a dominant component, typically diagonal, and a perturbative component. By truncating the series expansion associated with the perturbed matrix operations, we can obtain a highly accurate approximation with reduced computational cost. 

It is important to note that the perturbative approach is only preferred when the off-diagonal elements (i.e., the $1/f$ noise covariance) are much smaller than the diagonal elements (which include the white noise variance and the $1/f$ autocorrelation). However, the method, when truncated at first order, offers significantly higher computational efficiency. Therefore, we also present it in this appendix for reference.

\subsection{Levinson Algorithm}
\label{Appendix: numerical treatment Ncov}

The noise covariance matrix $\Ncov$ is the sum of the white noise covariance matrix $\NCov{w}$ and the gain variation covariance matrix $\NCov{\text{corr}}$. 
$\NCov{w}$ is a simple constant diagonal matrix [see Eq.~(\ref{eq: white noise})], while $\NCov{\text{corr}}$ is a non-diagonal but diagonal-constant matrix
\be
\label{eq: explicit form of Ncorr}
\NCov{\text{corr}}
=
\begin{pmatrix}
    \xi_0 & \xi_1 & \xi_2 & \cdots & \xi_{n-1} \\
    \xi_1 & \xi_0 & \xi_1 & \cdots & \xi_{n-2} \\
    \xi_2 & \xi_1 & \xi_0 & \cdots & \xi_{n-3} \\
    \vdots & \vdots & \vdots & \ddots & \vdots \\
    \xi_{n-1} & \xi_{n-2} & \xi_{n-3} & \cdots &\xi_0
\end{pmatrix},
\ee
where $\xi_i = \xi(|t_a - t_{a'}|)$ for any $a, a'$ which satisfies $|a'-a|=i$. The correlation function $\xi(|t_a - t_{a'}|)$ is given by Eq.~(\ref{eq: 1/f noise correlation}).
The structured matrix of type Eq.~(\ref{eq: explicit form of Ncorr}) is known as a symmetric Toeplitz matrix.
The linear system given by the Toeplitz matrix can be solved quickly with $O(n^2)$ time complexity using the Levinson-Durbin algorithm.
We find that the quadratic form can be transformed to a linear-solve problem up to the inner product operation.
We also note that the same recursion can be used to compute the determinant by realising
\begin{align*}
    \Gamma & =
    \begin{pmatrix}
        A & B \\
        C & D 
    \end{pmatrix},
    &
    \text{det}(\Gamma)
    &=
    \text{det}(A) \text{det}(D - CA^{-1}B),
\end{align*}
which share most of the intermediate steps with the quadratic form.
Taking advantage of this, we wrote a dedicated code to compute the log-likelihood, namely {\tt comat.logdet\_quad}.
The numerical workflow is explained in Algorithm~\ref{alg:log_like}, where the sub-loops are designed in a way that makes the best use of the {\tt openmp}.

In summary, {\tt comat.logdet\_quad} is a fast, scalable code for likelihood analysis when the parameterised noise covariance is non-diagonal but has a symmetric Toeplitz structure.

\begin{algorithm}
\caption{An algorithm for fast computation of the log-likelihood. It is adapted from the Levinson recursion algorithm in \citet{golub2013matrix}. }\label{alg:log_like}
\begin{algorithmic}
\State \[
\textbf{Input: } a = [\xi_0, \xi_1, \dots, \xi_{n-1}], \quad b = [d'_0, d'_1, \dots, d'_{n-1}]
\]

\State Initialisation: 
\State
\(
r_i  = \frac{\xi_{i+1}}{\xi_0}, \, i = 1, \dots, n-1 ,   \)
\Comment{Normalise the matrix.}
\State
\(
x_0 = b_0 ,  \; y_0 = 1 ,  \;
\beta = 1 ,  \;
\alpha = -r_0 ,  
\)
\State
\(
\text{logdet} = 0  
\)
\Comment{Initialise the logarithmic determinant.}

\State
\For{\( k = 1 \) to \( n-1 \)}
    \State \( \beta \gets \frac{\beta}{1 - \alpha^2} \)
    \State
    \State
    \(
    y[j] \gets y[j] + \alpha y[k-j-2], \quad \forall \, j < k-1
    \)
    \Comment{Sub-loop 1}
    \State \( y[k-1] \gets \alpha \)
    \State
    \State 
    \(
    \mu \gets \left(b_k - \sum_{j=0}^{k-1} r[j] x[k-j-1]\right) \times \beta
    \)
    \State 
    \(
    \text{logdet} \gets \text{logdet} + \log\left(1 + \sum_{j=0}^{k-1} r[j] y[j]\right)
    \)
    \State
    \State 
    \(
    x[j] \gets x[j] + \mu y[k-j-1], \quad \forall \, j < k
    \)
    \State \( x[k] \gets \mu \)
    \State
    \State 
    \(
    \alpha \gets -\left(r[k] + \sum_{j=0}^{k-1} r[j] y[k-j-1]\right) \times \beta
    \)
\EndFor
\State 
\State
\(
\text{logdet} \gets \text{logdet} + n \log(a_0)
\)
\Comment{Final result for \(\ln{[\det{(\mathbf{N})}]}\)}
\State
\(
\text{quad} \gets (x^\top b) /{a_0}
\)
\Comment{Final result for \(\Tr(\mathbf{N}^{-1}\mathbf{D})\)}
\State
\State \textbf{Output: } \text{logdet} + \text{quad} \Comment{The log-likelihood [see Eq.~(\ref{eq: likelihood function})]}
\end{algorithmic}
\end{algorithm}

\subsection{Perturbed Matrix Operations}
\label{Appendix: perturbed matrix inverse and det}

In this paper, we often need to calculate the inverse and determinant of matrices of this form
\be
\mathbf{\Sigma} = \mathbf{\Lambda} + \mathbf{P}
\ee
where $\mathbf{P}$ is considered as a small perturbation to $\mathbf{\Lambda}$.
Quantitatively, this perturbation scenario requires that $|| \mathbf{P} \mathbf{\Lambda}^{-1}||<1$ in any norm convention.
In this section, we provide details on how to expand the inverse and determinant into power series of $\mathbf{P}$. 

\subsubsection{Inverse of perturbed matrix}

Using the Sherman–Morrison–Woodbury formula, we have
\be
\label{eq: woodbury}
(\mathbf{\Lambda} + \mathbf{P})^{-1}
=
\mathbf{\Lambda}^{-1}
-
\mathbf{\Lambda}^{-1}
\mathbf{P}\left(\mathbf{I} + \mathbf{P} \mathbf{\Lambda}^{-1}\right)^{-1}\mathbf{\Lambda}^{-1}
\ee
which separates the inverse into contributions from $\mathbf{\Lambda}$ and $\mathbf{P}$.
The term $\left(\mathbf{I} + \mathbf{P} \mathbf{\Lambda}^{-1}\right)^{-1}$
accounts for the effect of the perturbation. 
Realizing that $\mathbf{\Tilde{P}} = \mathbf{P} \mathbf{\Lambda}^{-1}$ is also a perturbation matrix with a small norm, we can expand the inverse matrix into the Neumann series
\be
\label{eq: Neumann series}
\left(\mathbf{I} + \mathbf{\Tilde{P}} \right)^{-1}
= 
\sum_{k=0}^{\infty} \left(-\mathbf{\Tilde{P}}\right)^k.
\ee
Substituting Eq.~(\ref{eq: Neumann series}) into Eq.~(\ref{eq: woodbury}), the inverse matrix up to the $m$th order of $\mathbf{P}$ is given by
\be
\label{eq: expandsion of the inverse}
(\mathbf{\Lambda} + \mathbf{P})^{-1}
\simeq
\mathbf{\Lambda}^{-1}
-
\sum_{k=0}^{m-1}
\mathbf{\Lambda}^{-1}
\mathbf{P} \left[-\mathbf{P} \mathbf{\Lambda}^{-1}\right]^{k}
\mathbf{\Lambda}^{-1}.
\ee
\balance

\subsubsection{Determinant of perturbed matrix}

Before computing $\det(\mathbf{\Sigma})$, we recall two useful properties of matrix operators. 
First, the determinant of a general matrix $\mathbf{\mathbf{\Sigma}}$ satisfies the following property:
\be
\ln \left[\det(\mathbf{A})\right]
=
\Tr\left[\ln(\mathbf{A}) \right].
\ee
And second, for $\mathbf{A}=\mathbf{I}+\mathbf{B}$ with small $\mathbf{B}$, we can expand $\ln(\mathbf{A})$ into a Taylor series
\be
\ln(\mathbf{I}+\mathbf{B})
=
\sum_{k=0}^{\infty} \frac{{(-1)}^k \mathbf{B}^{k+1}}{k+1}
\ee
Now we are ready to compute the determinant of $\mathbf{\Sigma}$. We start by factoring out $\mathbf{\Lambda}$ so that $\mathbf{\Sigma} = \left(\mathbf{I} + \mathbf{P}\mathbf{\Lambda}^{-1}\right)\mathbf{\Lambda}$. Then the logarithmic determinant of $\mathbf{\Sigma}$ is given by
\be
\begin{split}
    \ln[\det(\mathbf{\Sigma})]
    &=
    \ln[\det(\mathbf{I} + \mathbf{P}\mathbf{\Lambda}^{-1} )] + \ln[\det(\mathbf{\Lambda})]  \\
    &= \sum_{k=0}^{\infty} \frac{{(-1)}^k }{k+1} \Tr\left[\left(\mathbf{P}\mathbf{\Lambda}^{-1} \right)^{k+1}\right] + \ln[\det(\mathbf{\Lambda})] ,
\end{split}
\ee
Truncating up the $m$th order with respect to $\mathbf{P}$, the logarithmic determinant is given by
\be
\label{eq: expandsion of the determinant}
    \ln[\det(\mathbf{\Sigma})]
    \approx
    \sum_{k=0}^{m-1} \frac{{(-1)}^k }{k+1} \Tr\left[\left(\mathbf{P}\mathbf{\Lambda}^{-1} \right)^{k+1}\right] + \ln[\det(\mathbf{\Lambda})] .
\ee

\section{Numerical specifications of the Gibbs sampler}
\label{sec: sampler numerical spec}

This appendix describes the numerical setups for the samplers.

\subsection{MCMC 1/f Sampler Setup} 
To sample the two noise parameters, we employ an ensemble MCMC sampler using the \texttt{emcee} package with 6 walkers. The sampler runs iteratively for up to 5 rounds of a specified number of steps (default 200 per walker per round), with each successive round continuing from the final walker positions of the previous round, accumulating chain length until convergence is achieved. After each round, burn-in and thinning parameters are adaptively determined based on the estimated autocorrelation time $\tau$ from the entire accumulated chain: burn-in is set to $3 \times \max(\tau)$, and thinning to $0.5 \times \min(\tau)$. The algorithm terminates early when sufficient chain length is achieved relative to the autocorrelation time. The sampler supports optional Jeffreys priors computed numerically using the Hessian of the log-likelihood function, and returns either the final sample from the chain (for single samples) or the maximum (conditional) posterior sample from the post-burn-in chain (for parameter estimation mode).

The analysis presented in this paper shows that this 5-round setup is sufficient to ensure that the burn-in length is much shorter than the total length.

\subsection{Iterative GLS Sampler}
We implement an iterative generalised least squares (GLS) solver for heteroskedastic noise models using MPI parallelisation (see Section~\ref{sec: ite GLS sampler} for detailed discussion). The algorithm addresses data models of the form $\mathbf{d} = \mathbf{U}\mathbf{p}(1 + \mathbf{n})$, where $\mathbf{n}$ represents Gaussian noise with covariance $\mathbf{N}$. Starting from an ordinary least squares initialisation, the solver iteratively refines parameter estimates by updating the heteroskedastic noise covariance structure $\boldsymbol{\Sigma}_\varepsilon = \text{diag}(\mathbf{U}\mathbf{p})\mathbf{N}\text{diag}(\mathbf{U}\mathbf{p})$ at each iteration. 

Convergence is determined by monitoring the fractional norm error between successive parameter estimates, $\|\mathbf{p}^{(k+1)} - \mathbf{p}^{(k)}\|/\|\mathbf{p}^{(k)}\|$, against a configurable tolerance (default $10^{-10}$). The algorithm enforces a minimum of 5 iterations and terminates either when the convergence criterion is satisfied or upon reaching the maximum of 100 iterations, reporting convergence diagnostics in the latter case.
The MPI implementation distributes data across processes and employs collective operations (\texttt{MPI\_Reduce}, \texttt{MPI\_Bcast}) to compute the global normal equations $\mathbf{U}^T \boldsymbol{\Sigma}_\varepsilon^{-1} \mathbf{U} \mathbf{p} = \mathbf{U}^T \boldsymbol{\Sigma}_\varepsilon^{-1} \mathbf{d}$ and solve the resulting linear system. The solver supports multiple linear algebra backends (conjugate gradient, direct solvers) and provides diagnostic output, including matrix rank and condition number assessments. Upon convergence, the algorithm returns the final parameter estimates along with the precomputed matrices required for efficient posterior sampling.

In practice, we find that the gain coefficients converge rapidly to the specified tolerance, typically within 5 iterations. However, for the higher-dimensional system temperature parameters, the algorithm usually reaches the maximum iteration limit with final convergence metrics ranging between $10^{-5}$ and $10^{-2}$, indicating slower but still meaningful convergence for these more complex parameter spaces.

\subsection{Gibbs Iterations}

\begin{figure}
    \centering
    \includegraphics[width=\linewidth]{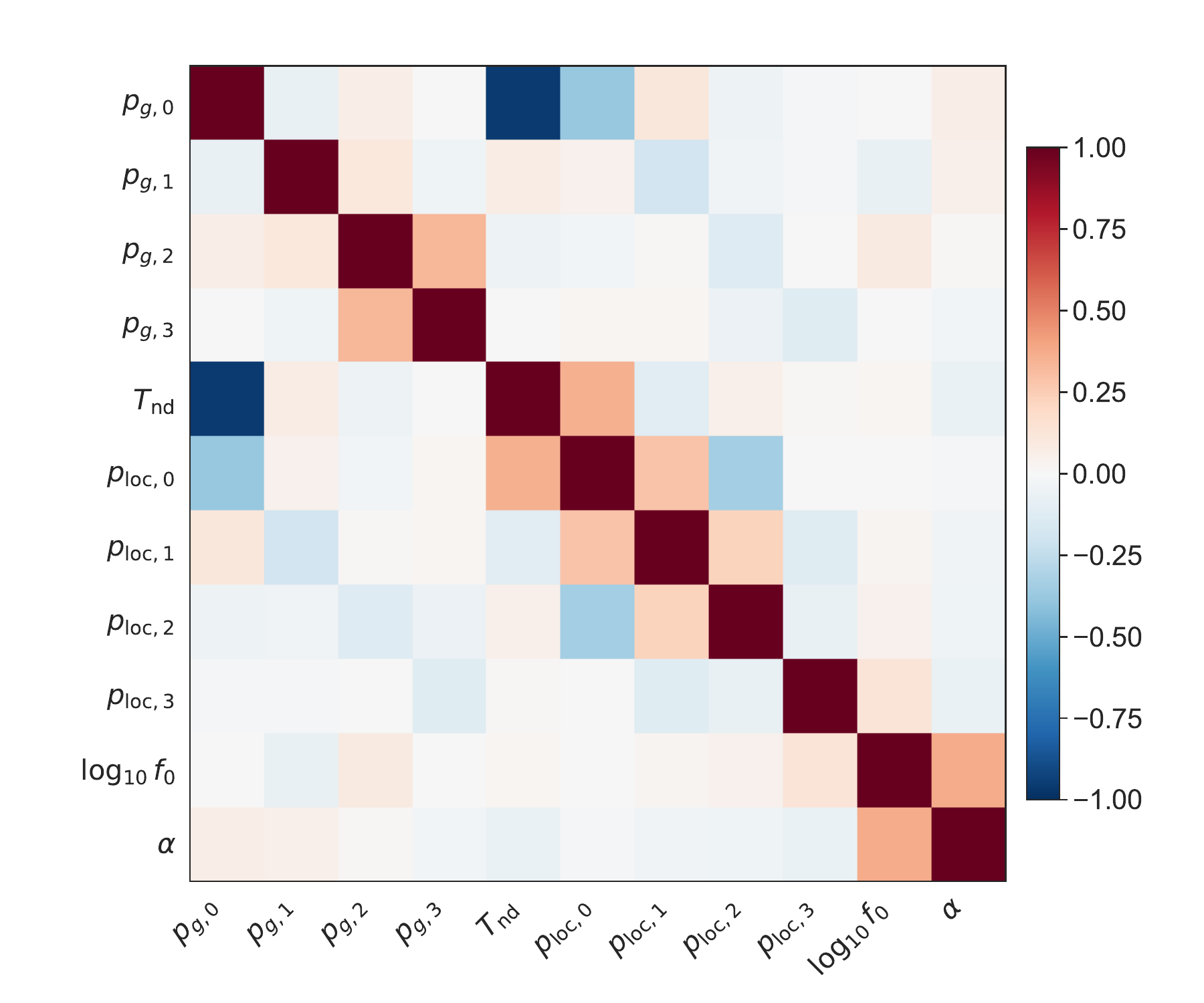}
    \caption{Correlation matrix of the nuisance parameters for the ``1×TOD; 1 CalSrc'' configuration.}
    \label{fig: GS1 nuisance corr}
\end{figure}

In this work, the initial parameter values are close to the true values, so no warm-up phase is required. For the scenario considered, $2,000$ samples are sufficient to obtain stable sky samples. Most nuisance parameters also show good convergence and stability, except for a subset of strongly correlated nuisance parameters (DC gain mode, noise diode temperature, DC receiver temperature; for an example, see Figure~\ref{fig: GS1 nuisance corr}, which shows correlation of nuisance parameters of the ``1$\times$TOD; 1 CalSrc''), which exhibit a large sampling correlation length.
Nevertheless, this does not affect map-making with multiple TODs, since these correlations occur only among the nuisance parameters and do not significantly interact with the sky parameters. This can be inferred from the off-diagonal blocks in Figure~\ref{fig: correlation matrix}. 
A more formal validation is provided by dividing the Gibbs iterations into four equal segments and comparing the residual sky maps from each segment. The high level of consistency confirms the stability of the sky estimation.


\section{Posterior means and credible intervals for the nuisance parameters}
\label{append: nuisance results}

\begin{table}
\centering
\caption{Posterior means and 68\% credible intervals for $2\times$TOD scenarios}
\label{table: posterior nuisance 2TOD}
\begin{tabular}{l|cccc}
\hline
Parameter & True & 1 CalSrc & 5 CalSrc  & 5 CalSrc + $1/f$   \\
\hline
$p_{\text{g},0}^{(1)}$ & $6.312$ & $6.315_{-0.003}^{+0.004} $ & $ 6.319_{-0.003}^{+0.003} $ & $6.317_{-0.010}^{+0.010}$ \\
$p_{\text{g},1}^{(1)}$ & $0.420$ & $0.421_{-0.003}^{+0.003}$ & $0.422_{-0.003}^{+0.003} $ & $0.421_{-0.003}^{+0.003}$ \\
$p_{\text{g},2}^{(1)}$& $0.264$ & $0.262_{-0.004}^{+0.004}$ & $0.262_{-0.004}^{+0.004} $ & $0.263_{-0.004}^{+0.004}$ \\
$p_{\text{g},3}^{(1)}$ & $0.056$ & $0.061_{-0.005}^{+0.004} $ & $0.062_{-0.004}^{+0.004} $ & $ 0.061_{-0.005}^{+0.004} $ \\
$p_{\text{g},0}^{(2)}$ & $6.85$ & $6.852_{-0.010}^{+0.008} $ & $6.848_{-0.006}^{+0.009}$ & $ 6.866_{-0.010}^{+0.009} $ \\
$p_{\text{g},1}^{(2)}$ & $0.142$ & $0.142_{-0.003}^{+0.004} $ & $0.141_{-0.004}^{+0.003} $ & $0.143_{-0.004}^{+0.004} $ \\
$p_{\text{g},2}^{(2)}$ & $0.744$ & $ 0.746_{-0.004}^{+0.004} $ & $0.747_{-0.004}^{+0.004}$ & $0.749_{-0.005}^{+0.005}$ \\
$p_{\text{g},3}^{(2)}$ & $0.779$ & $ 0.781_{-0.005}^{+0.006}$ & $ 0.779_{-0.006}^{+0.006} $ & $0.784_{-0.005}^{+0.005}$ \\
$T_{\rm nd}^{(1)}$ &$15.0$ & $14.994_{-0.010}^{+0.009}$ & $14.984_{-0.009}^{+0.009}$ & $14.988_{-0.023}^{+0.023} $ \\
$p_{\text{loc},0}^{(1)}$ & $12.6$ & $ 12.600_{-0.031}^{+0.030} $ & $12.561_{-0.019}^{+0.020} $ & $12.571_{-0.044}^{+0.043} $ \\
$p_{\text{loc},1}^{(1)}$ & $0.5$ & $0.493_{-0.012}^{+0.012} $ & $0.489_{-0.012}^{+0.012} $ & $0..493_{-0.011}^{+0.012} $ \\
$p_{\text{loc},2}^{(1)}$ &$0.5$ & $0.515_{-0.017}^{+0.017} $ & $0.515_{-0.014}^{+0.016}$ & $0.513_{-0.016}^{+0.017} $ \\
$p_{\text{loc},3}^{(1)}$ &$0.5$ & $0.482_{-0.017}^{+0.019} $ & $0.481_{-0.018}^{+0.017} $ & $ 0.488_{-0.018}^{+0.018} $ \\
$T_{\rm nd}^{(2)}$ & $15.0$ & $14.985_{-0.018}^{+0.024} $ & $14.993_{-0.020}^{+0.013} $ & $14.953_{-0.019}^{+0.022} $ \\
$p_{\text{loc},0}^{(2)}$ & $12.6$& $12.587_{-0.041}^{+0.045} $ & $12.582_{-0.035}^{+0.027} 
$ & $12.513_{-0.038}^{+0.042} $ \\
$p_{\text{loc},1}^{(2)}$ & $0.5$& $0.504_{-0.013}^{+0.013} $ & $ 0.501_{-0.014}^{+0.014} $ & $ 0.502_{-0.013}^{+0.014} $ \\
$p_{\text{loc},2}^{(2)}$ & $0.5$& $ 0.493_{-0.016}^{+0.015} $ & $0.487_{-0.014}^{+0.014} $ & $0.487_{-0.017}^{+0.017} $ \\
$p_{\text{loc},3}^{(2)}$ & $0.5$& $0.508_{-0.020}^{+0.019}$ & $0.514_{-0.022}^{+0.021} $ & $0.505_{-0.019}^{+0.018} $ \\
$\log_{10} f_0^{(1)}$ & $-4.875$& $-4.956_{-0.668}^{+0.636}$ & $-4.893_{-0.682}^{+0.660}$ & $-4.956_{-0.200}^{+0.195} $ \\
$\alpha^{(1)}$ & $2.0$ & $2.511_{-0.665}^{+0.780}$ & $2.486_{-0.673}^{+0.788} $ & $2.000_{-0.002}^{+0.002} $ \\
$\log_{10} f_0^{(2)}$ & $-4.875$ & $-4.961_{-0.649}^{+0.613} $ & $-4.965_{-0.648}^{+0.619}$ & $-4.924_{-0.194}^{+0.202} $ \\
$\alpha^{(2)}$ & $2.0$ & $2.463_{-0.662}^{+0.788} $ & $2.420_{-0.693}^{+0.767}$ & $2.000_{-0.002}^{+0.002}$ \\
\hline
\end{tabular}
\end{table}

\begin{figure*}
    \centering
    \includegraphics[width=\linewidth]{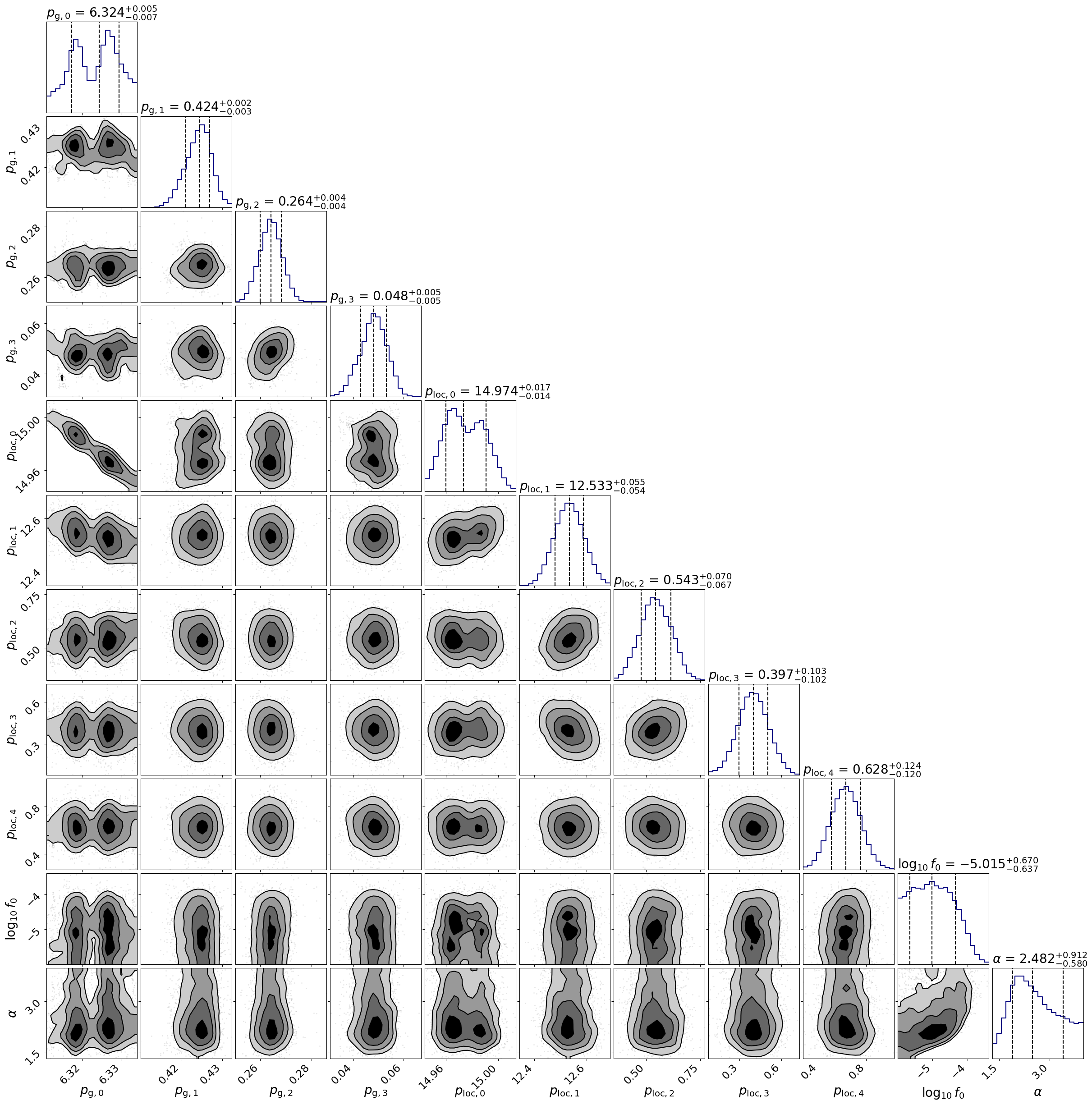}
    \caption{Corner plot of the nuisance parameters for the $1\times$TOD scenario with 1 CalSrc.}
    \label{fig: GS1 corner}
\end{figure*}

\begin{figure*}
    \centering
    \includegraphics[width=\linewidth]{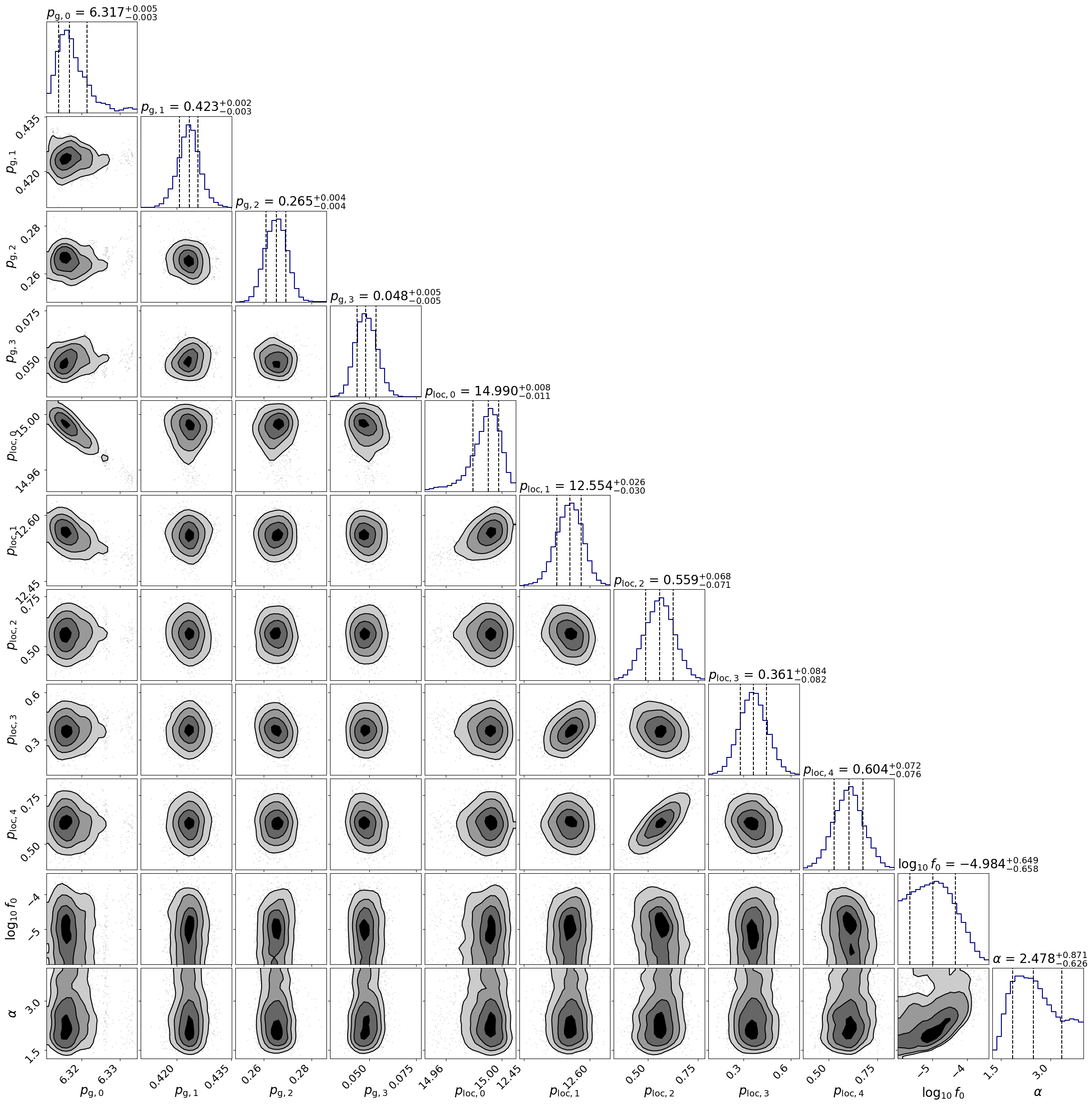}
    \caption{Corner plot of the nuisance parameters for the $1\times$TOD scenario with 5 CalSrc.}
    \label{fig: GS5 corner}
\end{figure*}

\begin{figure*}
    \centering
    \includegraphics[width=\linewidth]{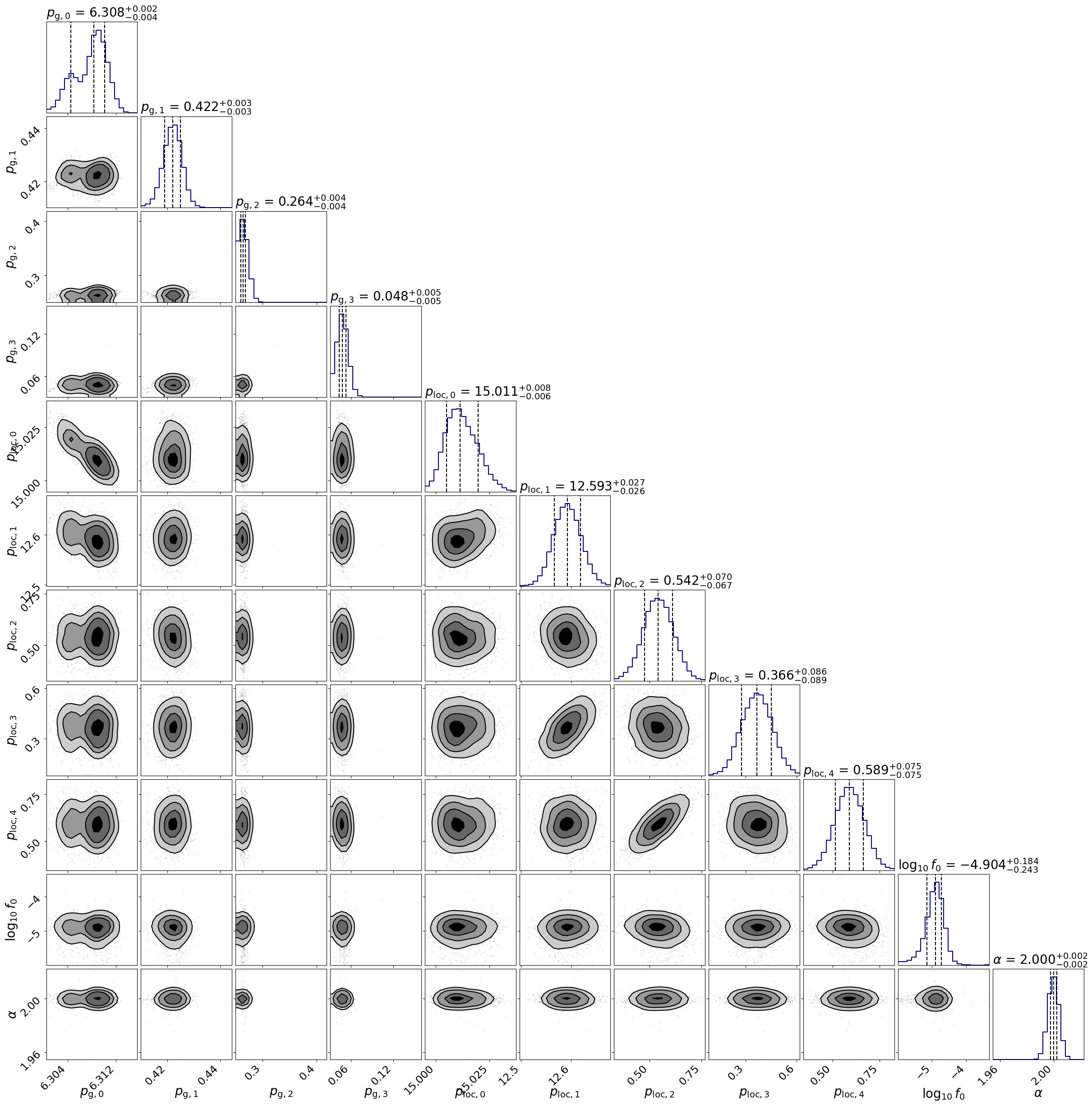}
    \caption{Corner plot of the nuisance parameters for the $1\times$TOD scenario with 5 CalSrc plus $1/f$ prior.}
    \label{fig: GSF5 corner}
\end{figure*}

In this appendix, we present the posterior means and 68\% credible intervals for the nuisance parameters across different experimental setups and prior assumptions. Figure~\ref{fig: GS1 corner}-\ref{fig: GSF5 corner} show the corner plots for the $1\times$TOD scenarios, while Table~\ref{table: posterior nuisance 2TOD} provides the corresponding results for the $2\times$TOD configurations.

\end{document}